\documentclass[12pt,preprint]{aastex}

\received{}
\revised{}
\accepted{}
\journalid{}{}
\articleid{}{}
\paperid{}
\cpright{}{}
\ccc{}

\newcommand{\beq}{\begin{equation}}
\newcommand{\eeq}{\end{equation}}
\newcommand{\tny}[1]{\mbox{\tiny #1}}
\newcommand{\dvr}{\mathbf{\nabla} \cdot}
\newcommand{\grad}{\mathbf{\nabla}}
\newcommand{\vvec}{\mathbf{v}}
\newcommand{\vect}[1]{\mathbf{#1}}
\newcommand{\vdot}{\mathbf{\cdot}}

\begin{document}

\title{Gamma-Rays from Intergalactic Shocks}

\author{Uri Keshet\altaffilmark{1}, Eli Waxman\altaffilmark{1}, 
Abraham Loeb\altaffilmark{2}, Volker Springel\altaffilmark{3} and 
Lars Hernquist\altaffilmark{2}}
\altaffiltext{1}{Department of Condensed Matter Physics, Weizmann
Institute, Rehovot 76100, Israel; waxman,keshet@wicc.weizmann.ac.il}
\altaffiltext{2}{Harvard-Smithsonian CFA, 60 Garden Street, Cambridge, 
MA 02138, USA}
\altaffiltext{3}{Max-Planck-Institut f\"{u}r Astrophysik, 
Karl-Schwarzschild-Stra\ss{}e 1, 85740 Garching bei M\"{u}nchen, Germany}

% --------------------------------------------------------------------------
% Version Notes:
% Corrected version after referee report. 
% Last updated - 24/10/02 by Uri. 
% --------------------------------------------------------------------------

% --------------------------------------------------------------------------
%                                Abstract 
% --------------------------------------------------------------------------

\begin{abstract}

Structure formation in the intergalactic medium (IGM) produces
large-scale, collisionless shock waves, where electrons can be accelerated
to highly relativistic energies.  Such electrons can Compton scatter
cosmic microwave background photons up to $\gamma$-ray energies.  We study
the radiation emitted in this process using a hydrodynamic cosmological
simulation of a $\Lambda \mbox{CDM}$ universe.  The resulting radiation,
extending beyond TeV energies, has roughly constant energy flux per decade
in photon energy, in agreement with the predictions of Loeb \& Waxman (2000).  
Assuming that a fraction $\xi_e=0.05$ of the shock thermal energy is
transferred to the population of accelerated relativistic electrons, 
as inferred from collisionless non-relativistic shocks in the interstellar
medium, we find that the energy flux of this radiation, $\epsilon^2
(dJ/d\epsilon) \simeq 50-160 \mbox{ eV cm}^{-2} \mbox{ s}^{-1} \mbox{
sr}^{-1}$, constitutes $\sim 10\%$ of the extragalactic $\gamma$-ray
background flux.  
The associated $\gamma$-ray point-sources are too faint to account for 
the $\sim 60$ unidentified EGRET $\gamma$-ray sources, 
but GLAST should detect and resolve several $\gamma$-ray sources associated 
with large-scale IGM structures for $\xi_e \simeq 0.03$, and many more 
sources for larger $\xi_e$. 
The intergalactic origin of the shock-induced radiation can be verified 
through a cross-correlation with, e.g., the galaxy distribution that traces 
the same structure. 
Its shock-origin may be tested by cross-correlating its properties with
radio synchrotron radiation, emitted as the same accelerated electrons
gyrate in post-shock magnetic fields.
We predict that GLAST and \v{C}herenkov telescopes such as MAGIC, VERITAS 
and HESS should resolve $\gamma$-rays from nearby (redshifts $z \la 0.01$) 
rich galaxy clusters, perhaps in the form of a $\sim 5-10\mbox{ Mpc}$ 
diameter ring-like emission tracing the cluster accretion shock, 
with luminous peaks at its intersections with galaxy filaments detectable 
even at $z \simeq 0.025$. 

\end{abstract}


% --------------------------------------------------------------------------
%                                Introduction
% --------------------------------------------------------------------------

\section{Introduction}
\label{sec:Introduction}

\indent

In the past three decades, clear evidence has emerged indicating the 
existence of an extragalactic $\gamma$-ray background (EGRB).  
The origin of this radiation, however, has remained highly speculative. 
The first unambiguous detection of isotropic extragalactic
$\gamma$-ray emission was obtained by the SAS-2 satellite at 1977.
Subsequent experiments, especially the EGRET instrument aboard the
\emph{Compton Gamma Ray Observatory} (CGRO), have confirmed the
existence of this radiation.  These measurements indicate
a generally isotropic (fluctuation amplitude $\la 20\%$ on $\ga 20^\circ$
scales) spectrum in the energy range 30 MeV-120 GeV, well fitted by a
single power law, with photon number density per energy interval
$dJ/d\epsilon \sim \epsilon^{-\left(2.10 \pm 0.03\right)}$
\cite{Sreekumar98}.  Known extragalactic sources, such as blazars,
account for less than 7\% of the EGRB, and unidentified blazars
probably contribute no more than 25\% of the radiation (Mukherjee \& Chiang 
2000; but for a different model see Stecker \& Salamon 2001).  
Thus, most of the EGRB flux originates from a source yet to be identified.

Recently, Loeb \& Waxman (2000) have proposed a model which attributes most
of the EGRB flux to emission from large-scale structure-forming regions in
the universe.  
Strong collisionless, non-relativistic shock waves characteristic of these 
regions accelerate particles to highly relativistic energies by the 
Fermi-acceleration mechanism, first discussed as sources of ultra-high energy 
cosmic rays by Norman, Melrose \& Achterberg (1995) and by 
Kang, Ryu \& Jones (1996). 
Electrons, accelerated in such shocks to highly relativistic energies 
with Lorentz factors up to $\gamma_{\tny{max}}\sim 10^7$,  
scatter a small fraction of the cosmic microwave background (CMB) photons up 
to $\gamma$-ray energies, as inverse-Compton radiation.  
The estimated energy flux from this process per decade in photon energy, 
$\epsilon^2 (dJ/d\epsilon) \approx 1.5 \mbox{ keV}
\mbox{ cm}^{-2} \mbox{ s}^{-1} \mbox{ sr}^{-1}$, is in good agreement
with the detected EGRB, for typical cosmological parameters, provided that
the shocked matter accumulated a present-day mass-average temperature
$T\sim 10^7 \mbox{ K}$, and that a fraction $\xi_e\approx0.05$ of the
associated thermal energy was transferred to the relativistic electrons.
In this model, nearby rich galaxy clusters are expected to be bright
$\gamma$-ray sources, as they host strong, dense shocks
\cite{LoebWaxman2000,Totani2000}.  

Kawasaki \& Totani (2001) report a correlation
between a sample of high-latitude steady unidentified EGRET sources and
pairs or groups of galaxy clusters, where strong accretion and merger
shocks are expected.  However, this sample is too small and the
correlation too weak for it to be statistically significant.
En\ss{}lin et al. (2001) suggest that various features of the asymmetric 
radio plasma ejected from the giant radio galaxy NGC 315 can provide the 
first indication of a structure-formation shock in a filament of galaxies. 
According to the above model, such a shock should be accompanied by 
$\gamma$-ray emission, yet to be detected. 

The model presented above has several interesting implications,
other than an identification of the origin of the EGRB: $\gamma$-ray
emission could be the first tracer of the undetected warm-hot intergalactic
medium (WHIM), gas at temperatures $10^5\mbox{ K} \la T \la 10^7 \mbox{ K}$ 
which is believed to contain a large fraction of the baryons in the low 
redshift universe (e.g. Dav\'e et al. 2001a).  
The angular fluctuations in the EGRB should, according to the model, 
reflect the underlying forming structure, and thus serve as a diagnostic of 
intergalactic shocks.  
The relativistic electrons emit synchrotron radiation, as they gyrate in
intergalactic magnetic fields.  The resulting radio radiation, although
weaker than the CMB, could have a fluctuating component dominating the CMB
fluctuations at low ($\nu \la 10 \mbox{ GHz}$) 
frequencies \cite{WaxmanLoeb2000}.
Inverse-Compton emission from shock-accelerated electrons in the hard X-ray
(HXR) and extreme ultra-violet (EUV) bands was recently studied by Miniati
et al. (2001).  They find that such emission could account for a small
fraction of the measured EUV excess, and for the entire HXR excess in a few
reported galaxy clusters, provided that the latter contain relatively
strong merger or accretion shocks.

In order to test this model, calibrate its free parameters and extract 
useful predictions, one must follow the evolution of structure in the 
intergalactic medium (IGM) in detail.  
The complex, non-linear processes and many scales involved, 
make this a difficult problem, rendering numerical simulations a powerful 
and indispensable tool.  In this paper, we apply 
cosmological simulations to the problem of non-thermal emission from 
shock--accelerated electrons, and demonstrate how various EGRB predictions 
can be efficiently extracted from them.  For this purpose,
we have developed the necessary tools to localize intergalactic shocks
in the simulations, inject the corresponding shock-accelerated electrons
into the gas, and calculate the resulting radiation.  We consider a
hydrodynamic simulation of a $\Lambda \mbox{CDM}$ universe, currently
considered the most successful theory of structure formation, which
describes a flat universe containing dark matter and presently dominated by
vacuum energy. 
The relativistic electrons are assumed to carry a fraction $\xi_e$ of the 
shock thermal energy. 
We adopt the value $\xi_e=0.05$, which we show is inferred from collisionless 
non-relativistic shocks in the interstellar medium. 

Overall, we find an emitted inverse-Compton spectrum well-described by a 
mildly broken power law, $dJ/d\epsilon \sim \epsilon^{-2.04}$ 
in the energy range $10 \mbox{ keV}-1 \mbox{ GeV}$ and 
$dJ/d\epsilon \sim \epsilon^{-2.13}$ 
in the energy range $10 \mbox{ GeV}- 5\mbox{ TeV}$, 
in good agreement with the prediction of Loeb \& Waxman (2000).  The
calculated energy flux (per decade in photon energy), $\epsilon^2
(dJ/d\epsilon) = 50-160 \mbox{ eV} \mbox{ cm}^{-2} \mbox{ s}^{-1}
\mbox{ sr}^{-1}$, constitutes $\sim 10\%$ of the EGRB flux and is smaller
by a factor of $\sim 10$ than estimated above, mainly due to the lower
present-day temperature reached by the simulation, 
$T_0\sim 4 \times 10^6 \mbox{ K}$,
and its slower heating rate.  We predict the existence of $\gamma$-ray
halos, associated with large-scale structure.  
Although too faint to account for the EGRET unidentified $\gamma$-ray sources, 
several of these halos should be detectable by future experiments, such as the 
\emph{Gamma-Ray Large-Area Space Telescope} (GLAST, planned
\footnote{See http://glast.gsfc.nasa.gov} to be launched in 2006), 
provided that $\xi_e \ga 0.03$. 
The number of detectable sources is sensitive to the fraction of 
shock-energy transferred to the relativistic electrons, 
with $\sim 15$ well-resolved GLAST sources predicted for $\xi_e = 0.05$. 
Detection of intergalactic shock-induced $\gamma$-ray sources may be used to 
calibrate the free parameter $\xi_e$. 
We examine images of such an object, as predicted to be produced by GLAST 
and by the \emph{Major Atmospheric Gamma-ray Imaging \v{C}herenkov} (MAGIC) 
telescope (expected \footnote{See http://hegra1.mppmu.mpg.de/MAGICWeb} to 
become operational during 2002). 
These images reveal an elliptic ring-like feature, 
of diameter corresponding to $5-10\mbox{ Mpc}$, surrounding a nearby 
($z\simeq 0.012$) rich galaxy cluster. 
This ring, tracing the cluster accretion shock, has luminous peaks along 
its circumference, probably indicating the locations of intersections with 
galaxy filaments, channeling gas into the cluster. 

We describe the Loeb-Waxman model in \S 2.  We first consider the gross
characteristics of intergalactic shocks based on order of magnitude
estimates. 
In this context, we parameterize the distribution of accelerated electrons, 
and discuss the inverse-Compton and synchrotron radiation they emit. 
The simulation we chose to study and its basic
predictions are described in \S 3.  
We discuss the underlying theory and cosmological model,
present some characteristics of the simulated universe and examine its
structure.  Section 4 describes how EGRB predictions can be extracted from
cosmological simulations in general and then focuses on the method applied
to the simulation discussed in \S 3.  In \S 5, we present our results
regarding the radiation predicted from the simulation.  We describe the
resulting inverse-Compton spectrum, analyze the corresponding $\gamma$-ray
sky maps and study the features and distribution of the simulated $\gamma$-ray 
sources.  
In \S 6 we discuss the implications of our results for future experiments and
models.  Appendix A contains some formulae used to calculate and integrate
the radiation emitted by the accelerated electrons.  Appendix B presents an
algorithm devised to identify point sources in our simulated maps of
the $\gamma$-ray sky.

% --------------------------------------------------------------------------
%                                 Theory
% --------------------------------------------------------------------------

\section{Theory}
\label{sec:GRBPhenomenologyModel}

\indent

This section describes the model for non-thermal emission from
structure formation shocks proposed by Loeb and Waxman (2000), using
simple order-of-magnitude estimates.  First, we estimate the
parameters of the large-scale shock waves involved in structure
formation.  
Next, we discuss the high-energy electrons accelerated by such shocks. 
We evaluate the power index and cutoff of their distribution, and evaluate 
its normalization by estimating the efficiency of electron acceleration in 
intergalactic shocks.  
We conclude by calculating the radiation emitted by
these electrons, mainly through inverse-Compton scattering off CMB
photons.  The estimates presented in this section needed for our
numerical simulation are discussed in more detail in \S \ref{sec:Method}.

\subsection{Intergalactic Shock Waves}
\label{sec:structure_formation_shock_waves}
\indent

Large-scale structure in the universe is believed to have evolved by
the gravitational collapse of initially over-dense regions.  
As such an initial seed accreted increasing amounts of mass, large converging 
flows of matter were accelerated towards it, inevitably brought to a 
violent halt.  This process resulted in large-scale shock waves,
forming as massive gas flows collided with opposite flows or with the
newly formed object.  In the following, we concentrate our attention
on the 'recent' stages of structure formation, taking place at
moderate to low redshifts ($z \la 3$), and deduce some order of magnitude
estimates of the resulting shock wave properties, following Cen \&
Ostriker (1999).

Linear theory of structure formation predicts that the length and mass
scales of an object, forming (becoming non-linear) at these low
redshifts, are roughly given by:
\beq \label{eq_linear_estimates}
        \lambda_{NL} (z) = \lambda_0 \mbox{ } \frac{f(z)}{1+z}
        \quad \mbox{and} \quad
        M_{NL} (z) = M_0 \mbox{ } f(z)^3 \mbox{ ,}
\eeq
where $\lambda_0$ and $M_0$ are the length and mass scales of
structures forming at the present time and $f(z)$ is a slowly varying
function of the redshift $z$, satisfying $f(z=0)=1$, such that smaller scales 
become non-linear first. 
These parameters are sensitive to the cosmological model. 
We use a 'concordance' $\Lambda\mbox{CDM}$ model of 
Ostriker \& Steinhardt (1995) - a flat universe with normalized 
vacuum energy density $\Omega_\Lambda=0.7$, matter energy density
$\Omega_M=0.3$, baryon energy density $\Omega_B=0.04$, 
Hubble parameter $h=0.67$, and initial perturbation spectrum 
of slope $n=1$ and normalization $\sigma_8=0.9$ - to find: 
\beq \lambda_0 \simeq 10 h_{67}^{-1} \mbox{ Mpc}
\quad \mbox{and} \quad M_0 \simeq 1.5 \times 10^{14}
h_{67}^{-1} \left( \frac{\Omega_M}{0.3} \right) M_\odot \mbox{ ,}
\eeq
where $h_{67}\equiv h / 0.67$.  The order of magnitude estimates
obtained from linear theory in this subsection should be regarded with
caution, as non-linear effects become important in the forming
objects, and the phases of the perturbations can no longer be ignored.
The distribution of the forming objects according to linear theory may 
be better estimated using the approach of Press \& Schechter (1974). 

Geometric considerations dictate that the gas in an object of size
$\lambda$, collapsing over a time period $t_{\tny{coll}}$, will
achieve velocities of order $v \sim \lambda/t_{\tny{coll}}$.  Since a
structure that became non-linear at redshift $z$ will collapse over a
time period of order $t_{\tny{coll}} \sim H(z)^{-1}$, where $H(z)$ is
the Hubble parameter, the formation of such an
object is expected to involve shocks with velocity:
\begin{equation}
        v_{\tny{sh}}(z) \simeq \mbox{ } k \mbox{ } \lambda_{NL}(z) \mbox{ } 
        H(z) = v_0 \mbox{ } \frac{f(z) h(z)} {1+z} \mbox{ ,} 
\end{equation}
where $k$ is a dimensionless number of order unity, $v_0 \equiv k
\lambda_0 H_0$ is the typical shock velocity at $z=0$ and
$h(z) \equiv H(z)/H(z=0)$.  For the $\Lambda \mbox{CDM}$ model
described above we find:
\begin{equation}
        v_0 \simeq 700 \mbox{ km s}^{-1}
        \quad \mbox{and} \quad
        h(z) = \sqrt{\Omega_\Lambda+\Omega_M(1+z)^3} \mbox{ .}
\end{equation}  
Hence, the shock waves resulting from converging flows during structure 
formation in the IGM have a non-relativistic velocity similar to those of 
shocks surrounding supernova remnants (where accelerated particles are 
observed) and they are steadily growing in both scale and velocity. 

The baryonic component may be treated as an ideal gas with an adiabatic index
$\Gamma=5/3$, and is sufficiently rarefied such that the shocks concerned
are collisionless.  
Using the Rankine-Hugoniot shock adiabat \cite{Landau},
we can relate the shocked gas temperature to the shock velocity, 
\begin{equation} \label{eq_T_avr}
        T(z) \simeq T_0 \left[ \frac{f(z)h(z)} {1+z} \right]^2 , 
\end{equation}
where $T_0 \equiv [m_p / (\Gamma k_B)] (\alpha v_0)^2$ is the
temperature of gas, shocked at $z=0$, 
$k_B$ is the Boltzmann constant, 
and $\alpha\equiv c_{s,d} / v_{\tny{sh}} $ is the ratio between the 
post-shock (downstream) sound velocity and the shock velocity (relative to 
the unshocked gas), approaching $\sim 0.56$ for a strong shock.  For the
$\Lambda \mbox{CDM}$ model above we find:
\begin{equation}
        T_0  \simeq 10^7 \mbox{ K} \mbox{ .} 
\end{equation}
As we show in \S 3.2, this estimate of $T_0$ is indeed valid only to
order of magnitude and appears to exceed the true value by a factor 
$\sim 3$.

\subsection{Shock-Accelerated Electron Population \label{EGRB_elec}}
\indent

\subsubsection{Acceleration Mechanism}

Collisionless, non-relativistic shock waves are generally known to
accelerate a power-law energy distribution of high-energy particles.  This
phenomenon has been observed in astrophysical shock waves on various scales, 
such as in shocks forming when the supersonic solar wind collides with
planetary magnetospheres, in shocks surrounding supernovae (SNe) remnants in 
the interstellar medium, and probably also in shocks in many of the most active
extragalactic sources: quasars and radio galaxies 
\cite{Drury83,Blandford87}.
These power law distributions extend up to energies $\sim 100 \mbox{ keV}$
in the Earth's bow shock, to $\sim 100 \mbox{ TeV}$ in SNe remnants
\cite{Tanimori98} where shock velocities are similar to those of
intergalactic shocks, and over $10^6 \mbox{ TeV}$ in Galactic cosmic rays.

Magnetic fields of amplitude $B \simeq 0.1-1 \;\mu \mbox{G}$ are measured in
halos of galaxy clusters \cite{Kronberg94} and in the Coma super-cluster
\cite{Fusco99}, suggesting that the IGM in which the shocks propagate is
significantly magnetized. The measured magnetic field energy constitutes a
fraction $\xi_B \simeq 0.01$ of the typical post-shock thermal energy in
cluster environments \cite{WaxmanLoeb2000}.

Electrons are expected to be accelerated in intergalactic shocks by the
Fermi acceleration mechanism.  The velocity difference between the
two sides of a shock front acts as a first-order acceleration mechanism
($dE / dt \sim E$), provided that slow-moving elastic scattering centers
are embedded in the gas.  Alfv\'en waves or magnetic inhomogeneities can
function as such centers, scattering electrons through small angles and
causing them to repeatedly diffuse through a shock front.  Diffusion of
the energetic electrons downstream, away from the shock front, provides a
Poissonian escape route.  These effects lead to a characteristic power-law
distribution of the high-energy electrons steady-state number density:
\begin{equation} 
        \frac{dn_e}{dE_e}(E) \simeq A E^{-s} \mbox{ ,}
\end{equation}
where $A$ is a normalization constant.  The vast size and perpetual
duration of these shocks enable them to dissipate significant
fractions of their energies in this process.

The resulting distribution of high energy electrons can be calculated
using the test-particle approximation \cite{Drury83,Blandford87}, 
where the test particles, accelerated by the shock, are assumed to have 
no effect on its structure.  This approximation, valid for the
acceleration of a small population of high energy electrons, predicts
that the power-law index of the resulting population depends only on
the kinematic structure of the shock front, and is given by:
\begin{equation}
        s=\frac{r+2}{r-1} \mbox{ ,}
\end{equation}
where $r$ is the compression factor of the shock. For a shock in an
ideal gas, $r$ is given by:
\begin{equation} \label{eq_r_vs_M}
        \frac{1}{r}=\frac{\Gamma-1}{\Gamma+1} + \frac{2}{(\Gamma+1)M^2} 
        \mbox{ ,}
\end{equation}
where $M\equiv v_{\tny{sh}}/c_{s,u}$ is the Mach number, 
and $v_{\tny{sh}}$ and $c_{s,u}$ denote the shock velocity and the 
pre-shock (upstream) sound velocity, respectively.  
The shock waves of interest are generally strong, i.e. involve high 
($M\gg1$) Mach numbers. Hence, we find 
$r \approx (\Gamma+1) / (\Gamma-1 ) = 4$, and thus $s\approx 2$.

In order to completely define the power-law distribution discussed, we must
specify its normalization and the energy range over which it extends.  The
Fermi acceleration mechanism per-se does not impose stringent limits on the
electron energies achieved, as long as the total energy carried by the
accelerated electrons is small compared to the shock energy and the shock
structure remains intact.  
The e-folding time for electron acceleration is given \cite{LoebWaxman2000} 
by: 
\beq \tau_{\tny{acc}}\sim \frac{r_{\tny{L}} c}
{v_{\tny{sh}}^2} \approx 2\times 10^4 \gamma_7 (B_{-7} T_{d,7})^{-1}
\mbox{ yr} \mbox{ ,} \eeq 
where $r_{\tny{L}}$ is the electron Larmor radius, $\gamma_7$ is the electron 
Lorentz factor - $\gamma_e$ - in units of $10^7$, $B_{-7}$ is the magnetic 
field in units of $0.1\;\mu\mbox{G}$, and $T_{d,7}$ is the downstream 
temperature of the gas, in $10^7 \mbox{ K}$.
This time-scale is much shorter than the shock lifetime.  The limit on
electron energy is therefore not determined by the acceleration
process, but rather by cooling processes: electrons will stop
accelerating once their cooling rate overcomes their acceleration
rate.  The dominant cooling process for the relevant parameters is
inverse-Compton scattering of the electrons off CMB photons, with
a characteristic cooling time:
\beq \label{eq_IC_t_cool}
        \tau_{\tny{IC}} = \frac{3m_e c}{4 \sigma_T \gamma_e 
        u_{\tny{CMB}} }
        \simeq 2 \times 10^5 \gamma_7^{-1} (1+z)^{-4} \mbox{ yr} \mbox{ ,}
\eeq
where $u_{\tny{CMB}}$ is the CMB energy density. 
Synchrotron cooling of the electrons is less efficient, with a 
characteristic cooling time 
$\tau_{\tny{syn}} = (u_{\tny{CMB}} / u_{\tny{B}} ) \tau_{\tny{IC}} \simeq 
10^3 (B_{-7})^{-2} (1+z)^4 \tau_{\tny{IC}}$, 
where $u_{\tny{B}}$ is the magnetic field energy density. 
Thus, inverse-Compton cooling is faster, as long as the downstream magnetic 
field satisfies $B < 3 (1+z)^2 \; \mu \mbox{G}$.
Equating $\tau_{\tny{acc}}$ and $\tau_{\tny{IC}}$ 
leads to the maximal Lorentz factor: 
\begin{equation} \label{eq_gamma_max}
        \gamma_{\tny{max}} = 3.3 \times 10^7 
        \left( B_{-7} T_{d,7} \right)^{1/2}     (1+z)^{-2} \mbox{ .}
\label{eq:gmax}
\end{equation}

Since the electron energy loss is dominated by scattering of CMB photons,
the flux of high energy photons produced by shock accelerated electrons
depends only weakly on the magnetic field strength. The magnetic field
determines, nevertheless, the highest energy of accelerated electrons, and
hence the energy to which the inverse-Compton spectrum extends.  
For $B\simeq 0.1\mbox{ }\mu\mbox{G}$, as observed in cluster halos and which 
corresponds to a fraction $\xi_B \simeq 0.01$ of the post-shock thermal 
energy, the highest energy electrons up-scatter CMB photons to energy 
$\epsilon \simeq 1\mbox{ TeV}$.  
It should be pointed out here, that the observed sub-$\mu\mbox{G}$ fields may 
be representative of only the downstream (post shock) magnetic field
strength, and that the upstream field may be significantly lower.  
In this case, the electron acceleration time will be longer, and the 
highest energy of accelerated electrons and of inverse-Compton photons may 
be significantly lower than estimated from equation~(\ref{eq:gmax}). 
The strength of the upstream magnetic field depends on the processes leading 
to IGM magnetization, which are not yet known. Magnetic fields may be 
produced by turbulence induced by the large scale structure shocks themselves
\cite{Kulsrud97}, by ``contamination'' by the first generation of stars
\cite{Rees87} or radio sources \cite{Daly90,Furlanetto01}, or by galactic
winds (Kronberg, Lesch, \& Hopp 1999). 
Under all these scenarios, the upstream magnetic
field strength is not expected to be much smaller than the downstream
field strength. Moreover, near equipartition fields are expected to be
produced in collisionless shocks through the growth of electromagnetic
instabilities, as indicated, for example, by observed supernovae remnants
\cite{Helfand87,Cargill88} and $\gamma$-ray bursts \cite{Gruzinov99,Medve99}, 
and there is evidence from supernovae remnant observations for an enhanced 
level of magnetic waves ahead of the shock front 
(Achterberg, Blandford, \& Reynolds 1994) 
We therefore adopt for our estimates of the maximum electron energy magnetic 
field strengths corresponding to fixed $\xi_B\approx0.01$, which reproduce 
the observed large scale field strengths.

We can now parameterize the normalization of the relativistic electron 
distribution, by assuming that a fraction $\xi_e$ of the thermal energy 
density induced by the shock, $\Delta u_{\tny{th}}$, 
is transferred to these electrons: 
\beq u_e = \xi_e \Delta u_{\tny{th}} 
= \frac{3}{2} \xi_e k_B \left( n_d T_d - n_u T_u \right) \mbox{ ,}  \eeq
where 
$u_e \equiv m_e c^2 \int_1^{\gamma_{\tny{max}}} 
\gamma (dn_e / d\gamma) \,d\gamma$ 
is the total energy density of relativistic electrons and 
$u$ and $d$ subscripts indicate upstream (pre-shock) and downstream 
(post-shock) values, correspondingly.
The gas number density, $n$, is related to the baryon number density, 
$n_{\tny{bar}}$, by $n \langle m \rangle = n_{\tny{bar}} m_p$, 
where $\langle m \rangle \approx 0.59 m_p$ is the average particle mass in an 
ionized gas with a hydrogen mass fraction $\chi =0.76$. 
For strong shocks $s\simeq2$ and $n_u T_u \ll n_d T_d$, thus 
$u_e \simeq A \ln{\gamma_{\tny{max}}}$, and we find: 
\beq \label{eq_normalization}
A \simeq 
\frac{3 \xi_e n_d k_B T_d}{2 \ln{\gamma_{\tny{max}}}} 
\mbox{ .}       \eeq

\subsubsection{Acceleration Efficiency}

The efficiency of particle acceleration in structure formation shocks, 
parameterized above by $\xi_e$, is an important parameter bearing linearly 
on our results, and thus deserves a special discussion. 
Although no existing model credibly calculates the acceleration efficiency, 
we can evaluate $\xi_e$ in intergalactic shocks by estimating the 
acceleration efficiency of other astrophysical shock waves. 

The best analogy to intergalactic shocks may be found in supernova remnant 
(hereafter SNR) shocks, drawing on the similarity between the velocities of 
intergalactic shocks and SNR shocks, both of the order of 
$\sim 10^3 \mbox{ km s}^{-1}$. 
The physics of shock waves is essentially determined by three parameters: 
the shock velocity $v_{\tny{sh}}$, the pre-shock density $n_u$ and the 
pre-shock magnetic field $B_u$. 
Although the plasma density is very different in intergalactic shocks and 
in SNR shocks, the density may be extracted from the problem by measuring 
time in units of $\nu_{p,i}^{-1}$, where $\nu_{p,i}$ is the plasma 
frequency of the ions. 
The pre-shock magnetic field, parameterized by the cyclotron frequency of 
the ions, $\nu_{c,i}$, may not be scaled out of the problem as well. 
However, by comparing $\nu_{c,i}$ to the growth rate of electromagnetic 
instabilities in the shocked plasma, 
$\nu_{\tny{e.m.}} \equiv \nu_{p,i} v_{\tny{sh}} /c$,
we find that for strong shocks 
$(\nu_{c,i}/\nu{\tny{e.m.}})^2 = 
\frac{B_u^2/8\pi}{(1/2)m_p n_u v_{\tny{sh}}^2} \ll 1$. 
We thus assume there is a well behaved limit when this ratio approaches zero, 
suggesting that the pre-shock magnetic field has little effect on the 
characteristics of strong shocks. 
With this assumption, we expect to find much similarity between strong 
shocks in different environments, as long as their shock velocities are 
comparable. 

The energy of the relativistic electrons accelerated in SNR shocks was 
calculated by several authors, using the multi-frequency emission 
from SNR's, preferably remnants with TeV detection such as SN1006 
(G$327.6-1.4$; Tanimori et al. 1998). 
Dyer et al. (2001) have modeled the multi-frequency emission from SN1006, 
finding a shock efficiency of $\xi_e=5.3\%$. 
The total energy they find in relativistic electrons constitutes a fraction 
$\eta_e \simeq 1.4\%$ out of the \emph{total supernova explosion energy} 
($W\simeq 5 \times 10^{50} \mbox{ erg}$, as often quoted in the literature). 
Other authors \cite{Mastichiadis96,Aharonian99} have estimated 
$\eta_e\simeq 1\%-2\%$ in SN1006, which corresponds to $\xi_e$ values 
in the range of $3.8\%-7.6\%$.  
Ellison, Slane and Gaensler (2001) have modeled the emission from supernova 
remnant $\mbox{G}347.3-0.5$, estimating the energy of relativistic electrons 
to constitute at least $1.2\%$ - and more likely $2.5\%$ - of the shock 
kinetic energy flux, corresponding to $2.5\% \la \xi_e\simeq 5\%$.   

An independent, less reliable method for estimating the electron acceleration 
efficiency relies on the observation, that relativistic ions accelerated in 
SNR shocks are dynamically significant \cite{Reynolds92,Baring99}, and thus 
carry a large fraction (tens of percents, e.g. $\sim 50\%$ according to 
Ellison et al. 2001) of the post shock thermal energy. 
The efficiency of electron acceleration may then be found by estimating the 
ratio $\eta$ between the relativistic electron energy and the energy of 
relativistic ions.
One may easily estimate the value of $\eta$ in interstellar medium cosmic 
rays: using the electron-to-proton ratio in a given energy, e.g. $1\%$ at 
GeV energies \cite{Barwick98}, correcting for electron cooling and 
integrating over energies, 
we find $\eta \simeq 3\%-4\%$, i.e. $\xi_e \simeq 1\%-2\%$. 
However, since there is no clear relation between $\eta$ in the interstellar 
medium and its value immediately behind the $\sim 10^3 \mbox{ km s}^{-1}$ 
shocks found in SNR's and in the intergalactic medium, this estimate of 
$\xi_e$ is controversial. 

To summarize, a shock efficiency of $\xi_e \simeq 0.05$ is found from studies 
of supernovae remnants, characterized by shock velocities similar to those 
of intergalactic shocks and thus expected to exhibit similar behavior. 
Since there is some uncertainty in this parameter, values of $\xi_e$ in the 
range $0.02-0.10$ are considered plausible.

\subsection{Resulting Radiation}
\indent 

In this subsection we calculate the radiation emitted by the relativistic
electron population described above.  As mentioned earlier, most of the
energy is produced when the electrons inverse-Compton scatter CMB photons,
resulting in a spectrum that extends up to TeV energies. 
A secondary radiation process is synchrotron emission from the
electrons, gyrating in the magnetic field, resulting in a similarly sloped
spectrum extending up to the infrared.  Other radiative processes
experienced by the electrons, such as Bremsstrahlung and \v{C}herenkov
radiation, are less important.

We begin by calculating the diffuse radiation, resulting from the various 
intergalactic shocks at recent epochs (moderate to low redshifts). 
We then focus our attention on a single forming structure and 
the angular distribution of relevant sources.

\subsubsection{Diffuse Radiation}
\label{sec:EGRB_diffuse_radiation}
\indent 

Inverse-Compton scattering describes the interaction of a free electron
with radiation, when the electron kinetic energy exceeds the photon energy.
In the ultra-relativistic limit, $\gamma_e \gg 1$, the average relative
increase in photon energy is given by $\Delta\epsilon_\gamma /
\epsilon_\gamma \simeq (4/3) \gamma_e^2$.  Hence, an electron with energy
$E_e = \gamma_e m_e c^2 \gg m_e c^2$ will scatter a CMB photon to an
average energy: 
\beq 
\label{eq_e_gamma_vs_E_e} 
\langle \epsilon_\gamma \rangle \simeq 
\frac{4}{3} \gamma_e^2 \epsilon_{\tny{CMB}} = \zeta E_e^2
\mbox{ ,} 
\eeq 
where we have defined $\zeta \equiv (4/3) (m_e
c^2)^{-2}\epsilon_{\tny{CMB}}$, and $\epsilon_{\tny{CMB}}\simeq
3.83 k_B T_{\tny{CMB}}$ is the average energy of a CMB photon.  The
extent of the spectrum is determined by the electron range of energies: The
maximal electron Lorentz factor, $\gamma_{\tny{max}} \simeq 3.3 \times
10^7$, is given by equation~(\ref{eq_gamma_max}), and
equation~(\ref{eq_IC_t_cool}) imposes a minimal energy threshold, as
electrons with Lorentz factors smaller than $\gamma_{\tny{min}}\simeq
200$ hardly radiate during a Hubble time $\sim 10^{10} \mbox{ yr}$.
Plugging this range of Lorentz factors into
equation~(\ref{eq_e_gamma_vs_E_e}) shows that the emitted photon spectrum
extends from $\sim 50 \mbox{ eV}$ up to $\sim \mbox{TeV}$ energies.

The emitted spectrum may be found by assuming that each 
relativistic electron scatters a photon only once. 
This approximation, discussed in \S \ref{sec:sim2EGRB_emitted_spec}, 
gives qualitatively correct results and is exact for strong shocks. 
With this simplification, we find that the emitted energy density per unit 
photon energy is given by:
\beq \label{eq_spectrum_approximation}
         \frac{du_\gamma}{d\epsilon_\gamma}(\epsilon) = \frac{1}{2\zeta} 
         \frac{dn_e}{dE_e}\left(\sqrt{\epsilon / \zeta}\right) = 
        \frac{A}{2} \zeta^{s/2-1} \epsilon^{-s/2}
\mbox{ .} \eeq
For a strong shock, where the electron number density satisfies 
$dn_e(E) / dE_e \simeq AE^{-2}$, this reduces to: 
\beq \label{eq_strong_spectrum}
\epsilon \frac{du_\gamma}{d\epsilon_\gamma}(\epsilon) = A/2 \mbox{ .} \eeq
For weak shocks, the approximation in
equation~(\ref{eq_spectrum_approximation}) 
introduces a small numerical error ($\sim 2$), because the temporal 
evolution of the electron distribution affects the normalization of the 
resulting spectrum. 

Determining the present-day photon spectrum requires integration over
different shocks at various times.  This, in turn, is determined by the
redshift-dependence of the IGM parameters, and requires the use of
non-linear structure formation theory.  We note that the contribution of a
shock that developed at redshift $z$ to the observed value of $\epsilon
[du_\gamma (\epsilon) / d\epsilon_\gamma ]$, is roughly proportional to
$(1+z)^{-4}$.  
Thus, the main contribution to the spectrum observed today,
especially at high photon energies, is expected to originate from recent
shocks at $z\la 0.5$.  Assuming that a fraction $f_{\tny{sh}}$ of
the baryons in the universe were recently 
shocked to a temperature $T\simeq 10^7
\mbox{ K}$, when $\gamma_{\tny{max}}$ already approached
$10^7$, we find from equation~(\ref{eq_normalization}) and
equation~(\ref{eq_strong_spectrum}): 
\beq \label{eq:diffuse_IC_spec}
\epsilon \frac {du_\gamma^{\tny{IC}}} {d\epsilon_\gamma} (\epsilon) \simeq
6.4 \times 10^{-7} \mbox{ } T_7 \left(\frac{\xi_e
f_{\tny{sh}}}{0.05}\right) \left(\frac{\Omega_B h_{67}^2}{0.04} \right)
\mbox{ eV cm}^{-3} \mbox{ .} 
\eeq 
The inverse-Compton flux per decade in photon energy thus becomes: 
\beq \label{eq:diffuse_IC_spec_flux} \epsilon^2 \frac
{dJ_\gamma^{\tny{IC}}} {d\epsilon_\gamma} (\epsilon) \simeq 1.5 \mbox{ }
T_7 \left(\frac{\xi_e f_{\tny{sh}}}{0.05}\right) \left(\frac{\Omega_B
h_{67}^2}{0.04} \right) \mbox{ keV cm}^{-2} \mbox{ s}^{-1} \mbox{ sr}^{-1}
\mbox{ .} \eeq

This result is in excellent agreement with observational data, when 
supplemented by the contribution from unresolved point sources 
\cite{LoebWaxman2000}. 
One can employ the Press-Schechter mass function in order to carry out 
the time-integration over the radiation resulting from intergalactic shocks 
at various epochs. 
This approach gives, for slightly different cosmological 
parameters ($\Omega_\Lambda=0.65$, $\Omega_B=0.05$, $h=0.7$), 
a similar result \cite{LoebWaxman2000}: 
\beq \label{eq:diffuse_IC_spec_PS}
\epsilon^2 \frac{dJ_\gamma^{\tny{IC,PS}}}{d\epsilon_\gamma}(\epsilon) \simeq 
1.5 \left( \frac{\xi_e}{0.05} \right) 
\mbox{ keV cm}^{-2} \mbox{ s}^{-1} \mbox{ sr}^{-1} \mbox{ .} \eeq

The synchrotron radiation emitted by the accelerated electrons may now be 
estimated. 
A relativistic electron with a Lorentz factor $\gamma_e$, 
gyrating in a magnetic field of amplitude $B$, 
emits photons with characteristic energy 
$\epsilon_\gamma \sim [(\hbar e B)/(m_e c)] \gamma_e^2$. 
Synchrotron radiation is thus expected in photon energies ranging from 
$10^{-12}\mbox{ eV}$ and up to $1\mbox{ eV}$. 
The emitted flux can be crudely estimated as: 
\begin{eqnarray} \nonumber
\epsilon^2 \frac{dJ_\gamma^{\tny{syn}}}{d\epsilon_\gamma}(\epsilon) 
& \simeq &  \epsilon^2 \frac{dJ_\gamma^{\tny{IC}}}{d\epsilon_\gamma}
        (\epsilon) \cdot \frac{u_B}{u_{\tny{CMB}}} \\
& \simeq & 1.5 \mbox{ } T_7 B_{-7}^2 
\left( \frac{\xi_e f_{\tny{sh}}}{0.05} \right) 
\left(\frac{\Omega_B h_{67}^2}{0.04} \right) 
\mbox{ eV cm}^{-2} \mbox{ s}^{-1} \mbox{ sr}^{-1} \mbox{ ,}
\end{eqnarray}
whereas a more careful approach is to assume that the energy density 
of the post-shock magnetic field constitutes a fraction $\xi_B$ of the 
shock-induced thermal energy. 
This assumption, combined with the Press-Schechter mass function, leads to a 
similar (somewhat higher) flux: 
\beq 
\epsilon^2 \frac{dJ_\gamma^{\tny {syn,PS}}}{d\epsilon_\gamma}(\epsilon) \simeq 
3 \left( \frac{\xi_e}{0.05} \right) \left( \frac{\xi_B}{0.01} \right) 
\mbox{ eV cm}^{-2} \mbox{ s}^{-1} \mbox{ sr}^{-1} \mbox{ ,} \eeq
where a typical value $\xi_B = 0.01$ produces magnetic fields consistent 
with observations of the outer halos of X-ray clusters \cite{WaxmanLoeb2000}.

\subsubsection{Radiation From Forming Objects, Angular Statistics}
\indent

We start with the general case of an astrophysical object formed 
recently, and calculate its inverse-Compton luminosity.  
We assume that the object accretes mass at a rate $\dot{M}$, heating 
it up to a temperature $T$ by strong shocks.
The baryon number accretion rate is thus given by 
$\dot{M} (\Omega_B / \Omega_M) \langle m \rangle^{-1}$. 
We assume, as previously, that a fraction $\xi_e$ of the 
post-shock thermal energy is converted into relativistic electrons. 
The maximal electron Lorentz factor $\gamma_{\tny{max}}$ is determined 
from equation~(\ref{eq_gamma_max}), and depends on the assumed magnetic 
field. 
Electrons with cooling time longer than the virialization time of the 
object, $t_{\tny{vir}}$, will not have sufficient time to radiate a 
significant fraction of their energy. 
This introduces a low-energy cutoff in the emitted spectrum, at photon energy 
$\epsilon_{\tny{min}} = 
(4/3) \gamma_{\tny{min}}^2 \epsilon_{\tny{CMB}}$, where 
$\gamma_{\tny{min}} \simeq 2 \times 10^{12} \mbox{ yr} / t_{\tny{vir}}$.   
This effect reduces the total luminosity by a factor 
$\ln{\left(\gamma_{\tny{max}} / \gamma_{\tny{min}}\right)} 
/ \ln{\gamma_{\tny{max}}}$.
Combining these considerations, we find that the total and specific 
luminosities of the object are given by: 
\beq  
L^{\tny{IC}} = \dot{M}\frac{\Omega_B}{\Omega_M} \langle m \rangle^{-1}
\mbox{ } \xi_e \frac{3}{2} k_B T \mbox{ }
\frac {\ln{\left(\gamma_{\tny{max}}/\gamma_{\tny{min}}\right)}} 
{\ln{\gamma_{\tny{max}}}} \mbox{ ,}
\eeq
and:
\beq 
\epsilon_\gamma \frac{d L^{\tny{IC}}}{d \epsilon_\gamma} = 
\frac{L^{\tny{IC}}} 
{2 \ln{\left(\gamma_{\tny{max}} / \gamma_{\tny{min}}\right)}} 
\quad \mbox{for energies} \quad 
1 \left(\frac{\gamma_{\tny{min}}}{10^3} \right)^2 \mbox{ keV} \la  
\epsilon_\gamma \la 
0.1 \left(\frac{\gamma_{\tny{max}}}{10^7} \right)^2 \mbox{ TeV} 
\mbox{ .} \eeq

Structure formation at the current epoch produces dense, 
large-scale sheets and filaments, entangled in space. 
The densest regions appear at the intersections of these features, 
around the locations of galaxy clusters. 
As an example, we consider such a galaxy cluster. 
The typical gas mass involved in the formation of a cluster is 
$M_{\tny{gas}} \simeq 10^{14} M_\odot$, 
accreting over a virialization time 
$t_{\tny{vir}} \sim 10^9 \mbox{ yr}$ 
and shocked to a temperature $T_{\tny{gas}} \ga 10^7 \mbox{ K}$. 
A magnetic field of amplitude $B \simeq 0.1 \;\mu\mbox{G}$, as
suggested by observations, 
gives $\gamma_{\tny{min}} \simeq 2000$ and 
$\gamma_{\tny{max}} \simeq 3.3 \times 10^7$. 
Whence, 
\beq L_{\tny{cluster}}^{\tny{IC}} \simeq 2.2 \times 10^{45}
\left( \frac{\xi_e}{0.05} \right) 
\left( \frac{k_B T_{\tny{gas}}} {5 \mbox{ keV}}\right)
\left( \frac{M_{\tny{gas}}} {10^{14} M_\odot} \right) 
\left( \frac{t_{\tny{vir}}} {10^9 \mbox{ yr}} \right)^{-1} 
\mbox{ erg s}^{-1} \mbox{ ,} \eeq 
and: 
\beq \epsilon_\gamma \frac{d L_{\tny{cluster}}^{\tny{IC}}}{d\epsilon_\gamma} 
\simeq 0.05 L_{\tny{cluster}}^{\tny{IC}} 
\quad \mbox{for energies} \quad 
5 \mbox{ keV} \la \epsilon_\gamma \la 1 \mbox{ TeV} \mbox{ .} \eeq
A comprehensive discussion of particle acceleration in galaxy clusters and 
the resulting inverse-Compton radiation up to hard X-rays, may be found in 
Sarazin (1999). 

The angular statistics of the radiation predicted by the model can be
investigated by simplifying assumptions regarding the mass distribution. 
Waxman and Loeb (2000) calculated the angular correlation of the spectrum 
using the Press-Schechter mass function.  On sub-degree angles they find
fluctuations $\ga 40\%$ in the inverse-Compton spectrum and fluctuations
$>100\%$ in the synchrotron spectrum.

% --------------------------------------------------------------------------
%                                 Simulation
% --------------------------------------------------------------------------

\section{Simulation}
\label{sec:Simulation}

\indent 

The previous section presented some order of magnitude estimates,
leading from linear or non-linear structure formation theories to an
EGRB spectrum, that roughly matches present-day observations.  
However, further
progress along this route is severely compromised by the complexity and
non-linearity of structure formation.  This situation naturally calls
for the use of cosmological simulations that incorporate
hydrodynamical effects.  For this project, we analyze a simulation of
a $\Lambda\mbox{CDM}$ cosmology.  In the following, we describe the
cosmological model and background theory of this simulation, and
discuss some characteristics of the simulated universe and the
predicted underlying structure.

\subsection{Model and Theory}
\indent

The cosmological model adopted in the simulation is the 'concordance'
$\Lambda\mbox{CDM}$ model of Ostriker \& Steinhardt (1995), presented
in \S \ref{sec:structure_formation_shock_waves}.  An initial
Harrison-Zel'dovich spectrum of fluctuations was assumed, and linear
structure formation theory \cite{EisensteinHu} employed to predict the
resulting spectrum at a certain redshift $z_{\tny{start}}$, providing
initial conditions for the simulation.  The various parameters of the
cosmological model are summarized in Table \ref{tab:CosmoParams} (see
e.g. Springel, White \& Hernquist 2001b).

The simulation was performed using the TreeSPH code {\small GADGET}
(Springel, Yoshida, \& White 2001a).  This code, designed for both
serial and parallel computers, is intended for collisionless and
gas-dynamical cosmological simulations.

Dark matter is modeled by the code as a self-gravitating collisionless
fluid, represented by a large number $N_{\tny{DM}}$ of particles.
Newton's gravitational potential is modified by introducing a spline
softening at small separations, equivalent to treating each particle
as a normalized spline mass distribution, instead of as a point mass
(see, e.g. Hernquist \& Katz 1989).  This modification is necessary in
order to suppress large angle scattering in two-body collisions.
Poisson's equation is solved using a hierarchical tree algorithm
\cite{BarnesHut}.

The baryonic component of the universe is modeled by the simulation 
as an ideal gas, 
with an adiabatic index $\Gamma=5/3$, and is represented by 
a large number of particles, similar to the dark matter. 
The gas is governed by the continuity equation:
\beq \frac{d\rho}{dt}+\rho \dvr \vvec = 0 \mbox{ ,} \eeq
the Euler (momentum) equation: 
\beq \rho \frac{d\vvec}{dt} = -\grad P - \rho \grad \Phi + 
\dvr \overline{\Sigma} \mbox{ ,} \eeq
and the thermal energy equation (the first law of thermodynamics): 
\beq \frac{du}{dt} = -\frac{P}{\rho} \dvr \vvec  
- \frac{1}{\rho} \overline{\Sigma}_{ij} \partial_i \vvec_j
- \frac{\Lambda(u,\rho)}{\rho} \mbox{ ,} \eeq
where $P$, $\rho$, $u$ and $\vvec$ are the pressure, mass density, 
thermal energy per unit mass and velocity of a gas element; 
$\Phi$ is the gravitational potential, found from Poisson's equation: 
\beq \nabla^2 \Phi(r,t) = 4\pi G \rho(r,t) \mbox{ ;} \eeq
and we have used Lagrangian derivatives: 
\beq \frac{d}{dt}=\frac{\partial}{\partial t} + \vvec \vdot \grad 
\mbox{ .} \eeq
An artificial viscosity \cite{Steinmetz} term $\overline{\Sigma}$ was added 
to the above equations, in order to capture shocks. 
A cooling function $\Lambda(u,\rho)$ may be included in the energy equation, 
to incorporate radiative cooling processes.

These hydrodynamical equations are solved using the SPH - Smoothed
Particle Hydrodynamics - technique (Lucy 1977; Gingold \& Monaghan
1977; Monaghan 1992), well suited for non-axisymmetric
three-dimensional astrophysical problems.  
The version of the code used for the simulation employs adaptive SPH 
smoothing lengths, such that each particle has a constant number $N_s$ 
of neighbors lying within its kernel.  
The hydrodynamical equations and their solutions in the formalism of this 
version of SPH appear in Springel et al. (2001a). 
We note that the simulation analyzed in this paper
employed a formulation of SPH which does not rigorously conserve
entropy in adiabatic parts of the flow (for a discussion, see
Hernquist 1993, Springel \& Hernquist 2002a).

The code simulates the evolution of dark matter and gas (hereafter:
SPH) particles within a finite simulation box of comoving size
$L_{\tny{com}}^3$, using periodic boundary conditions.  Due to
technical reasons, to be discussed in the next section, we have chosen
to employ a simulation that did \emph{not} include radiative cooling
processes.  Radiative cooling, dominated by Bremsstrahlung and line
emission at the epoch of interest, is sub-dominant relative
to inverse-Compton
emission from relativistic electrons in all but the densest
regions within the cores of clusters where the baryon number density exceeds 
$\sim 10^{-3} \mbox{ cm}^{-3}$, and in cold ($T \la 10^4 \mbox{ K}$), 
dilute regions.  
We discuss the significance of radiative cooling towards the end of this 
section.  
The simulation neglected feedback from star formation and supernovae.  
The code parameters used in the simulation are presented in 
Table \ref{tab:AlgoParams}.

\subsection{Resulting Structure}
\indent

The simulation chosen for the project, as most cosmological simulations,
predicts a highly structured IGM 
in the present-day universe, as illustrated in Figure \ref{fig:map3D}.  
Several hot and dense objects are
spread throughout space, separated by large empty voids.  The largest of
these structures are sheet-like objects (pancakes) and filaments, entangled
in space, some as long as $\sim 100\mbox{ Mpc}$.  At the intersection of
the filaments appear denser, spheroidal objects, with typical size $\sim 1
\mbox{ Mpc}$, where galaxy clusters are formed.  Simulations predict that
the pancake-resembling objects are the first large structures to appear.
At the intersections of these pancakes, filaments emerge, which in turn
intersect and drain mass into spheroidal objects.  At the large scales
discussed, dark matter and baryons essentially trace the same structure.

The objects described above are significantly denser than their
surroundings.  Thus, the last stages of their evolution are dominated
by non-linear processes.  These processes are mainly gravitational,
though other physical processes, such as non-linear gas dynamics and
radiative cooling (when included), influence these stages of structure
evolution as well.  

It is instructive to examine the thermodynamic evolution of the
universe, by applying some averaging schemes to the temperature and
density at different epochs, as shown in Figure \ref{fig:Sim_Stat}.
The mass averaged temperature is notably higher than the volume 
averaged temperature, because hot regions are typically much denser, 
leaving most of the volume with a dilute, cooler gas (e.g. Dav\'e et 
al. 1999).  The present-day averaged temperature reached by the 
simulation, $T(z=0)=3.9\times 10^6\mbox{ K}$ (see Figure \ref{fig:Sim_Stat}), 
is significantly lower than the order-of-magnitude estimate presented earlier. 
(Note that the approximation of eq.~[\ref{eq_T_avr}] is in rough 
agreement with the mass averaged temperature, although deviating towards 
higher temperatures at recent times.)  Furthermore, the simulation result for 
the present-day mass averaged temperature 
is lower than that predicted by Cen \& Ostriker (1999) by a factor
$\sim 2.5$, but in good agreement with the results of e.g. Refregier
et al. (2000), Croft et al. (2001), and Refregier \& Teyssier (2000).
R. Cen (private communication) informs us that revised estimates for
the temperature evolution of the Cen \& Ostriker (1999) simulation,
correcting for an error in the integration
of the thermal energy equation, now actually lie below the other
simulation results.

The volume averaged density is proportional to $(1+z)^3$, due to the
expansion of the universe, whereas the approximately constant mass averaged 
density indicates that much of the mass has virialized. 
The temperature averaged density is higher than the volume averaged density, 
due to the above mentioned correlation between hot and dense regions. 
This (temperature averaged) density decreases in time, growing closer to the 
volume averaged density, as the dilute gas surrounding virialized structures 
heats up. 

We examine the effect of radiative cooling by super-imposing
cooling-time contours on the distribution of mass in the
temperature-density phase-space (Figure \ref{fig:Cool_Stat}, left).
The figure suggests that radiative cooling has important effects on
the present-day gas, residing in the most dense, hot regions and in
the most cold, dilute regions.  The cooling time in these volumes due
to Bremsstrahlung and line emission, $t_{\tny{cool}}$, is comparable
to or smaller than the dynamical time, $t_{\tny{H}} \sim H^{-1}$.  
Such regions, where $t_{\tny{cool}}<t_{\tny{H}}$, 
presently contain approximately $25\%$ of the gas
mass, and a progressively larger mass fraction at
higher redshifts (Figure \ref{fig:Cool_Stat}, right).  The dense and
hot regions, where radiative cooling is significant, are associated
with the cores of clusters, well beyond the accretion shock fronts of
interest and should thus have little effect on our results, although
the weaker merger and flow shocks may be affected.  Radiative effects
in the cold, dilute regions cools the collapsing gas and may result in
slightly stronger accretion shocks.  However, as these regions are
cold to begin with, this effect should be unimportant.

% --------------------------------------------------------------------------
%                                 Method
% --------------------------------------------------------------------------

\section{Method}
\label{sec:Method}

% section -  Extracting EGRB from the Simulation
% ----------------------------------------------

\indent

Next we discuss how cosmological simulations can be exploited to yield
various predictions regarding the EGRB, according to the Loeb-Waxman model.
Parts of the discussion are specific to the cosmological simulation chosen
for the project, introduced in the previous section, and to the EGRB
extraction schemes used.  
Predictions of shock-induced radiation are obtained from the
simulation in three basic steps: First, the simulation is used to keep
track of the forming structures and locate the shocks involved in
their virialization.  Second, relativistic electron populations are
injected into the shocked gas, according to the predictions of the Fermi
acceleration mechanism.  And third, the radiation emitted by the
relativistic electrons is calculated and integrated to give measurable EGRB
features.  The organization of this section corresponds to these three
steps.

\subsection{Extracting Shocks}
\label{subsection_shock_extraction} 
\indent

Since shock waves are discontinuities in the thermodynamic quantities,
locating shocks involves identifying large gradients in the properties
of the gas.  Gradients of quantities such as
pressure, temperature or velocity, can be calculated directly from a
simulation.  In our case, this requires SPH kernel interpolation, as
described in the previous section.  Next, one can search for gradients
larger than some imposed thresholds.  These would imply the presence of
shocks, stronger than a corresponding shock strength.  The shock
front could then be traced out, yielding the shock topology and the jump
conditions across the shock, which can in turn be checked to see if 
they satisfy the Rankine-Hugoniot adiabat. 

Such an approach was adopted by Miniati et al. (2000), using simulations
based on a version of the total variation diminishing scheme (TVD),
designed to capture strong shocks \cite{Ryu}.  They identified very strong
accretion shocks with Mach numbers $M$ of up to $10^3$ around the filaments
and halos of their simulations.  These halos exhibit, in addition, weaker
shocks ($M\sim 3-10$), originating from the merger of structures or from
accretion shocks into old structures, already embedded within newly formed
halos.  Combined, these shocks lead to a complicated pattern of
inter-winding shocks.  
Miniati et al. find that the jump conditions, imposed by a shock, 
are altered by adiabatic gravitational compression, especially when 
the shock is close to the core of a cluster.  Since the limited simulation 
resolution renders these two effects inseparable, shock parameters were 
extracted under the assumption that gravitational effects are 
isothermal.

\subsubsection{Shock-Extraction Scheme}
\indent

Since we are interested more in the shocked gas than in the topology of
shocks, we find SPH simulations sufficient for our needs, although they do
provide limited shock front resolution.  This enables us to adopt a
simpler approach for locating shocks, appropriate for Lagrangian
simulations such as SPH.  This method involves following a gas element
throughout the simulation, searching for rapid increases in its
\emph{entropy}.  In cosmological simulations, entropy changes of the gas
should result only from shock waves or from cooling processes.  This is the
reason we have chosen, for simplicity, to work with a simulation in
which radiative cooling is neglected, 
shown in the previous section to yield reliable shocks.  In
such a simulation, the entropy of the gas should only increase, and do so
in discrete steps, decisively indicating the presence of shocks.  Ideally,
one could then identify all entropy 'jumps' as shock waves, determining
their space-time coordinates.  The thermodynamic jump conditions across a
shock could then be found, by comparing the thermodynamic parameters of the
gas element before ($T_i$, $\rho_i$) and after ($T_f$, $\rho_f$) passing
through the shock, directly, or using the entropy-change, $\Delta S =
C_V\ln{\left[(T_f/T_i)(\rho_f/ \rho_i)^{(1-\Gamma)} \right]}$, induced by
the shock.  Using the Rankine-Hugoniot adiabat, we find that the
compression factor $r$ of a shock - related to its Mach number by
equation~(\ref{eq_r_vs_M}) - can be determined from $\Delta S$ according
to: \beq \label{eq_S_vs_r} \Delta S_* \equiv C_V^{-1} \Delta S = \ln{
\left[\frac{r(\Gamma+1)-(\Gamma-1)}{(\Gamma+1)-r(\Gamma-1)} r^{-\Gamma}
\right]} \mbox{ .} \eeq

In SPH simulations, one can choose the SPH particles \emph{themselves}
as the gas elements to be followed, since they are of constant mass
and the simulation is Lagrangian.  This technically simplifies our
shock-locating scheme, as kernel interpolation becomes unnecessary.
It is instructive to study the distribution of mass - i.e. SPH
particles - that accumulated various entropy-changes during recent
epochs. 
Figure \ref{fig:dS_Stat} (left) gives this distribution in terms of the 
mass fraction $f(\Delta S_*^{\prime})= m_{\tny{tot}}^{-1} 
dm(\Delta S_*^{\prime}) / d\Delta S_*$ of SPH particles that accumulated 
a normalized entropy change $\Delta S_*^{\prime}$ between $z=2$ and $z=0$, 
where $m_{\tny{tot}}$ is the total mass of SPH particles. 
First, note that a small fraction of the gas experienced
\emph{negative} entropy changes, most probably as a result of
lack of strict entropy conservation in the absence of shocks, owing
to the particular formulation of SPH used for this simulation
(see, e.g. Springel \& Hernquist 2002a).
The entropy-difference distribution among such
particles resembles a normal distribution, consistent with
accumulating noise.  The shape of the remaining distribution has an
interesting structure, suggesting a pronounced population of 'weak'
shocks, with $M\sim 3-10$, and a large population of stronger shocks,
with Mach numbers up to a few $10^3$, in agreement with Miniati et al.
(2000).

In order to determine the location and temporal duration of shock waves, 
one must examine the simulated gas in sufficiently short time intervals. 
This is also necessary in order to identify multiple shocks suffered
by an SPH particle, as the effect of consecutive shocks is not
additive, although the entropy change they induce is.  
We chose to examine snapshots of the simulated gas, separated by the
simulation-box light-crossing time, 
as illustrated in Appendix \ref{app:formulae}. 
For the mass resolution of the simulation, the
entropy-increase experienced by most shocked particles is spread
over several of these snapshots, indicating that shorter time
intervals are not required.  With the time intervals that were chosen,
the entropy-difference distribution between any two snapshots (Figure
\ref{fig:dS_Stat}, right) is fairly constant throughout the period of
interest.  Based on this distribution, we impose a minimal threshold
$\Delta S_*^{\tny{min}}$, such that only particles experiencing a
larger entropy-increase between two consecutive snapshots are
considered shocked.  
A threshold value $\Delta S_*^{\tny{min}}=0.4$ was found to be reasonable, 
considering numerical noise, although it entails the danger of eliminating 
some of the mildly weak shocks ($M \la 10$), if an SPH particle requires many 
snapshots to cross such a shock. 
The fact that only $\sim 10\%$ of the gas suffers entropy-changes larger
than $\Delta S_*^{\tny{min}}$ simplifies the computational work
considerably.  
Demanding that the entropy increases at least by $n_s\Delta S_*^{\tny{min}}$ 
during $n_s=2$ consecutive snapshots, and dismissing shocked particles that 
did not reach a temperature $T_{\tny{min}}\equiv 10^5 \mbox{ K}$ by the 
present time, further simplifies the analysis.  
Our final results do not strongly depend upon
the choice of these threshold values near those specified above. 
Note that with finite resolution, SPH particles may pass through a shock in 
a period longer than the snapshot temporal resolution. 
The simulation must thus be traced down to $z<0$ values in order to capture 
all recent shocks. 

The temperature jump across a shock determines the thermal energy (per
particle) it adds to the gas, $\Delta U_{\tny{th}}/N= (\Gamma-1)^{-1}
k_B (T_d-T_u)$, and is thus vital for the calculation of shock-induced 
radiation. 
In principle, this jump could be found directly, by comparing the temperature 
of the shocked SPH particle before and after a shock takes place.  
However, with finite simulation resolution, this approach leads to
errors, due to significant adiabatic heating or cooling of the
particle in parallel to the shock.  In some extreme cases, we even
found the shocked particle cooler than it was prior to the shock.
(A similar problem when radiative cooling is active has been
identified by e.g. Hutchings \& Thomas [2000] and Martel \& Shapiro [2001].) 
A partial remedy for this problem is to assume that the entropy induced
by the shock - and thus the compression factor - was correctly
determined so that the state of the shocked particle can be deduced
from its state prior to the shock, or vice versa.  Indeed, neither the
pre-shock state of the particle nor its post-shock state are
accurately known, due to the inaccurate shock timing and ongoing
adiabatic processes, leading to some inevitable error.  We chose to
regard the temperature of the gas, before the shock was detected, as
the true pre-shock temperature: $T_u \equiv T_i$.  Hence, we take:
\beq T_d = T_i r^{(\Gamma-1)} e^{\Delta S_*} \mbox{ } \eeq as the
post-shock temperature, where $r$ and $\Delta S_*$ are related by
equation~(\ref{eq_S_vs_r}). 
This choice yields more realistic results than the alternative, 
$T_d=T_f$, as discussed in \S \ref{sec:ShockExtractionResults}.

\subsubsection{Shock-Extraction Results}
\label{sec:ShockExtractionResults}
\indent

The shock-extraction scheme, described above, identifies shocks in $\sim
40\%$ of the simulated gas mass during recent epochs ($z<2$), where about
$30\%$ of the baryons were shocked at least once (Figure
\ref{fig:Shock_Stat}, left).  Most identified shocks are concentrated
around the halos in the simulation, although filaments and sheets are
traced out as well.  In some cases, but not always, shock fronts can be
identified (Figure \ref{fig:ShockMaps}).

The distribution of entropy-differences, accumulated by
\emph{identified shocks} (Figure \ref{fig:Shock_Stat}, right),
suggests that most of the strong shocks with Mach numbers in the
range $M>10^2$ were located.  It confirms, however, our suspicion of
loosing most of the 'weak' shock population ($M\sim 3-10$).  In such
shocks, the entropy-change suffered by a single particle is 
spread over many snapshots, and is thus eliminated by the imposed
entropy-change threshold.  Figure \ref{fig:Shock_Stat} (right)
suggests that some of the very large accumulating entropy-increases,
suffered by an SPH particle, result from multiple shocks.  Separating
the shocks in such cases results in a population of weaker shocks,
artificially concentrated at the threshold shock strength ($M\sim 4$).

As a 'sanity check' of the proposed shock-location and
energy-evaluation schemes, we compare the thermal energy gain of the
simulation to the evaluated energy, induced by the identified shocks.
We find that the energy, injected by our shocks, constitutes a fraction 
$f_{\tny{sh}} \sim 2 / 3$ of the total thermal energy gained by the simulated 
gas (see Table \ref{tab:ProgParams} and Figure \ref{fig:Shock_Stat}, left). 
For comparison, in the estimates of 
\S \ref{sec:structure_formation_shock_waves} we took $f_{\tny{sh}}=1$.
Miniati et al. (2000) find, for a slightly different $\Lambda\mbox{CDM}$ model 
($\Lambda=0.55$, $h=0.6$), that the energy passing through shocks with
Mach numbers $M>10^2$ between redshift $z=1.5$ and today,
approximately equals the present-day thermal energy of their
simulation.

The above results were obtained by estimating the energy of an SPH particle, 
gained when passing through a shock, based on its temperature \emph{before} 
passing through the shock (i.e. taking $T_u=T_i$). 
Estimating this energy gain based on the temperature \emph{after} passing 
through the shock (i.e. taking $T_d=T_f$) results in shock energies lower, 
on average, by a factor $\sim 4.5$. 
This result is a combined effect of the low resolution of the simulation 
and adiabatic cooling, taking place in parallel to the shock.  

We find the shock-extraction results described above reassuring, as
they indicate that the proposed shock location-and-evaluation recipe
captures most of the strong shocks, and measures their energy
reasonably well.  
Better performance of this algorithm may be achieved
by using simulations with a higher mass resolution and with a more
accurate handling of adiabatic flows (e.g. Springel \& Hernquist
2002a).
It should be stressed that the elimination of weak shocks by imposing
an entropy threshold for shock identification has only a small effect
on the predicted diffuse $\gamma$-ray radiation, especially at high
energies.  First, the consistency check described above verifies that the
energy injected by weak shocks is small compared to the energy
injected by strong shocks: accounting for all weak shocks may increase
the integrated flux at most by a factor of $\sim 1.5$.  Second, the
contribution of weak shocks to the $\gamma$-ray emission, especially
at high photon energies, is small due to the steep distribution of the
electrons accelerated by such shocks, reducing their $\gamma$-ray
efficiency. At $10\mbox{ GeV}$, for example, shocks with $M=10$ have
$66\%$ efficiency (compared to strong shocks), dropping below $50\%$
for $M\la 8$.  Weak shocks could, nonetheless, increase the
$\gamma$-ray brightness of individual nearby structures such as galaxy 
clusters experiencing merger events, mainly in low photon energies, 
thereby slightly altering the appearance of their inner regions, and as 
a result may somewhat enhance the predicted source number counts.

\subsection{Accelerated Electron Population}
\indent

Here we rely on the arguments of \S \ref{EGRB_elec}, deriving
expressions more general and more precise than the order of magnitude
estimates presented earlier.  These results are used to calculate the
relativistic electron distributions produced by shocks, and their
temporal evolution.  Appropriate electron populations can thus be
'injected' into the shocked SPH particles to simulate radiative
processes.

\subsubsection{Production}
\indent

We postulate that a fraction $\xi_e$ of the post-shock density of the 
thermal energy induced by a shock, 
$\Delta u_{\tny{th}}=[1/(\Gamma-1)] n_d k_B (T_d-T_u)$, 
is converted into a power law distribution of relativistic electrons, 
$dn_e(E) / dE_e = A E^{-s}$. 
The power law index is related to the compression factor 
by $s=(r+2)/(r-1)$, 
where $\Gamma=5/3$ was used here and is assumed henceforth. 
Imposing a distribution cutoff at the maximal energy attainable by the 
electrons, 
$E_{\tny{max}}\equiv \gamma_{\tny{max}}m_e c^2$, fixes the normalization:
\beq    
A=\xi_e \Delta u_{\tny{th}} \times \cases{ 
1/\ln{\gamma_{\tny{max}}} & for $s=2$ ; \cr
(s-2)(m_e c^2)^{s-2}/\left[1-\gamma_{\tny{max}}^{-(s-2)}\right]	
& for $s>2$ . \cr }
\eeq

We estimate the maximal Lorentz factor of the electrons by equating their 
acceleration e-folding time to the cooling time due to inverse-Compton 
scattering, yielding:
\beq    \gamma_{\tny{max}}^2 =  \frac{3(\Gamma-1)}{8}
\frac{e B_d}{\langle m \rangle c^2 \sigma_T u_{\tny{CMB}}}
\left( k_B T_u + r \frac{\Gamma+1}{\Gamma-1} k_B T_d \right) \mbox{ .}\eeq
The average mass per particle is given by
\beq \langle m \rangle = \frac{4m_p}{3\chi+1+4\chi_e} \simeq 0.59 m_p 
\mbox{ ,} \eeq
where $\chi$ is the hydrogen mass fraction, $\chi_e$ is the ionization
ratio (per nucleon) and the second equality holds for a fully ionized gas, 
where $\chi_e=(1+\chi)/2$, with the hydrogen mass fraction of the simulation, 
$\chi=0.76$, assumed henceforth.

We parameterize the post-shock magnetic field, $B_d$, by assuming that a 
fraction $\xi_B$ of the post-shock thermal energy is transferred into 
the energy of this field, giving:
\beq B_d = 4.3 \times 10^{-8} \mbox{ } 
\left[T_{d,7} \left(\frac{\xi_B}{0.01}\right) 
\left(\frac{n}{10 n_{\tny{av}}(z)}\right)
\right]^{1/2} (1+z)^{3/2} \mbox{ Gauss ,} \eeq 
where $n_{\tny{av}}(z)$ is the average particle number density at 
redshift $z$. 
Hence, 
\beq \gamma_{\tny{max}}= 
1.1 \times 10^7 \mbox{ } (r T_{d,7} + T_{u,7}/4)^{1/2} 
\left[ T_{d,7} \left(\frac{\xi_B}{0.01}\right)
\left(\frac{n}{10n_{\tny{av}}(z)}\right)
\right]^{1/4} (1+z)^{-5/4} \mbox{ ,} \eeq
which for strong shocks, where $r\rightarrow 4$ and $T_u \ll T_d$, 
reduces to: 
\beq \gamma_{\tny{max}}^{(\tny{strong})} =
2.2 \times 10^7 \mbox{ } T_{d,7}^{3/4} 
\left[\left(\frac{\xi_B}{0.01}\right)
\left(\frac{n}{10n_{\tny{av}}(z)} \right) \right]^{1/4} 
(1+z)^{-5/4} \mbox{ .} \eeq

\subsubsection{Evolution}
\indent

After the relativistic electrons diffuse away from the shock, they cool,
mainly by inverse-Compton scattering off CMB photons. 
The electron distribution - initially a power law - may thus evolve into 
a different form, as energetic electrons cool faster: 
\beq P_{\tny{IC}} \equiv \frac{dE}{dt} = -C E^2 \mbox{ ,} \eeq
where we have defined 
$C\equiv 4 u_{\tny{CMB}} \sigma_T / 3 m_e^2 c^3$. 
This leads to a partial differential equation for the electron 
distribution:
\beq \frac{\partial}{\partial t} n_{\tny{E}}(E,t) = 
C E^2 \frac{\partial}{\partial E} n_{\tny{E}}(E,t) + 
2 C E n_{\tny{E}} (E,t) \mbox{ ,} \eeq
where the notation $n_{\tny{E}}\equiv \partial n_e / \partial E_e$ 
is introduced. We ignore adiabatic cooling due to the Hubble
expansion since the cooling timescales  relevant for our 
calculated $\gamma$--ray spectrum are all much shorter
than the Hubble time. 
For the initial (power-law) condition we find the solution: 
\beq \label{eq:elec_dist_evolution} n_{\tny{E}}(E,t)= \cases{ 
AE^{-s}(1-CEt)^{(s-2)} & 
for $E<$ min $\{ E_{\tny{max}},E_{\tny{cool}}(t) \}$ ; \cr
0 & for $E>$ min$\{ E_{\tny{max}},E_{\tny{cool}}(t) \}$ , \cr}
\eeq
where we have defined $E_{\tny{cool}}(t)\equiv 1/Ct$.
This solution implies that the electron distribution steepens
(if $s>2$) towards $E_{\tny{cool}}(t)$, 
beyond which all electrons have already cooled off. 
For strong shocks, where $s \rightarrow 2$, the electron distribution is 
stationary for $E<E_{\tny{cool}}(t)$, 
as equal numbers of electrons enter and leave each energy bin at any 
given time. 

The inverse-Compton cooling time of a relativistic electron is a function
of its Lorentz factor $\gamma$, approximately given by
$t_{\tny{cool}}(\gamma) \simeq 2.3 \times 10^{12} \gamma^{-1} (1+z)^{-4}
\mbox{ yr}$.  This implies that electrons with
$\gamma<\gamma_{\tny{min}}(z)$, where $\gamma_{\tny{min}}$ is calculated in
equation~(\ref{eq:gamma_min}), do not have sufficient time to cool between
redshift $z$ and today, and will not significantly contribute to the
emitted radiation.

\subsection{Emitted Spectrum}
\label{sec:sim2EGRB_emitted_spec}

\subsubsection{Inverse Compton Emission}
\indent 

In the previous subsection we have demonstrated how the electron
distribution, carried by each simulated SPH particle, can be calculated at
any given time.  Thus, it is straightforward, if somewhat lengthy, to
calculate the inverse-Compton spectrum, emitted by each SPH particle.  The
emitted energy density per unit photon energy is obtained by integrating
the single particle emission function, over both electron and photon
distributions {\cite{Rybicki}: 
\beq \label{eq:IC_power}
\frac{dP_\gamma^{\tny{IC}}}{d\epsilon_\gamma}(\epsilon,t) = \frac{3}{4}
\sigma_T c \int \frac{\epsilon}{\epsilon_0} \nu(\epsilon_0) \,d\epsilon_0
\int \gamma^{-2} \frac{dn_e}{d\gamma_e}(\gamma,t) f\left( \frac{\epsilon}{4
\gamma^2 \epsilon_0} \right) \,d\gamma \mbox{ ,} 
\eeq 
where
$\nu(\epsilon_0)$ is the specific number density of the incident photon
field, assumed constant (in the time period of interest) and isotropic,
$f(x)$ is the emission function, given by: \beq f(x) = 2x\ln{x} + x + 1 -
2x^2 \mbox{ ,} \eeq and we have assumed Thomson scattering in the electron
rest frame: $\gamma \epsilon_0 \ll m_e c^2$.  The incident CMB photon field
is approximately blackbody radiation, whence: 
\beq \nu(\epsilon_0) =
\frac{8 \pi \epsilon_0^2 / h^3 c^3 } {\exp{ \left( \epsilon_0 / k_B
T_{\tny{CMB}} \right) - 1 }} \quad \mbox{and} \quad 
T_{\tny{CMB}} \simeq 2.73 (1+z) \mbox{ K .} 
\eeq

The assumptions made above are all satisfied for the ultra-relativistic
electrons ($\gamma\gg100$) of interest: The CMB photon field is isotropic
and varies slowly in time, 
thus it can be treated as constant for the fast
emission of such electrons.  The Thomson condition is satisfied for
$\gamma_7 \ll 57(1+z)^{-1}$, rendering Klein-Nishina corrections
unnecessary.

A straightforward integration of equation~(\ref{eq:IC_power}) may be
replaced by a simpler approach.  The typical lifetime of a shock,
comparable to the Hubble time $t_{\tny{sh}} \simeq t_{\tny{H}}\equiv H^{-1}
= 1.5\times 10^{10} h(z)^{-1} \mbox{ yr}$, is much longer than the cooling
time of ultra-relativistic electrons.  Hence, a population of such
electrons, constantly produced by a well resolved shock, will resemble a
\emph{steady} source of radiation.  This enables us to approximate the
emission from these electrons as \emph{instantaneous}, i.e. we assume that all
ultra-relativistic electrons with $\gamma>\gamma_{\tny{min}}(z)$ loose all
their energy to radiation, \emph{immediately} after passing through the
shock front.  The error thus introduced is small, for electrons satisfying
$t_{\tny{cool}}(\gamma)\ll t_{\tny{sh}} \Leftrightarrow \gamma \gg 155
h(z)/(1+z)^4$.  Further justification for this approximation may be found
in the temporal resolution of the snapshots used.  The time between
consecutive snapshots examined is determined by the simulation-box
light-crossing time, of order $t_{\tny{res}} \simeq 6.5 \times 10^8
(1+z)^{-1} \mbox{ yr}$.  Thus, it is obviously pointless to follow the
temporal evolution of electrons, satisfying $t_{\tny{cool}}(\gamma) <
t_{\tny{res}} \Leftrightarrow \gamma>3.5 \times 10^3 (1+z)^{-3}$.

In the 'instantaneous emission' approximation described above, we are
interested only in the \emph{time-integrated} specific power, emitted by
the accelerated electrons: 
\beq
\frac{du_\gamma^{\tny{IC}}}{d\epsilon_\gamma}(\epsilon) =
\int_0^{t_{\tny{sh}}}\frac{dP_\gamma^{\tny{IC}}}{d\epsilon_\gamma}
(\epsilon,t)\,dt \mbox{ .} 
\eeq 
Since the only time-dependence in the
emitted power of equation~(\ref{eq:IC_power}) appears in the distribution
of electrons, $dn_e(\gamma,t) / d\gamma_e$, we may change the order of
integration and first calculate the time-integrated electron distribution:
\beq \label{eq:integrated_elec_dist} \frac{dn_e^{\tny{int}}}{dE_e}(E) =
\int_0^{t_{\tny{sh}}} n_{\tny{E}} (E,t) \,dt \simeq \frac{A}{C(s-1)}
E^{-(s+1)} \mbox{ ,} \eeq where the second equality is obtained by taking
the upper integration limit to infinity or to $\mbox{max} \{ t_{\tny{sh}},
t_{\tny{cool}}(E) \}$, which is well 
justified for ultra-relativistic electrons.
Integrating the power emitted by the \emph{evolving} electron distribution,
as in equation~(\ref{eq:IC_power}), may thus be replaced by calculating the
\emph{instantaneous} emission from the \emph{integrated} electron
distribution of equation~(\ref{eq:integrated_elec_dist}).

Since the integrated electron distribution is a simple power law, 
we may use the well known result for inverse-Compton emission of a 
power-law electron distribution, scattering a blackbody photon field 
\cite{Rybicki}. 
This leads to: 
\begin{eqnarray} \label{eq:exact_spectrum} \nonumber
\frac{du_\gamma^{\tny{IC}}}{d\epsilon_\gamma} (\epsilon) & = & 
Q(s) A \left[ k_B T_{\tny{CMB}} / \left( m_e c^2 \right)^2 \right]^{s/2-1} 
\epsilon^{-s/2} \\ 
& = & Q(s) A \epsilon_{\tny{IC}}^{-(s/2-1)} (1+z)^{s/2-1} \epsilon^{-s/2}
\mbox{ ,} \end{eqnarray} 
where $Q(s)$ is a numerical coefficient, defined by: 
\beq Q(s) \equiv \frac{135}{2\pi^4} 2^s 
\frac{s^2 + 6s + 16}{(s+4)^2 (s+6) (s+2)} 
\Gamma\left( 3+\frac{s}{2} \right) \zeta \left( 3+\frac{s}{2} \right) 
\mbox{ ,} \eeq 
$\zeta(s)$ is the Riemann zeta function, 
and we defined $\epsilon_{\tny{IC}}\equiv 
\left(m_e c^2\right)^2/\left[k_B T_{\tny{CMB}}(z=0)\right] = 1.1 \mbox{ PeV}$. 
This approximate result holds for emitted photon energies $\epsilon$ in the
range $4\gamma_{\tny{min}}^2 \epsilon_0 \ll \epsilon \ll
4\gamma_{\tny{max}}^2 \epsilon_0$.  Writing this as:
\beq 3.6 \left( \frac{\gamma_{\tny{min}}}{10^3} \right)^2 \mbox{ keV} \ll 
\epsilon \ll 
3.2 \left( \frac{\gamma_{\tny{max}}}{3\times 10^7} \right)^2 \mbox{ TeV} 
\mbox{ ,} \eeq 
indicates that the approximation holds for the entire relevant $\gamma$-ray
regime.

It is interesting to compare the above results to the approximation used in
\S \ref{sec:EGRB_diffuse_radiation}, where only the first photon emitted by
each electron was treated, thus neglecting the evolution of the electron
distribution.  The approximate result of
equation~(\ref{eq_spectrum_approximation}) agrees with the last equality in
equation~(\ref{eq:exact_spectrum}), up to a function of the power index
$s$.  The two would become identical, if one was to replace $Q(s)$ in the
latter by $1/2$, and take $\epsilon_{\tny{IC}}=0.22 \mbox{ PeV}$.  For a
strong shock, we indeed find that $Q(s=2)=1/2$, and precisely recover
equation~(\ref{eq_strong_spectrum}) - the early estimate for a strong
shock.  This is not surprising, recalling that the electron distribution
for $s=2$ is mostly stationary, rendering evolutionary effects
insignificant.  The numerical mismatch between
equation~(\ref{eq_spectrum_approximation}) and
equation~(\ref{eq:exact_spectrum}) varies as function of $s$, but remains
of order $2$ for relevant values of $s$ ($s \la 8$).

\subsubsection{Space-Time Integration}
\label{sec:space_time_int}
\indent

Two different space-time integration schemes were used.  
The first, \emph{line of sight} integration, calculates the specific 
intensity $I(\epsilon)$ at a given direction $(\theta, \phi)$ on the sky, 
where $\theta$ is its angle with the $z$-axis and $\phi$ is the angle of 
its projection on the $x$-$y$ plane with the $x$-axis. 
Such a scheme may be used to study the statistics of the predicted radiation, 
by calculating moments of the specific intensity: the mean intensity 
$J(\epsilon)=\langle I(\epsilon)\rangle$, 
the correlation of intensities at various angular separations,
etc.  The second scheme provides \emph{direction bin} integration,
calculating the flux predicted within a small region
$\sigma=(\theta_1,\theta_2;\phi_1,\phi_2)$ on the sky, defined by $\theta_1
\leq \theta < \theta_2$ and $\phi_1 \leq \phi < \phi_2$.  
The flux, related to the specific intensity by 
$F(\epsilon) = \int_\sigma I(\epsilon)\cos{\theta} \,d\Omega$, 
is required for sky maps and for source identification and number counts.

Both integration schemes involve simulating an integration volume (IV, for
short), backtracking the propagation of radiation towards an imaginary
observer.  The IV starts at an arbitrarily chosen observer location today,
and is moved away from the observer and back in time, at the speed of
light.  The radiation from sources (i.e. shocked SPH particles) within the
IV is calculated and summed up, to yield the total radiation detected by
the observer today.  Since the simulation-box light-crossing time is much
shorter than the Hubble time, periodic boundary conditions must be used to
simulate the motion of the IV.

We treat the crossing of a shock front by an SPH particle as a single source 
of radiation for our integration scheme and call it, for short, a shock. 
Let such a shock, with specific luminosity $L_{\tny{sh}}(\epsilon)$ and 
redshift $z$, lie at coordinates $(r, \theta_0,\phi_0)$ of a comoving 
coordinate system centered upon the observer. 
The contribution of this shock to the specific flux $F_{\tny{sh}}(\epsilon)$, 
detected by the observer today, satisfies: 
\beq \label{eq:source_flux}
\epsilon F_{\tny{sh}}(\epsilon) = 
\frac{\epsilon^{\prime} L_{\tny{sh}}(\epsilon^{\prime})}
{4\pi d_L(z)^2} 
\mbox{ ,} \eeq 
where $d_L(z)=r R_0 (1+z)$ is the luminosity distance between source and 
observer, $R_0$ is the present-day value of the cosmological scale factor, 
and we have defined $\epsilon^{\prime}\equiv (1+z)\epsilon$. 
With the spatial and temporal resolution used, many ($\gg 10$) shocks 
fall within the IV in each time step. 
This enables us to approximate the shocks as point sources, such that the 
contribution of a shock with space-time coordinates $(\vect{r},t)$ 
to the detected flux is determined by: 
\beq 
\Delta \left[ F(\epsilon, \sigma) \right] = \cases{
F_{\tny{sh}}(\epsilon) & if $(\vect{r},t) \in \mbox{ IV} (\sigma) $ ; \cr
0 & otherwise. \cr } \eeq

The specific intensity contributed by the shock is calculated by 
approximating the latter as a sphere of uniform specific brightness 
$B_{\tny{sh}}(\epsilon)$. 
The proper radius of this sphere is determined by the mass density $\rho(z)$ 
of the SPH particle, according to 
$d_{\tny{pr}}(z) = \left[ 3 M_{\tny{SPH}} / 4\pi \rho(z) \right]^{1/3}$, 
where $M_{\tny{SPH}}$ is the mass of an SPH particle. 
The brightness of the sphere is related to its detected flux by: 
\beq F_{\tny{sh}}(\epsilon) = \pi B_{\tny{sh}}(\epsilon) \left[ 
\frac{d_{\tny{pr}}(z)}{d_A(z)} \right]^2 \mbox{ ,} \eeq 
where $d_A(z)=r R_0 / (1+z)$ is the angular diameter distance between the 
source and the observer. 
Equating this result and equation~(\ref{eq:source_flux}) gives: 
\beq \epsilon B_{\tny{sh}}(\epsilon) = 
\frac{ \epsilon^{\prime} L_{\tny{sh}}(\epsilon^{\prime})} 
{4 \pi^2 d_{\tny{pr}}(z)^2 (1+z)^4} 
\mbox{ ,} \eeq
and the contribution of the source to the specific intensity at 
direction $(\theta,\phi)$ is given by: 
\beq \Delta \left[ I(\epsilon, \theta,\phi) \right] = \cases{
B_{\tny{sh}}(\epsilon) & 
if the ray $(\theta,\phi)$ intersects the sphere; \cr 
0 & otherwise. \cr } \eeq

We approximate the specific luminosity of a shock by: 
\beq L_{\tny{sh}} (\epsilon) \simeq \frac{1}{\Delta t_{\tny{sh}}} 
\left[\frac{M_{\tny{SPH}}}{\rho(z)}\right]
\frac{du_\gamma}{d\epsilon_\gamma} (\epsilon) \mbox{ ,} \eeq 
where $\Delta t_{\tny{sh}} \equiv t_f - t_i$ is the period of time during 
which the shock was detected. 
Since our shock locating-and-timing scheme is limited by the temporal
resolution of the snapshots, $\Delta t_{\tny{sh}}$ tends to overshoot the
true duration of the shock.  This error is partially compensated by the
integration schemes, as they are limited by the same snapshot resolution:
although the approximate luminosity will tend to undershoot its true
value, the probability of including the shocked particle in the IV is
enhanced proportionally.

% --------------------------------------------------------------------------
%                                 Results
% --------------------------------------------------------------------------

\section{Results}
\label{sec:Results}
\indent

Next we present various features of the radiation,
inverse-Compton emitted by the shock-accelerated electrons, according
to the cosmological simulation studied.  We begin by presenting the
predicted spectrum and discussing its implications.  
Next, we present some maps of the simulated $\gamma$-ray sky and discuss 
the sources they are composed of. 
We conclude with source number-counts extracted from such maps.

\subsection{Predicted Spectrum}
\indent

The inverse-Compton spectrum, predicted by the simulation, extends
from the optical band into the far $\gamma$-ray regime, with photon
energies in the range $1\mbox{ eV}<\epsilon<10^4\mbox{ TeV}$ (Figure
\ref{fig:IC_results}).  However, the high energy tail of this spectrum
is modified by photon-photon pair-production interactions, rendering
the universe non-transparent for $\epsilon \ga 1\mbox{ TeV}$ photons.
Photons of $\epsilon \ga 100\mbox{ TeV}$ thus interact with CMB photons
or, at much higher energies, with radio background photons.  Lower
energy ($1\mbox{ TeV}<\epsilon<100\mbox{ TeV}$) photons interact
mainly with the intergalactic infrared (IR) background. 
The $e^{+}e^{-}$ pairs, produced in such interactions, may inverse-Compton 
scatter CMB or IR photons, thus initiating a cascade process 
\cite{Nikishov62,Gould67,Protheroe}.  
As a result, the $\ga 1\mbox{ TeV}$ spectrum is significantly attenuated,
whereas at lower photon energies, piling-up of the cascading photons
slightly enhances (flattens) the spectrum. 

The resulting spectrum provides roughly equal energy flux per decade in photon 
energy, in the photon energy range 
$10\mbox{ keV} \la \epsilon \la 1 \mbox{ TeV}$. 
This flux ranges from 
$\epsilon^2 dJ_\gamma(\epsilon) / d\epsilon_\gamma \simeq 
160 \mbox{ eV cm}^{-2} \mbox{ s}^{-1} \mbox{ sr}^{-1}$ 
at $\epsilon=10\mbox{ keV}$, 
through $100 \mbox{ eV cm}^{-2} \mbox{ s}^{-1} \mbox{ sr}^{-1}$ 
at $\epsilon=1\mbox{ GeV}$, 
and down to 
$50 \mbox{ eV cm}^{-2} \mbox{ s}^{-1} \mbox{ sr}^{-1}$ 
at $\epsilon=1\mbox{ TeV}$.
The number flux of photons in the energy range 
$10\mbox{ keV} \la \epsilon \la 1 \mbox{ GeV}$ is well approximated by:
\beq \label{eq:IC_best_fit} 
\frac{dJ_\gamma}{d\epsilon_\gamma}(\epsilon) \simeq 1.1 \times 10^{-14} 
\left(\frac{\epsilon}{100\mbox{ MeV}} \right)^{-2.04} 
\left( \mbox{eV cm}^2 \mbox{ s sr} \right)^{-1}
\mbox{ ,} \eeq
whereas in the energy range $10\mbox{ GeV} \la \epsilon \la 5\mbox{ TeV}$ 
the spectrum is slightly steeper and may be approximated, 
before including the cascade process, by:
\beq \label{eq:IC_best_fit_GeV}
\frac{dJ_\gamma}{d\epsilon_\gamma}(\epsilon) \simeq 8.8 \times 10^{-19} 
\left(\frac{\epsilon}{10\mbox{ GeV}} \right)^{-2.13}
\left( \mbox{eV cm}^2 \mbox{ s sr} \right)^{-1}
\mbox{ .} \eeq
The slope of the resulting spectrum and the energy range over which it extends 
are in good agreement with the order of magnitude estimates, carried out in 
\S \ref{sec:EGRB_diffuse_radiation}. 
The resulting flux, on the other hand, falls short of the early 
estimates of equation~(\ref{eq:diffuse_IC_spec_flux}) 
and equation~(\ref{eq:diffuse_IC_spec_PS}), 
by a factor $\sim 10$ (Figure \ref{fig:IC_results}). 

The discrepancy between the above flux, calculated from the simulation, 
and the flux predicted in \S \ref{sec:EGRB_diffuse_radiation}, 
is due to a combination of several factors. 
First, the temperature reached by the simulation is a factor of $\sim 2.6$ 
lower than the order-of-magnitude temperature $10^7\mbox{ K}$ assumed 
previously.
The thermal energy gain in the epoch examined ($0<z<2$) is thus smaller than 
the thermal energy used in \S \ref{sec:EGRB_diffuse_radiation} by a factor 
of $\sim 3$.
Second, the energy induced by the identified shocks constitutes a fraction 
$f_{\tny{sh}}\sim 2/3$ of the thermal energy gain in the simulation, 
as opposed to the previous assumption $f_{\tny{sh}}=1$.  
Third, the simulation identifies only approximately \emph{half} of the 
shock-induced electron-energy in electrons, that had sufficient time to cool 
($\gamma>\gamma_{\tny{min}}$), 
in contrast to the factor 
$1-\ln{(\gamma_{\tny{min}})} / \ln{(\gamma_{\tny{max}})} \simeq 2/3$ 
of \S \ref{sec:EGRB_diffuse_radiation}, indicating the presence of weak shocks 
and very recent shocks. 
Finally, in \S \ref{sec:EGRB_diffuse_radiation} we neglected the redshift 
of the emitted energy, caused by the expansion of the universe, 
assuming that most the detected radiation was emitted recently ($z<0.5$). 
However, the energy accumulated by the detected shocks increases almost 
steadily for $z<2$ (Figure \ref{fig:Shock_Stat}, left), 
such that a significant fraction of the predicted flux originates at high 
redshifts: only $\sim 40\%$ of this flux was emitted at $z<0.5$. 
This effect reduces the predicted flux by a factor $\sim 5/3$. 
Summing up the effects listed above, found from Table \ref{tab:ProgParams},
yields the discrepancy factor: 
$3 \cdot (3/2) \cdot (4/3) \cdot (5/3) = 10$.

\subsection{Sky Maps, $\gamma$-ray Sources and Source Number-Counts}
\label{sec:sky_maps_and_SNC}
\indent

By integrating the predicted flux, arriving from different directions, 
one can construct maps of the simulated $\gamma$-ray sky. 
Maps of the entire sky (Figure \ref{fig:FullSkyMap}) and of 
two selected regions (Figure \ref{fig:regional_maps}) are presented in the 
following. 
They reveal a globally isotropic picture, with strong local fluctuations,  
e.g. flux variation over $2$ orders of magnitude at $\sim 1^{\circ}$ 
resolution and over more than $3$ orders of magnitude at $\sim 5^{\prime}$ 
resolution. 
Such maps of the predicted $\gamma$-ray sky can be directly compared to the 
maps produced by detectors such as EGRET \cite{Sreekumar98} aboard the CGRO 
satellite, detectors on board GLAST, planned for launch in 2006, 
and \v{C}herenkov telescopes such as MAGIC, 
VERITAS \footnote{See http://veritas.sao.arisona.edu/} 
and HESS \footnote{See http://mpi-hd.mpg.de/hfm/HESS/HESS.html}. 
Since such devices have finite angular resolution,
the simulated maps must be convolved with appropriate filter 
functions, as demonstrated in Figure \ref{fig:regional_maps}.  
The relevant parameters of EGRET, GLAST and MAGIC are summarized in 
Table \ref{tab:EGRETandGLAST}.

Various $\gamma$-ray sources, of different shapes and sizes, are observed 
in the simulated maps of the $\gamma$-ray sky.  
Some of these sources exhibit irregular shapes, revealing the underlying 
structure of their emitter. 
Two typical examples, depicted in Figure \ref{fig:regional_maps}, are strong  
emission from a semi-spherical source, and weak emission from a filamentary 
object. 
Some interesting features of the semi-spherical $\gamma$-ray source are shown 
in Figure \ref{fig:near_structure}. 
This source is the signature of a nearby rich galaxy cluster, lying 
$\sim 53\mbox{ Mpc}$ from the simulated observer, at a redshift of 
$z \simeq 0.012$. 
The $\gamma$-ray emission from the cluster appears approximately as an 
elliptic ring, corresponding to a diameter $5-10\mbox{ Mpc}$, surrounding the 
location of the cluster and indicating the position of its accretion shock. 
The luminosity has two evident peaks along the circumference of the ring, 
approximately at the locations of its intersections with two large galaxy 
filaments. 
This suggests that the strong emission from these two regions is due to 
shocked gas, channeled by the filaments into the cluster region. 
Other galaxy filaments may be responsible for less luminous peaks of emission 
along the circumference of the ring, on the right hand side of 
Figure \ref{fig:near_structure} images.

The simulated sky maps produced may be decomposed into discrete $\gamma$-ray
sources, yielding number counts of sources above given photon energies 
(Figure \ref{fig:number_count}).  
This is performed using a greedy source-identification algorithm devised for 
the purpose, discussed in Appendix \ref{app:SIA}.  
The resulting number counts of sources above $100\mbox{ MeV}$ and above 
$1\mbox{ GeV}$ fall short of the EGRET detection threshold.
Our simulation thus predicts that none of the $\sim 60$ unidentified 
extragalactic EGRET sources \cite{Ozel} can be attributed to 
intergalactic shock-induced emission. 
The number-counts are especially sensitive to the fraction of shock-energy 
transferred to the relativistic electrons.  
For $\xi_e=0.05$ our simulation predicts a few dozen sources detectable 
by GLAST at both photon energy ranges, among them $16$ well-resolved sources 
above $1 \mbox{ GeV}$ associated with different objects in the sky (see 
Figure \ref{fig:FullSkyMap}). 
For $\xi_e=0.02$, the simulation predicts only one source detectable by 
GLAST, whereas $\xi_e=0.10$ yields more than a hundred GLAST sources, 
well resolved above $1 \mbox{ GeV}$, yet too faint for EGRET detection. 

Since our source-identification algorithm was devised for spherical sources, 
very large, nearby objects tend to be broken down and identified as several 
separate, weaker ones. 
This implies that when comparing our source number counts with analytical 
predictions \cite{WaxmanLoeb2000,Totani2000}, 
one should impose a threshold on the angular size of the sources. 
As evident from Figure \ref{fig:number_count}, Our number counts fall short 
of those of Waxman \& Loeb (2000) by a factor of $\sim 6$. 
This may be explained, at least partially, by the same factors listed above
that reduce the simulated diffuse $\gamma$-ray flux compared to the analytic 
calculation:
the lower average temperature of the simulation suggests less hot, 
massive clusters, and combined with the thermal efficiency of the shocks 
and the fraction of electrons that had sufficient time to cool, 
both lower than assumed in the analytical predictions, 
may account for the discrepancy factor.

% --------------------------------------------------------------------------
%                                Discussion
% --------------------------------------------------------------------------

\section{Discussion}
\label{sec:Discussion}

\indent

We have studied the $\gamma$--ray emission from intergalactic shock waves,
using a hydrodynamic $\Lambda \mbox{CDM}$ cosmological simulation,
according to the model proposed by Loeb \& Waxman (2000).  A simple
approach for extracting large-scale shocks from a Lagrangian simulation,
based on following entropy changes of single simulated particles, was
employed.  This method was shown to identify most of the strong (Mach number 
$M \ga 100$) accretion shocks in the simulation containing a large fraction 
($f_{\tny{sh}} \sim 2/3$) of the gas thermal-energy gain, although some of 
the weaker ($M \la 10$) flow or merger shocks might be missed.  
Relativistic electron populations,
Fermi-accelerated by the shocks, were injected into the shocked gas,
assuming that a fraction $\xi_e=0.05$ of the shock thermal energy was
transferred into relativistic electrons.
The inverse-Compton radiation, emitted as these electrons scatter CMB photons,
was then calculated and integrated over space and time, to yield the
radiation detected by an imaginary observer.
  
This radiation was found to extend from the optical band and into the far
$\gamma$-ray regime, containing roughly equal energy flux per decade in
photon energy, $\epsilon^2 [dJ_\gamma (\epsilon) / d\epsilon_\gamma] \simeq
50-160 \mbox{ eV cm}^{-2} \mbox{ s}^{-1} \mbox{ sr}^{-1}$, in the photon
energy range $10\mbox{ keV} \la \epsilon \la 1\mbox{ TeV}$.  
The calculated spectral \emph{slope}, $dJ / d\epsilon \sim \epsilon^{-2.04}$ 
for $10 \mbox{ keV} \la \epsilon \la 1\mbox{ GeV}$ 
and $dJ / d\epsilon \sim \epsilon^{-2.13}$ 
for $10 \mbox{ GeV} \la \epsilon \la 5\mbox{ TeV}$ (before accounting for 
photon-photon pair-production cascade), 
is consistent with the predictions of Loeb \& Waxman (2000) and with the 
EGRET observations.  
The \emph{energy flux} constitutes $\sim 10\%$ of the EGRB flux, or up to 
$\sim 15\%$ of the EGRB flux after subtracting the expected contribution from 
unresolved point sources based on empirical modeling of the luminosity 
function of blazars \cite{Mukherjee}. 

The simulated $\gamma$-ray sources, mostly spheroidal halos associated with 
large-scale structure, fall short of the EGRET detection threshold, thus 
producing a diffuse background.  Hence, the simulation predicts that none of 
the $\sim 60$ unidentified EGRET extragalactic $\gamma$-ray sources can be 
attributed to intergalactic shock-induced radiation.  
Several sources do fall within the detection range of GLAST, 
planned for launch in 2006, provided that $\xi_e \ga 0.03$.  
The number of detectable sources is sensitive to the fraction of shock-energy 
transferred to the relativistic electrons: 
For $\xi_e=0.05$ we find several dozen sources detectable by GLAST, 
among them $16$ well-resolved sources 
(see Figures \ref{fig:FullSkyMap}, \ref{fig:number_count}),   
whereas $\xi_e = 0.10$ is still insufficient for EGRET detection, 
but leads to more than a hundred well-resolved GLAST $\gamma$-ray sources.  
Such sources could be identified as emission from intergalactic 
shock-accelerated electrons, by association with
large-scale structure, a characteristic $dJ / d\epsilon \sim \epsilon^{-2}$
spectrum, extending up to the far $\gamma$-ray regime, and a low cutoff at
photon energy $\epsilon_{\tny{min}} \simeq 5 \left( t / 10^9 \mbox{ yr}
\right)^{-2} \mbox{ keV}$, caused by low-energy electrons not having
sufficient time to cool during the lifetime $t$ of the source.  
Detection of intergalactic shock-induced $\gamma$-ray sources may be used 
to calibrate $\xi_e$, once their temperature has been determined.  
The shock-induced $\gamma$-ray emission, a first direct tracer of the 
warm-hot IGM ($10^5 \mbox{ K}\la T \la 10^7\mbox{ K}$), may thus be used to 
determine the baryon density in the present-day universe. 

We have examined the $\gamma$-ray signature of a nearby simulated rich galaxy 
cluster, lying at a redshift $z \simeq 0.012$. 
An elliptic emission ring of a diameter corresponding to $5-10\mbox{ Mpc}$ 
surrounds the cluster, tracing the location of its accretion shock
(see Figures \ref{fig:regional_maps}, \ref{fig:near_structure}). 
The luminosity peaks along the circumference of this ring, probably at the 
locations of intersections with galaxy filaments, channeling gas into the 
cluster region. 
Similar features should be detected by GLAST or by the MAGIC telescope 
in nearby rich galaxy clusters, at photon energies above $10 \mbox{ GeV}$, 
and by VERITAS and HESS above $200 \mbox{ GeV}$, if the background signal is 
assumed flat.  
A ring-like feature resembling the simulated emission should thus be 
detected in such galaxy clusters lying at redshifts $z\la 0.01$, 
whereas the brightest regions of emission along the circumference of the ring 
could be detected in clusters with redshifts up to $z \simeq 0.025$. 
These bright regions of emission lie above the background signal and should 
be detected at redshifts $z \la 0.015$ even for $\xi_e=0.02$, 
whereas $\xi_e=0.10$ will enable the detection of the emission ring itself at 
similar redshifts. 
Detection of such a ring will enable a direct study of the cluster accretion 
shock, as well as establish the validity of our model. 
Detected bright regions along the circumference of such a ring may be tested 
to comply with data regarding a nearby galaxy filament and be used to measure 
the density and velocity of the untraced gas in such a filament. 

As mentioned above, our algorithm for extracting shocks succeeded in 
capturing strong shocks, but missed some of the weaker ($M \la 10$) ones. 
Accounting for these weak shocks could only have a minute effect on the 
predicted diffuse $\gamma$-ray radiation, since most ($\sim 2/3$) of the 
thermal energy has been identified in strong shocks and because the steep 
spectrum of electrons accelerated by weak shocks reduces their $\gamma$-ray 
efficiency. 
Weak shocks could, however, increase the $\gamma$-ray brightness of nearby 
structures such as galaxy clusters encompassing ongoing merger processes, 
mainly in low photon energies, slightly altering the appearance of their 
inner regions and enhancing the predicted source number counts. 

In our work we have neglected the radiation resulting from the protons 
accelerated in the intergalactic shocks. 
These protons lead to $\gamma$-ray emission as they scatter off thermal 
protons to produce pions, most of the $\gamma$-ray flux originating from 
$\pi^0$ decay. 
Even with a large proton-to-electron energy ratio $\eta$, the radiation 
resulting from the relativistic protons is, on average, 
negligible compared to the emission from electrons. 
Only in the densest ($n > 10^{-3} \mbox{ cm}^{-3}$) regions in the cores of 
clusters, the relativistic protons may alter the $\gamma$-ray spectrum. 
This effect may make the cores of clusters even brighter $\gamma$-ray 
sources, provided that is not eliminated by diffusion of the relativistic 
protons out of the over-dense regions. 
For the values of $\xi_e$ and $\eta$ discussed above, the relativistic 
protons are dynamically significant and play an important role in shock 
dynamics and in particle acceleration. 
These effects, possibly detected in the radio spectra of some supernovae 
remnants, are small \cite{Reynolds92}, suggesting that the linear approach 
adopted in this paper is sufficiently adequate. 

One can confirm that at least part of the EGRB is associated with the
large-scale structure of the universe, by cross-correlating $\gamma$-ray
maps with other sources that trace the same structure, such as galaxies,
X-ray gas and the Sunyaev-Zel'dovich effect.  The intergalactic shock-origin 
of $\gamma$-ray radiation may be verified by cross-correlating the EGRB with
radio maps, partly attributed to synchrotron emission from the same
shock-accelerated electrons.  The synchrotron radio radiation is
sub-dominant to the CMB, although its strong fluctuations on sub-degree
scales were shown to contaminate CMB fluctuations at low 
($\nu \la 10\mbox{ GHz}$) frequencies \cite{WaxmanLoeb2000}.  
Detection of a radio --
$\gamma$-ray cross-correlation signal from nearby clusters may be used to
measure the intergalactic magnetic field.  
The auto-correlation of inverse-Compton or synchrotron radiation, as well as
their cross-correlation, may be calculated directly from our simulation. 

A study of the shock-induced component of the extragalactic radiation,
combined with a calculation of shock-induced emission according to
cosmological simulations as described in this paper, may provide an
important test of structure-formation models.  Angular statistics of the
$\gamma$-ray radiation, both on its own and cross-correlated with other
sources, may be measured and compared to the predictions of such
simulations.  This will enable a study of the large-scale structure of the
universe, and provide insight to the structure-formation process itself.
We plan to apply the tools, developed in this research, to other
cosmological simulations, including high-resolution simulations
incorporating radiative cooling processes and supernovae feedback
(e.g. Springel \& Hernquist 2002b).  Such research will permit an extensive
study of shock-induced radiation, in the framework of various cosmological
structure-formation models.  Other physical phenomena, such as the
acceleration of cosmic-rays by intergalactic shocks, may thus be
investigated as well.

Order-of-magnitude estimates, presented in \S
\ref{sec:EGRB_diffuse_radiation}, predict an inverse-Compton flux higher
by a factor of $\sim 10$ than obtained from the simulation we have
studied.  This is primarily attributable to the lower present-day temperature 
reached by the simulation, $T_0 \simeq 4 \times 10^6 \mbox{ K}$, and to its 
slower heating-rate, which seem to comply with other recent simulations. 
Our simulation neglected radiative
cooling, suffered from relatively low mass resolution ($\sim 10^{11}
M_{\odot}$), and did not include feedback from supernovae explosions,
photo-heating the surrounding matter.  The effect of radiative cooling
on the accretion shocks is expected to be small, as cooling is
efficient in hot, dense regions, beyond these shocks, and in cold,
dilute regions, where the temperatures are low to begin with.  Cooling
should have a non-negligible effect on merger and flow shocks, but
since these are much weaker than the accretion shocks, this should not
significantly alter our results.  In order to examine the sensitivity
of our results to the low resolution of the simulation, a preliminary
check of a similar adiabatic $\Lambda$CDM simulation, with identical
cosmological parameters but with eight times the mass resolution, was
carried out, exhibiting no substantial difference from the results
presented above.  Outflows from galaxies (due to starburst 
or quasar activities) could also result in shock acceleration of
relativistic electrons as they impact on the surrounding IGM. 
At the same time, preheating of the collapsing matter by, for example,
supernovae, may lead to weaker, cooler shocks, as recently pointed out
by Totani \& Inoue (2001), and thus result in softer, weaker photon
emission.  However, Dav\'{e} et al. (2001b) find that the
luminosity-temperature relation for galaxy clusters, which originally
motivated the addition of preheating to structure formation models,
may in fact be attributed to radiative cooling.  Therefore, preheating
may not have played an important role in structure formation after
all, and thus have little effect on our results.

% ------------------------------------------------------------------------
% Comments
% ------------------------------------------------------------------------
\emph{Note added in proof}: 
Recently, Scharf \& Mukherjee (2002) have pointed out a possible association 
of $\gamma$-ray emission with the locations of galaxy clusters, by 
cross-correlating high Galactic latitude EGRET data with Abell clusters. 
The correlation signal, identified at a confidence level of $3\sigma$, 
is broadly consistent with the extended emission and with the source number
counts discussed above, suggesting good prospects for the detection of 
extended $\gamma$-ray emission from clusters by future observations.

% ------------------------------------------------------------------------
% Acknowledgments
% ------------------------------------------------------------------------

\acknowledgments
\emph{Acknowledgments:}
This work was supported in part by grants from the Israel-US BSF
(BSF-9800343) and NSF (AST-9900877).  UK \& EW thank the Harvard-Smithsonian
Center for Astrophysics for its kind hospitality during a period where part
of this work was done. AL thanks the Minerva Einstein Center 
for partial support during his visit to the Weizmann Institute.
EW is the incumbent of the Beracha foundation career development chair.

% ------------------------------------------------------------------------
% Appendices
% ------------------------------------------------------------------------

\appendix

\section{Integration Parameters}
\label{app:formulae}
\indent

In order to perform the space-time integration schemes described in section 
\ref{sec:space_time_int}, one must find the redshift dependence of several 
parameters: 
the time that elapsed between redshift $z$ and the present, 
the coordinate distance, traversed by a photon in that time, 
and the minimal Lorentz factor of electrons that lost a significant fraction 
of their energy since redshift $z$. 
Formulae for these parameters are provided below.
	
Using the FRW equations, we find that the time that elapsed between redshift 
$z$ and the present is given by: 
\beq \Delta t = t_0 - t(z) 
 = \int_{R(z)}^{R(z=0)}\frac{1}{\dot{R}}\,dR 
 = H_0^{-1} \int_0^z \frac{1}{(1+z)h(z)}\,dz 
\mbox{ ,} \eeq 
where $h(z)\equiv H(z) / H_0$. 
The coordinate distance traversed by a photon in that period is given by: 
\beq r = g \left[ \int_{R(z)}^{R(z=0	)} \frac{c}{\dot{R}R} \,dR \right] 
 = g \left[ \frac{c}{R_0 H_0} \int_0^z h	(z)^{-1} \,dz \right] 
\mbox{ ,} \eeq
where $g(x)$ is a function of the curvature $k$, defined by: 
\beq g(x) \equiv \cases{
\sin{(x)} & for $k=1$ ; \cr 
x & for $k=0$ ; \cr
\sinh{(x)} & for $k=-1$ . \cr } \eeq

For our $\Lambda\mbox{CDM}$ model, where $\Omega_\Lambda+\Omega_M=1$ 
and $h(z)=\sqrt{\Omega_\Lambda+(1+z)^3 \Omega_M}$, 
the above equations are solved to give: 
\beq
\Delta t=\frac{2}{3H_0\sqrt{\Omega_{\Lambda}}}
\ln\left[{\frac{(1+\sqrt{\Omega_{\Lambda}})(1+z)^{3/2}}
{\sqrt{\Omega_{\Lambda}}+\sqrt{\Omega_{M}(1+z)^3+\Omega_{\Lambda}}}}
\right]
\mbox{ ,} \eeq
and: 
\beq r = \frac{c}{3R_0H_0}
        \frac{{(-1)}^{2/3}}{\Omega_{M}^{1/3}\Omega_{\Lambda}^{1/6}}
        \mbox{ }\beta\left[\frac{1}{\Omega_{\Lambda}},
        1+\frac{\Omega_{M}}{\Omega_{\Lambda}}\left(1+z\right)^3,
        \frac{1}{2},\frac{1}{3}\right]
\mbox{ ,} \eeq
where $\beta$ is the incomplete beta function, defined by:
\beq    \beta\left[t_1,t_2,a,b\right]\equiv
        \int_{t_1}^{t_2}t^{a-1}\left(1-t\right)^{b-1}\,dt
\mbox{ .} \eeq 

In a similar fashion, we find the redshift dependence of 
$\gamma_{\tny{min}}(z,p)$, the minimal Lorentz factor of an electron, 
that looses a fraction $p$ of its energy between redshift $z$ and today 
by inverse-Compton scattering. 
This parameter, which determines the low cutoff of the resulting spectrum, is 
given by: 
\beq \label{eq:gamma_min} \gamma_{\tny{min}}(z,p) = \frac{p}{(1-p)} 
\frac{3 m_e c H_0}{4 \sigma_T u_{\tny{CMB}}(z=0)} 
\left[ \int_0^z \frac{(1+z)^3}{h(z)}\,dz \right]^{-1} \mbox{ ,} \eeq
which for our $\Lambda\mbox{CDM}$ model can be solved to give: 
\begin{eqnarray}  
\gamma_{\tny{min}}(z,p) & = & 400 h_{67} \frac{p}{(1-p)} \Omega_M /
\left\{ (1+z)h(z)-1 + \sqrt{\Omega_\Lambda}
F\left(\frac{\Omega_M}{\Omega_\Lambda}\right) \right. \nonumber \\ 
& - & \left. (1+z)\Omega_\Lambda F\left[ (1+z)^3 
\frac{\Omega_M}{\Omega_\Lambda} \right] \right\} \mbox{ ,} \end{eqnarray}
where $F(x)$ is the hyper-geometric function 
$_2F_1\left(\frac{1}{3},\frac{1}{2}, \frac{4}{3}, -x\right)$, defined by 
$_2F_1\left(\frac{1}{3},\frac{1}{2}, \frac{4}{3}, -x\right) \equiv 
\sum_{k=0}^{\infty} \left(\frac{1}{3}\right)_k \left(\frac{1}{2}\right)_k 
(-x)^k / \left(\frac{4}{3}\right)_k k!$. 

The redshift dependence of the parameters calculated above, also demonstrating 
the criteria for snapshot selection (simulation-box light-crossing time), 
is depicted in Figure \ref{fig:Vals}.

\section{Source Identification Algorithm}
\label{app:SIA}
\indent
 
In order to count discrete $\gamma$-ray sources, predicted by the
simulation, we have devised a greedy algorithm for extracting point
sources from a sky map.  This algorithm is used to isolate such
sources, determine their location and flux, and measure their angular
size.  The algorithm operates iteratively, identifying sources and
removing them from the map, until the contrast of the remaining map is
smaller than some imposed threshold (of order $2$).  At each
iteration, the algorithm searches for the brightest point in the sky,
and attempts to identify the brightest source associated with it.  Our
candidate for the source is assumed to have its kernel positioned
about this bright point.  We determine the area around the kernel,
associated with the source by examining concentric rings about the
kernel and adding them to the 'growing' source.  This accumulation
process lasts as long as the flux within such a ring is sufficiently
higher than the background, throughout its circumference, and yet
monotonically decreasing, at least at one point along the
circumference.  Once the entire source has been identified, the flux
associated with it is carefully removed from the map. 
The flux subtracted from each ring is determined according to
the faintest point along its circumference, in order to prevent
upsetting adjacent sources.

The source number counts produced by the algorithm were analyzed in section
\ref{sec:sky_maps_and_SNC}, with respect to the experimental devices EGRET
and GLAST.  
In order to do so, algorithm parameters were tuned to mimic the
source identification process, used to analyze the experimental data.
Hence, the angular resolution of the analyzed sky maps was set to match the
point-source resolution of the experimental devices, and the angular size of 
each source was compared to the photon-position error of each experiment, as 
illustrated in Figure \ref{fig:number_count}.

Figure \ref{fig:SIA} demonstrates the results of the algorithm, when
applied to two regions in the simulated $\gamma$-ray sky.  
The sky maps displayed in this paper were produced by approximating the 
source, associated with each SPH particle, as a rectangle in $\theta-\phi$
space.  The resulting image of a source, produced by a single SPH
particle, thus assumes a trapezoidal-like shape when mapped to a
surface, centered at the source.  Similarly, The concentric 'rings'
discussed above are in fact rectangles in $\theta-\phi$ space.  The
identified sources are thus depicted in Figure \ref{fig:SIA} as
ellipses instead of rectangles for illustrative purposes only.

% -----------------------------------------------------------------------------
% ----------------------------    BIBLIOGRAPHY     ----------------------------
% -----------------------------------------------------------------------------

% -------------------------   END OF BIBLIOGRAPHY     -------------------------

% ------------------------------------------------------------------------
% Figures:
% ------------------------------------------------------------------------

\clearpage

% Slice through simulation box:
% -----------------------------
\begin{figure}
\plottwo{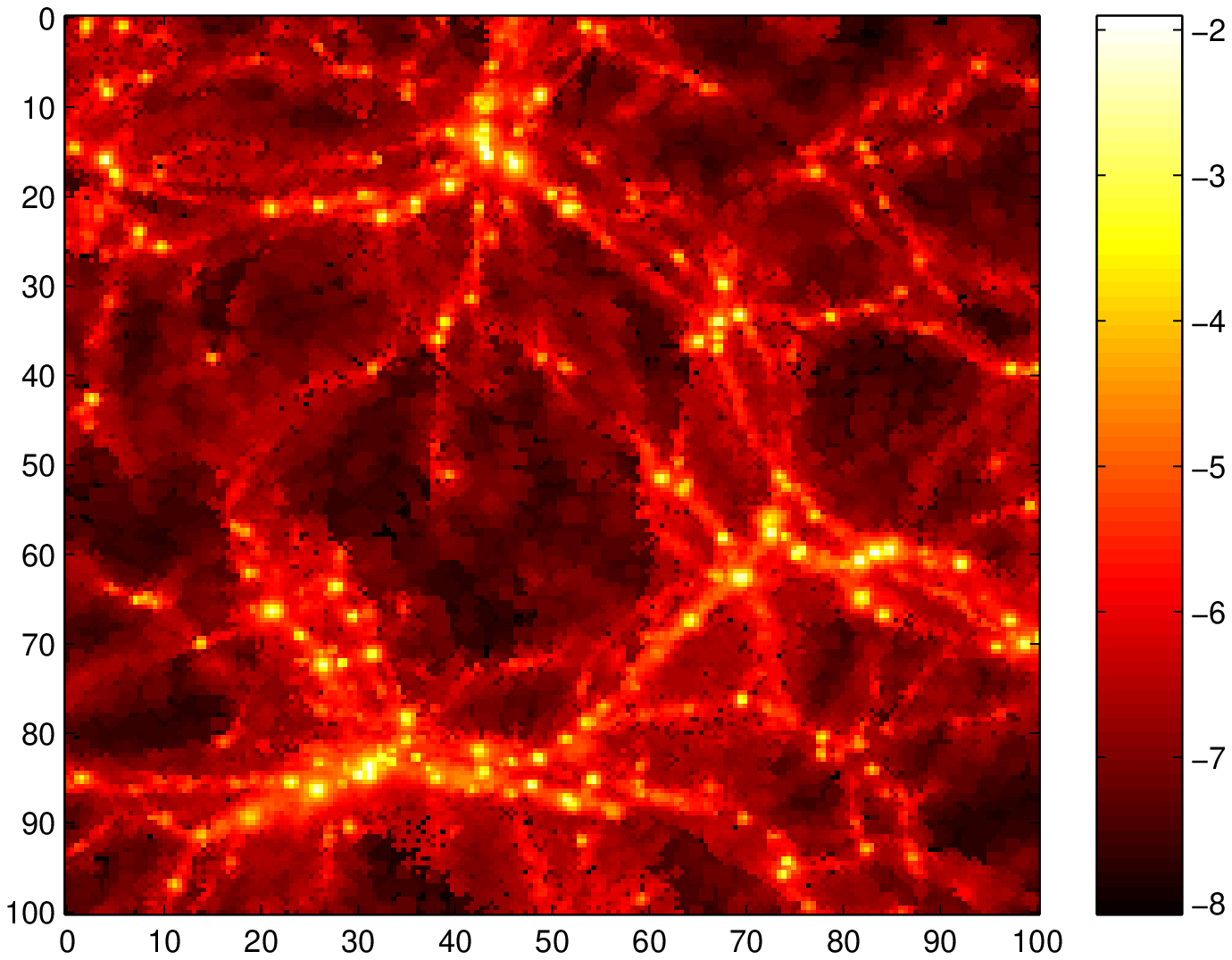}{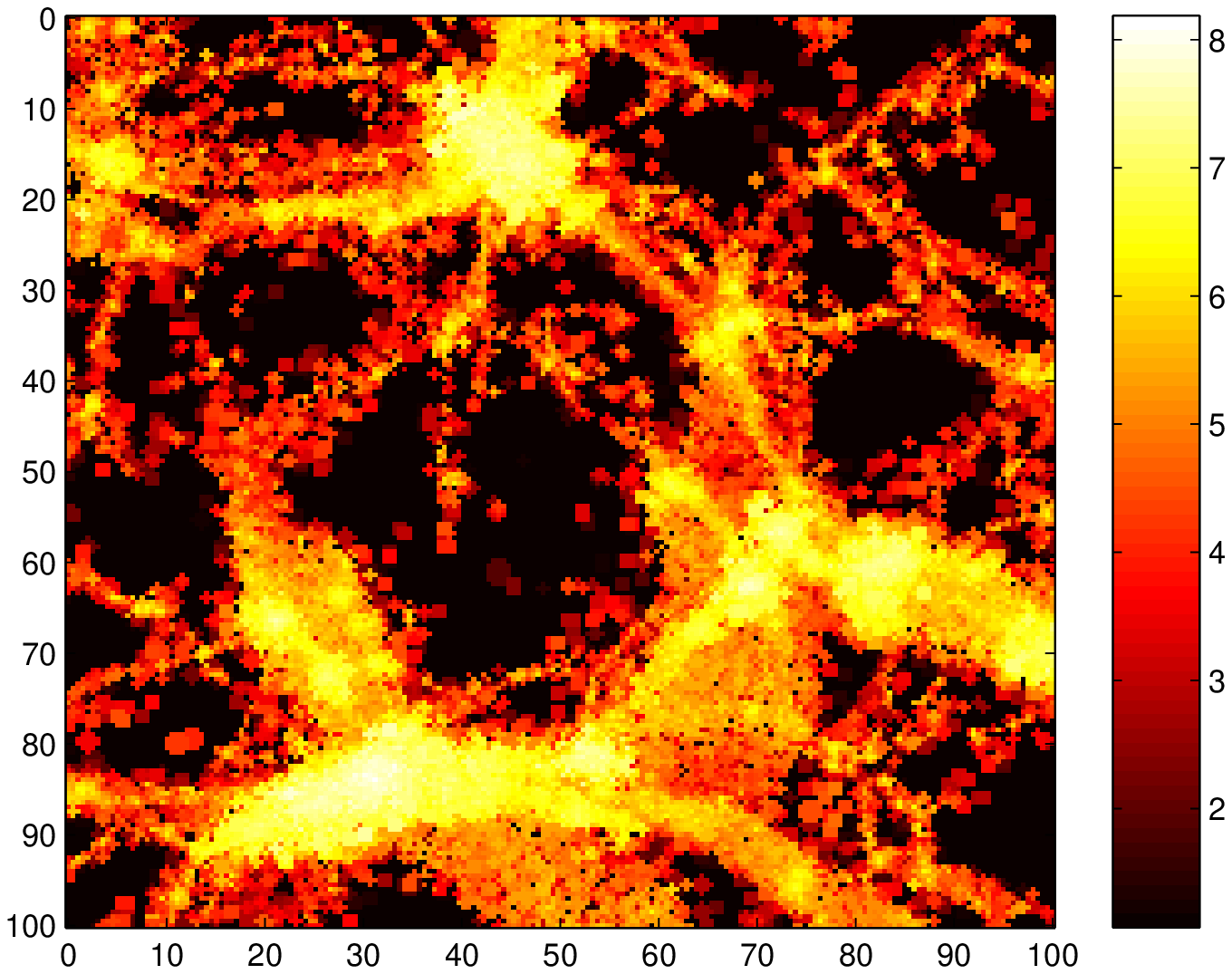}
\caption{
Imaging a partial slice of the simulation box at $z=0$. 
Width and height are $100 \mbox{ Mpc}$, depth is $10 \mbox{ Mpc}$. 
\newline
\emph{left}: Maximal baryon number density through the slice. 
Color scale: $\log_{10}{ \left( n \left[ \mbox{cm}^{-3} \right] \right) }$. 
\newline
\emph{right}: Maximal temperature through the slice. 
Color scale: $\log_{10}{\left(T \left[\mbox{K} \right] \right)}$.
}
\label{fig:map3D}
\end{figure}

% Thermodynamical evolution of the simulation:
% --------------------------------------------
\begin{figure}
\plottwo{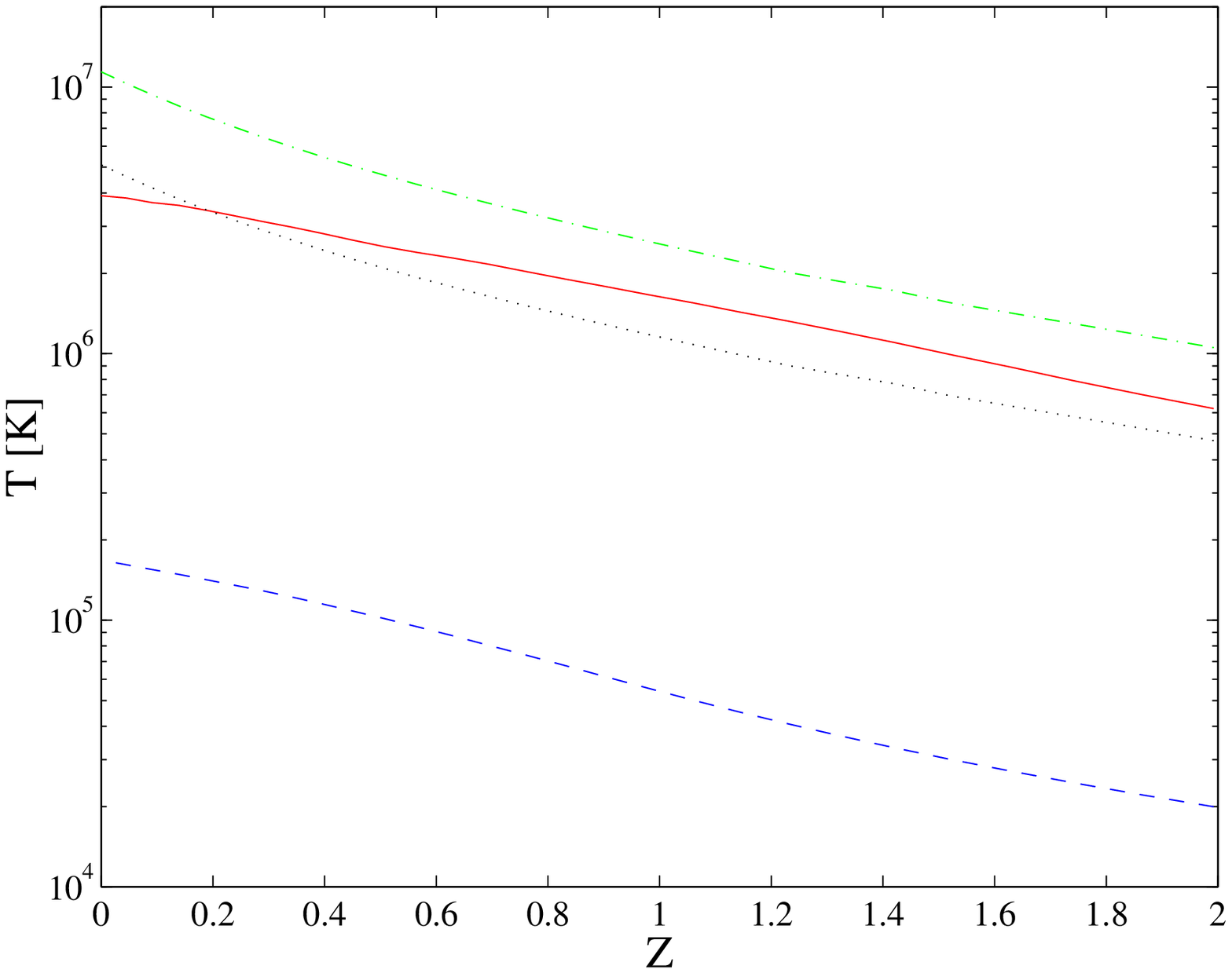}{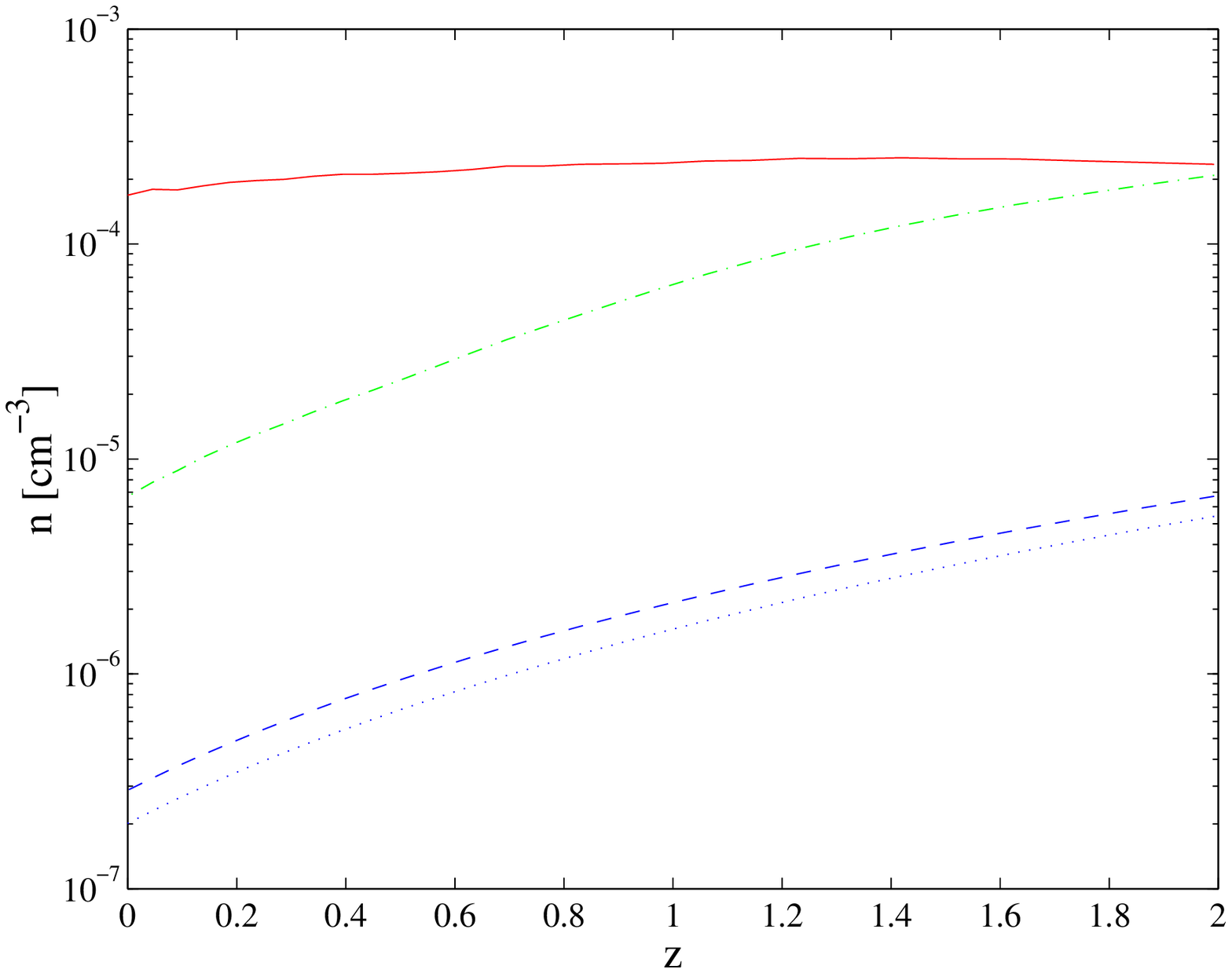}
\caption{
Thermodynamic evolution of the simulated universe at recent epochs. 
\newline
\emph{left}: 
The temperature averaged over mass ($\rho^1$ averaging, solid line) and over 
volume ($\rho^0$ averaging, dashed line). 
The mass averaged temperature according to the approximation of 
equation~(\ref{eq_T_avr}) is plotted as well, for $k=1$ (as assumed in 
\S \ref{sec:structure_formation_shock_waves}, dash-dotted line) 
and for $k=0.67$ (best fit to data, dotted line).  
\newline
\emph{right}: 
Baryon number density averaged over mass (solid line), over volume 
(dashed line) and over temperature (dash-dotted line). 
The volume-averaged density overshoots its predicted value 
($\sim \left[ 1+z \right]^3$, dotted line), probably due to limited 
simulation resolution in dilute regions. 
}
\label{fig:Sim_Stat}
\end{figure}

% Radiative cooling processes:
% ----------------------------
\begin{figure}
\plottwo{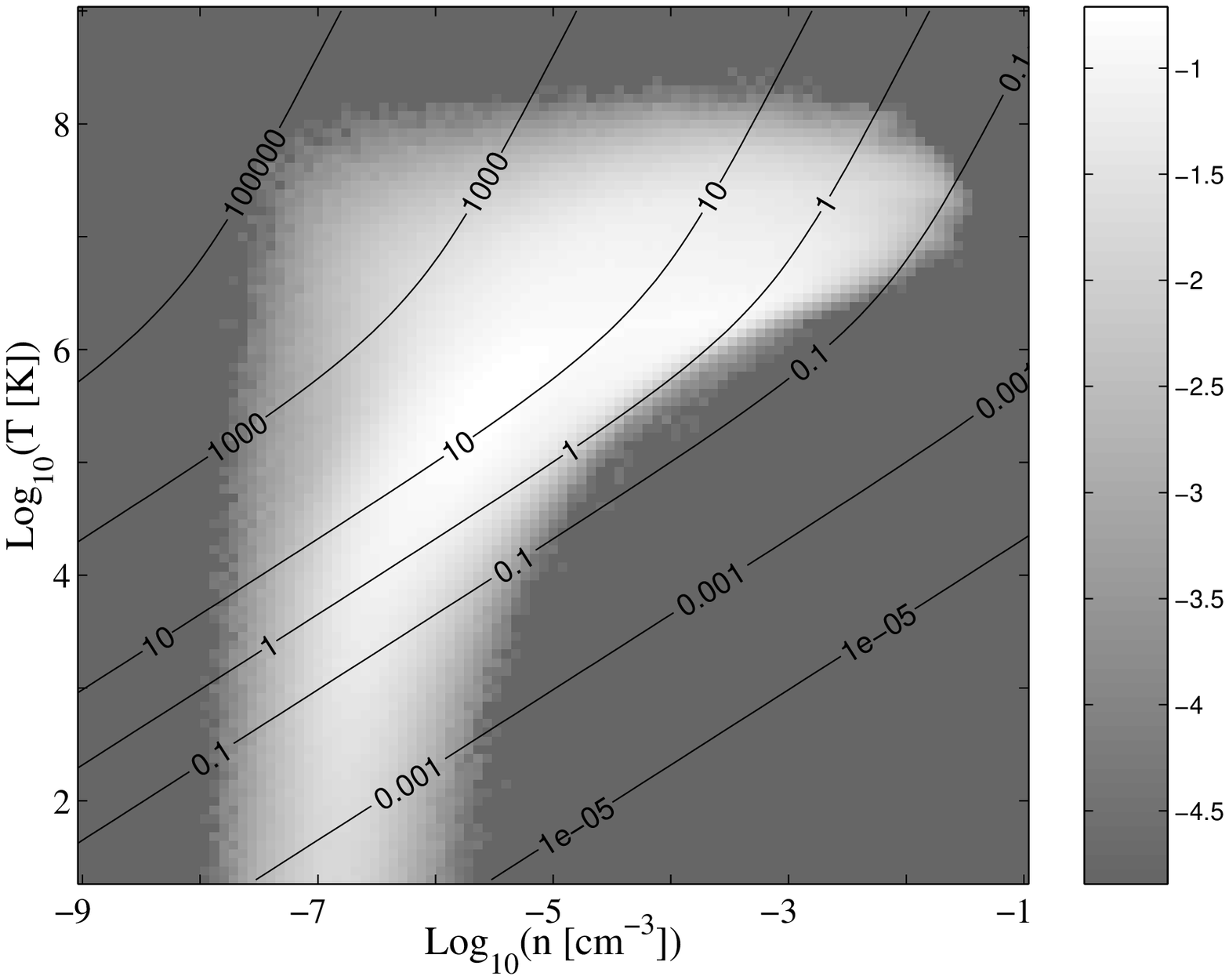}{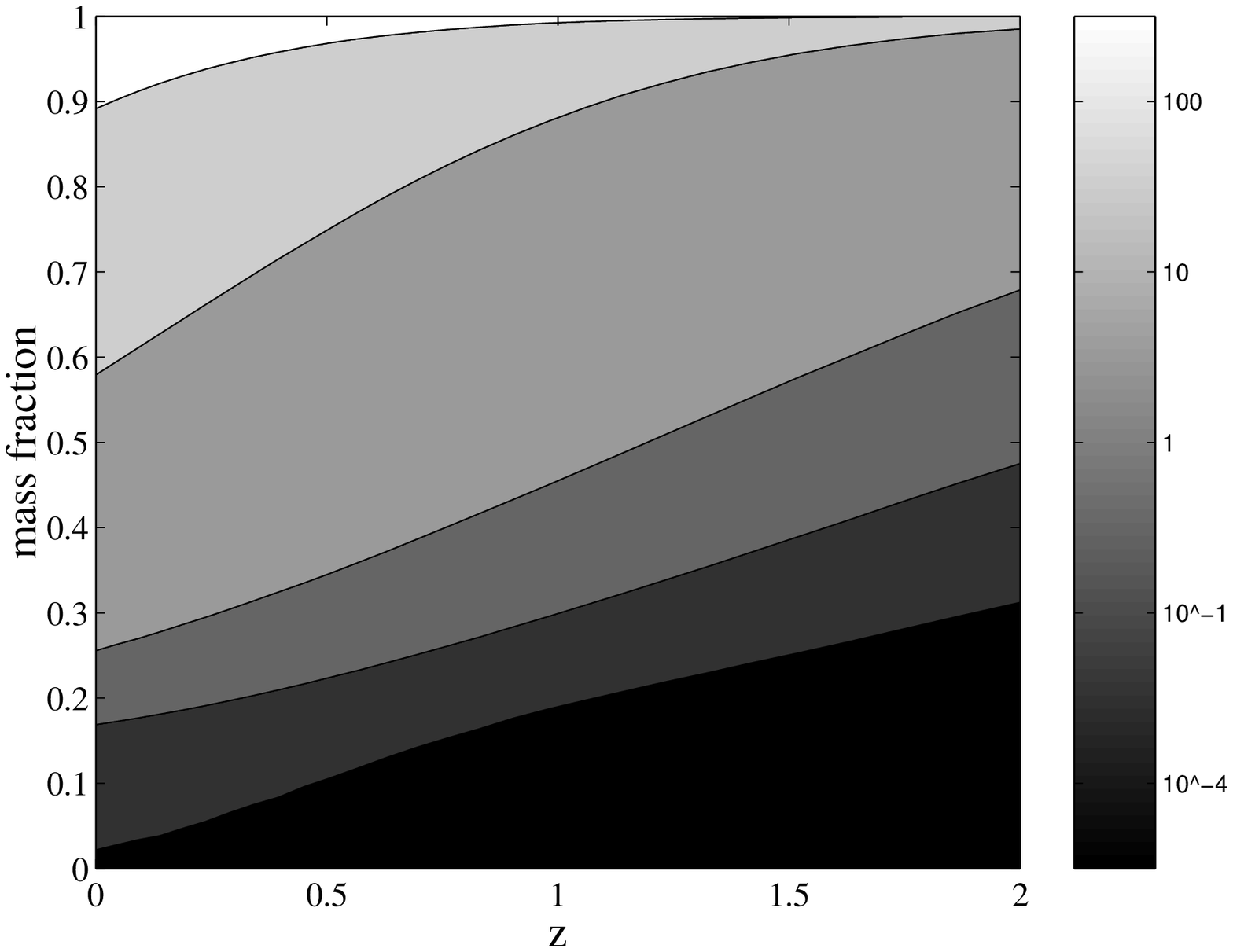}
\caption{
Radiative cooling of the simulated gas. 
\newline
\emph{left}: 
Mass distribution at $z=0$ in the phase space of temperature and baryon number 
density. 
Gray-scale: the logarithm (base $10$) of the mass fraction 
$m_{\tny{tot}}^{-1} \Delta m/ 
\left[ \Delta \log_{10}(T) \cdot \Delta \log_{10}(n) \right]$. 
Superimposed black contours denote the cooling time $t_{\tny{cool}}$
due to Bremsstrahlung and line emission \cite{Peacock}, 
normalized to the Hubble time $t_{\tny{H}}$. 
\newline
\emph{right}: 
Mass fraction of particles with various normalized cooling times 
in different epochs. 
Gray scale: $t_{\tny{cool}} / t_{\tny{H}}(z)$. 
}
\label{fig:Cool_Stat}
\end{figure}

% Entropy statistics + results of shock extraction scheme:
% --------------------------------------------------------
\begin{figure}
\plottwo{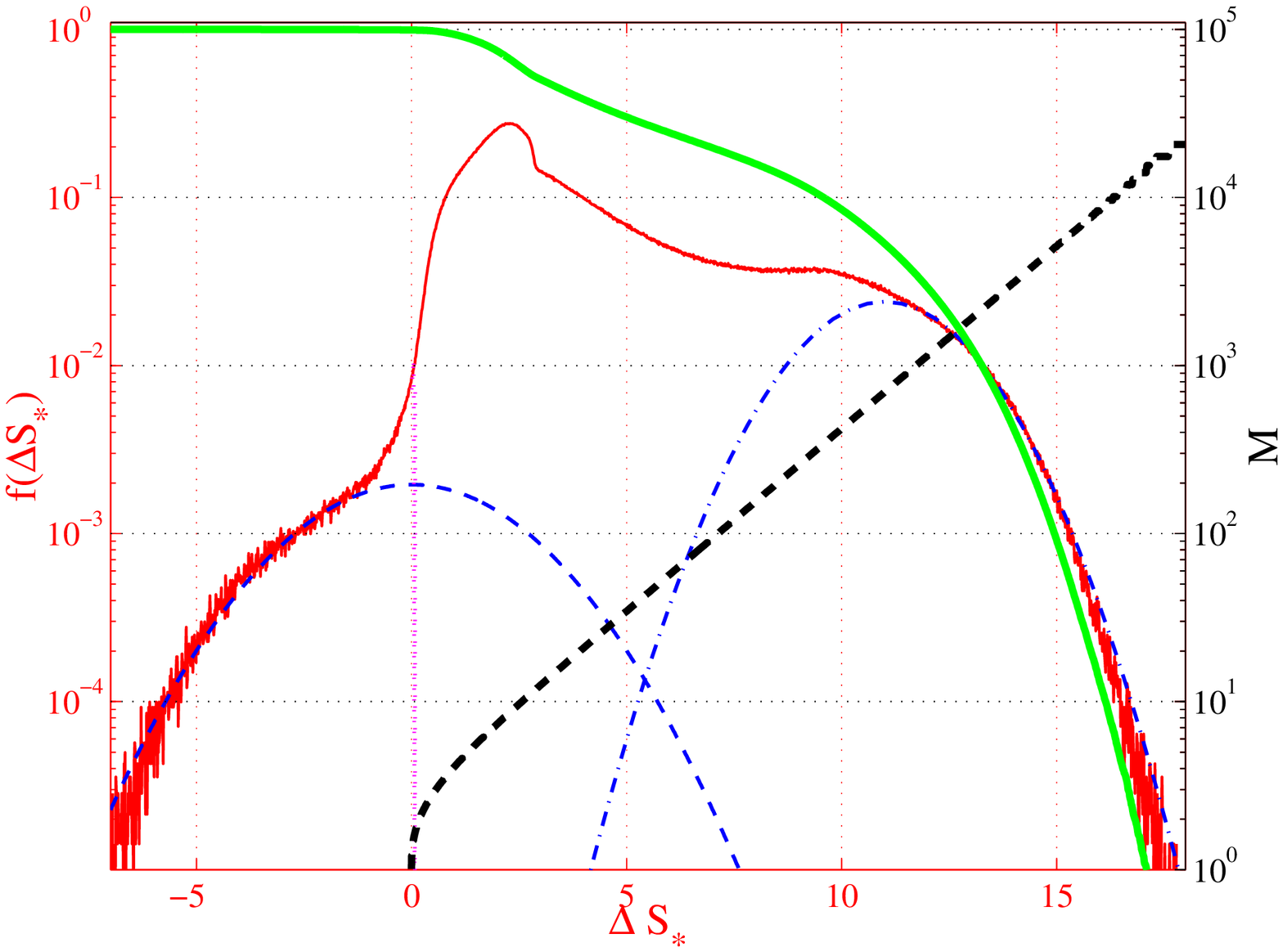}{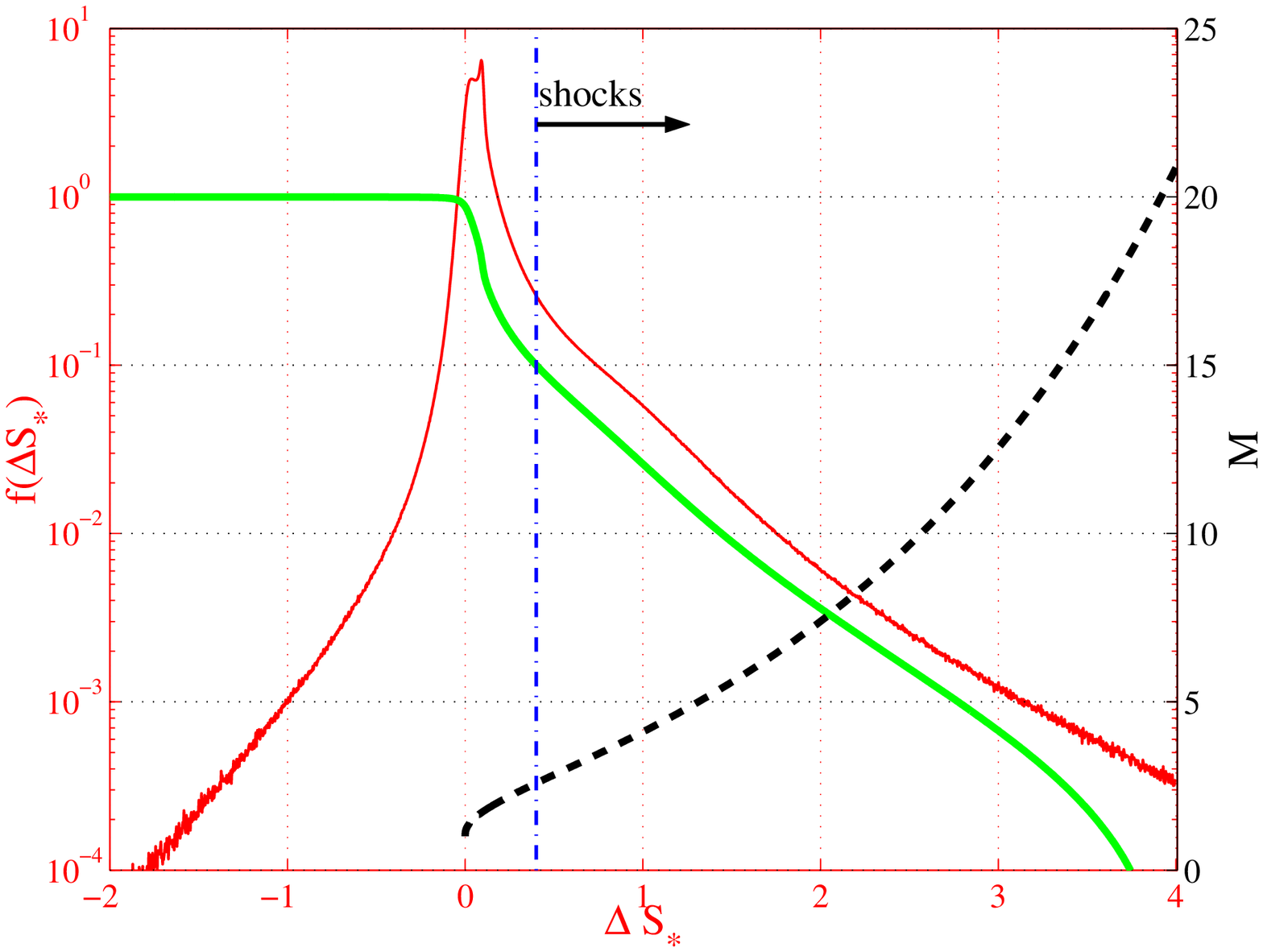}
\caption{
Statistics of normalized entropy-change, $\Delta S_*$,
accumulated by simulated particles. 
\newline 
Plotted are the mass fraction
$f(\Delta S_*^{\prime})= m_{\tny{tot}}^{-1} 
dm(\Delta S_*^{\prime}) / d\Delta S_*$ (thin solid line), 
integrated mass fraction
$m_{\tny{tot}}^{-1}m(\Delta S_* > \Delta S_{*}^{\prime})= \int_{\Delta
S_{*}^{\prime}}^{\infty} f(\Delta S_*) \,d\Delta S_*$ (thick solid line) 
and the shock Mach number, $M$, corresponding to positive $\Delta S_*$ 
values (thick dashed line, right axes).
\newline
\emph{left}: 
Total entropy accumulated in the recent epoch, 
 $\Delta S_* = S_*(z=0) - S_*(z=2)$.
The distribution of negative $\Delta S_*$ values, approximated by a 
normal distribution (thin dashed line), is likely due to 
numerical errors, leaving behind a narrower distribution 
(well-deconvolved up to $\Delta S_*\sim 1$, dotted line). 
High positive $\Delta S_*$ values are also well-fitted by a normal 
distribution (dash-dotted line). 
\newline
\emph{right}: Average $\Delta S_*$ between two consecutive snapshots.  A
minimal threshold for shock detection (dash-dotted line) was imposed at
$\Delta S_*=0.4$, where estimated mass fraction due to numerical noise is
$f(\Delta S_*) \sim 1\%$ (according to negative $\Delta S_*$ distribution).
On average, $10\%$ of the gas experiences higher entropy increases between 
consecutive snapshots.  }
\label{fig:dS_Stat}
\end{figure}

% Entropy statistics among shocks:
% --------------------------------
\begin{figure}
\plottwo{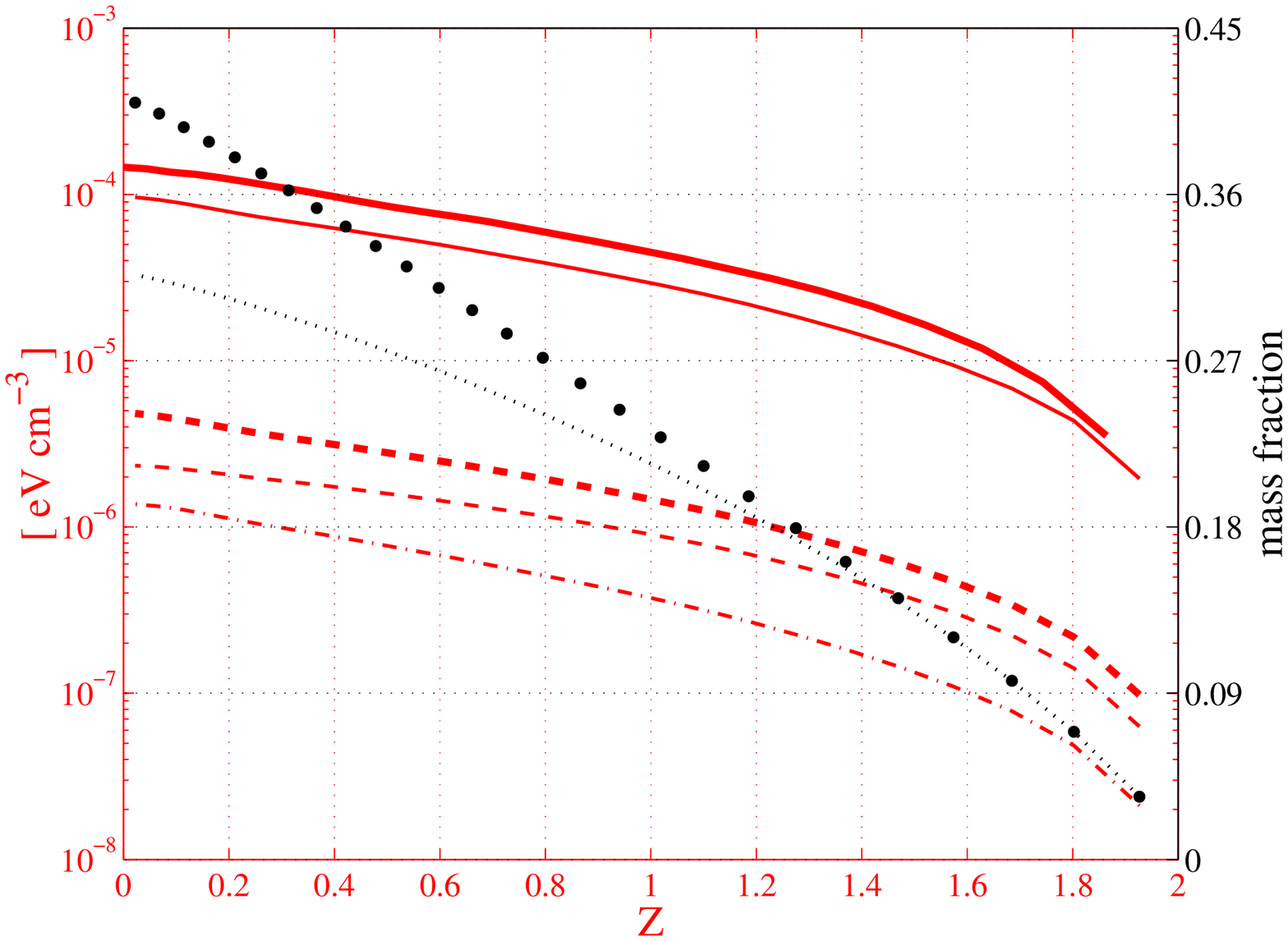}{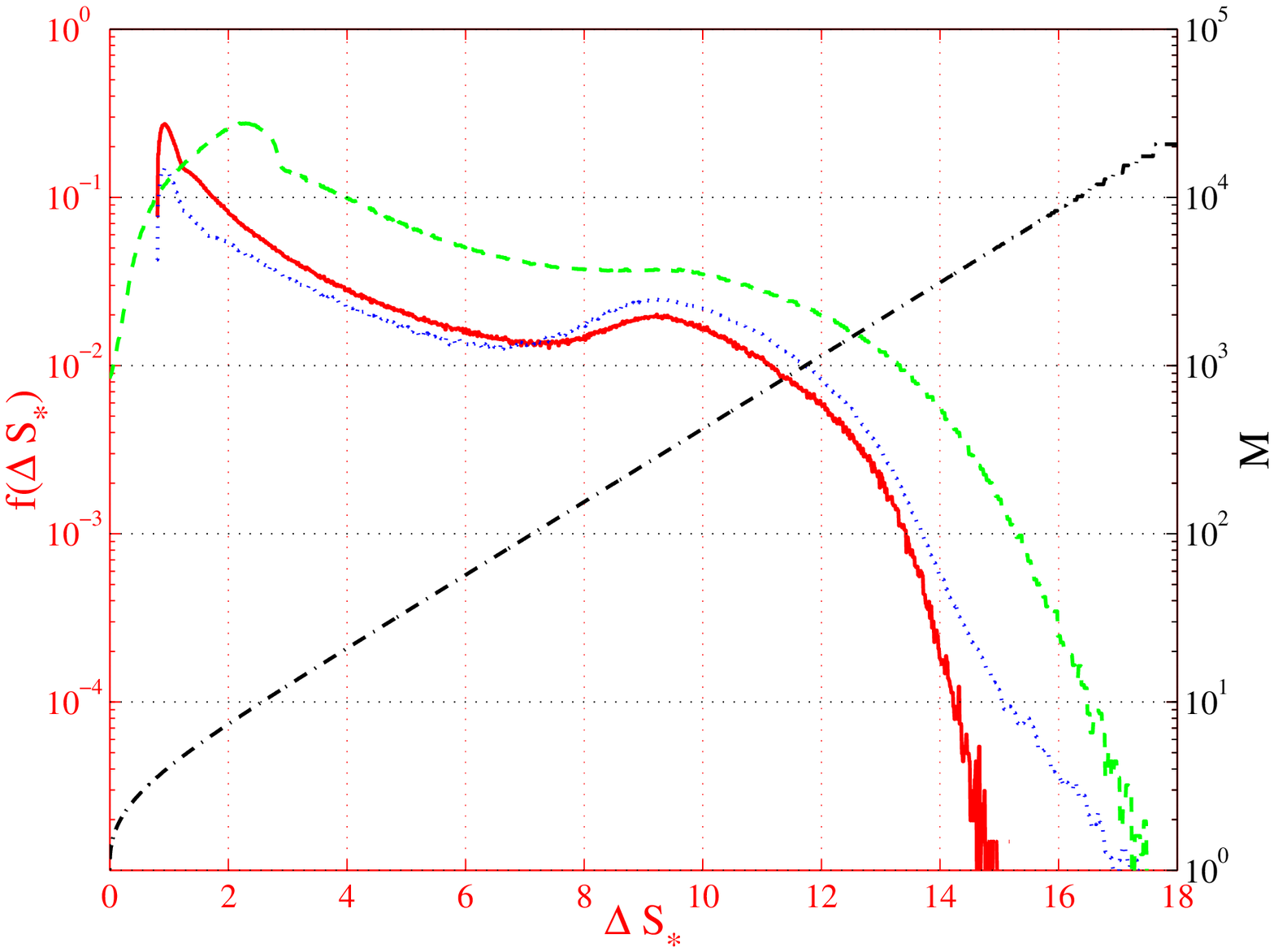}
\caption{ 
Results of shock extraction between $z=2$ and $z=0$. 
\newline
\emph{left}: Redshift dependence of the comoving energy densities 
(left axes), accumulated since $z=2$, of: 
the simulated gas (thick solid line), the identified shocks
(thin solid line), shock--accelerated electron population assuming
$\xi_e=0.05$ (thick dashed line), electrons within the relevant energy
range $\gamma_{\tny{min}} < \gamma < \gamma_{\tny{max}}$ (thin dashed line)
and the redshifted inverse-Compton radiation reaching an observer today
(dash-dotted line).  
Also shown (right axes) are the fraction of the mass that was shocked at 
least once since $z=2$ (thin dotted line), 
and the fraction of the mass processed by shocks ($n$ shocks experienced 
by the same particle counted as $n$ times its mass, thick dotted line).
\newline
\emph{right}: 
Mass fraction $f(\Delta S_*)$ with entropy change $\Delta S_*$ 
accumulated between $z=2$ and $z=0$, for all SPH particles before shock 
extraction (dashed line), 
among the particles identified as shocked (solid line), and among these 
particles before multiple shocks experienced by the particles were separated 
(dotted line). 
Also shown (right axes) is the Mach number $M$ corresponding to $\Delta S_*$ 
(dash-dotted line).
}
\label{fig:Shock_Stat}
\end{figure}

% Shock maps:
% -----------
\begin{figure}
\plottwo{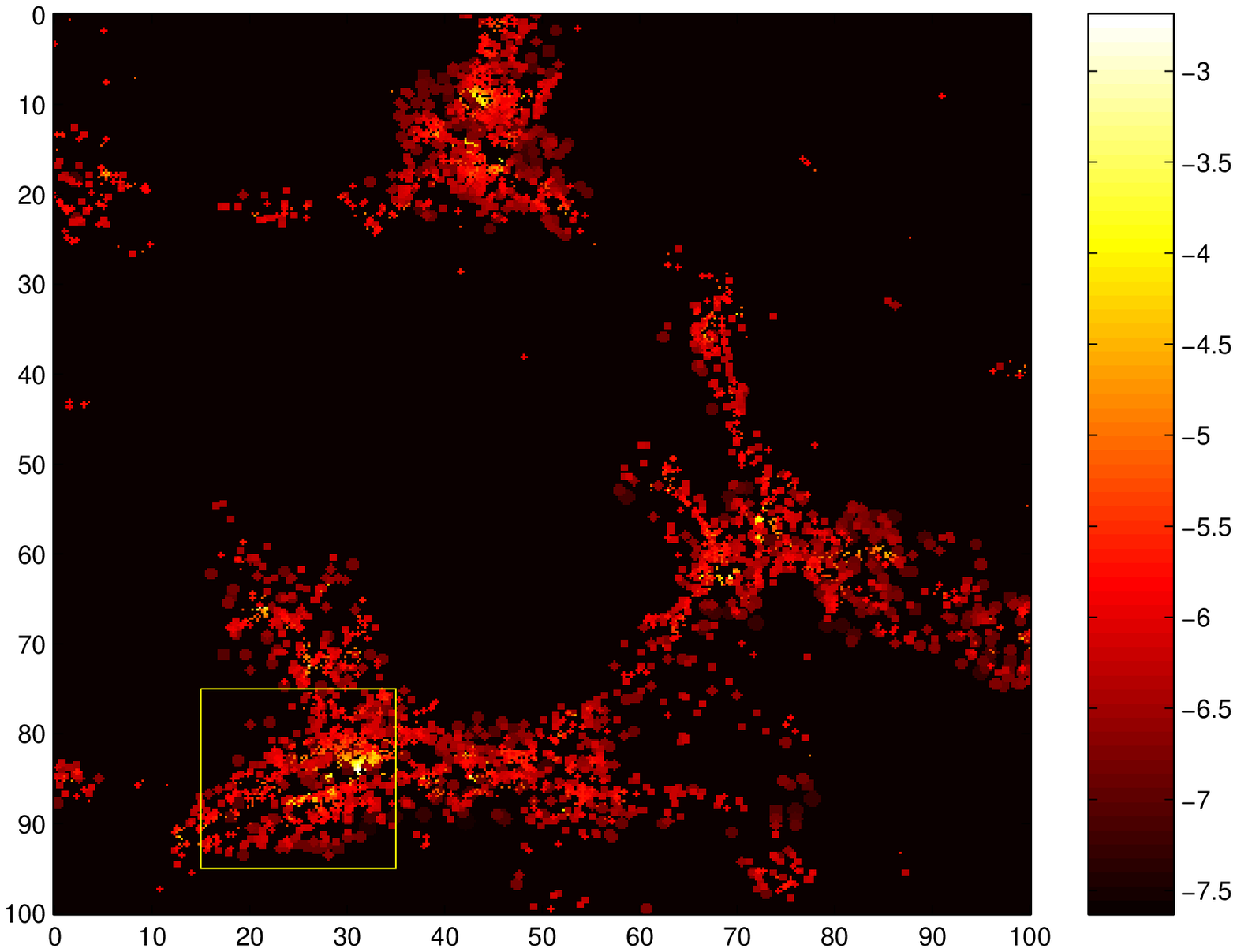}{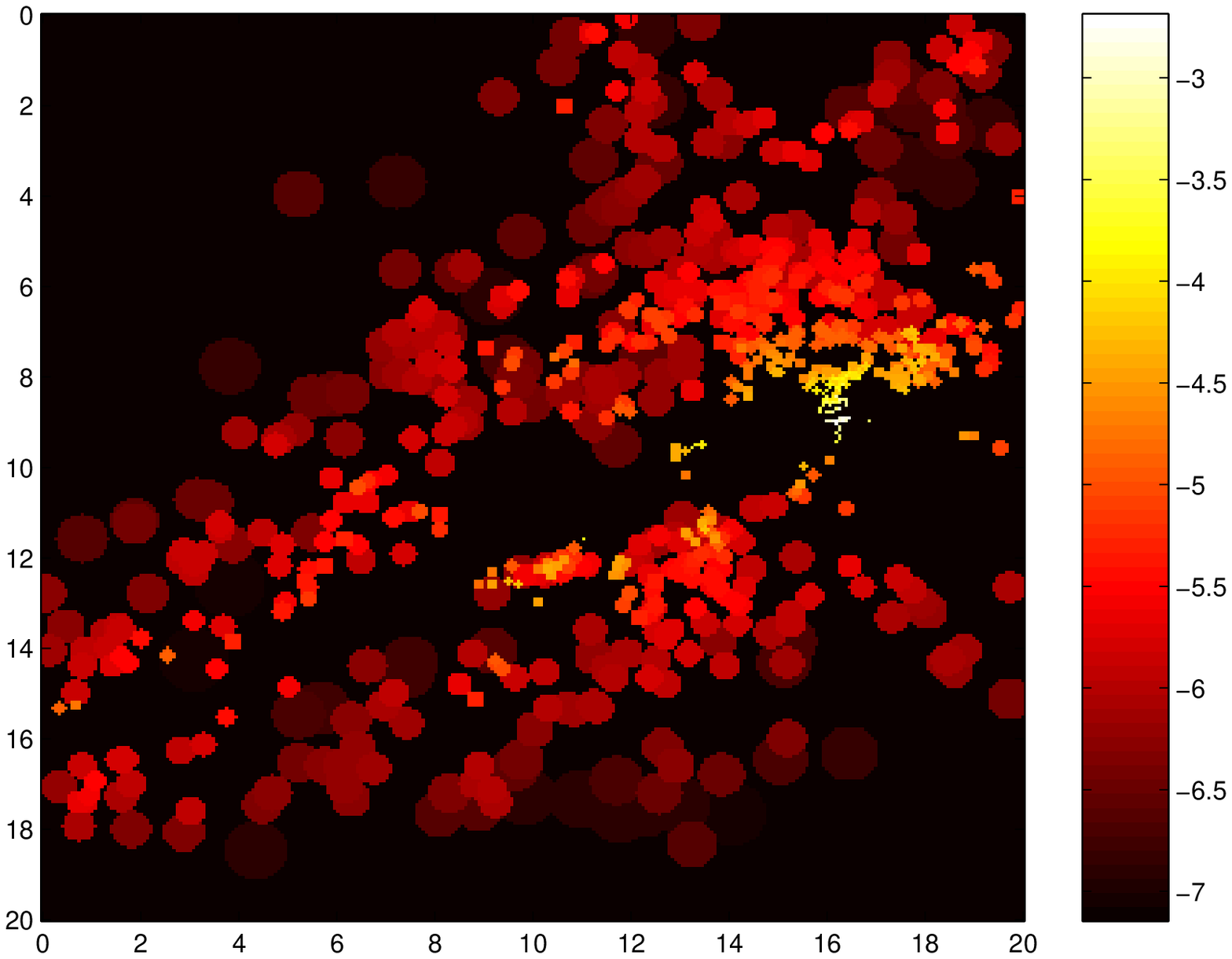}
\caption{ 
The gas shocked between two consecutive snapshots, 
taken at $z=0.14$ and at $z=0.09$. 
Color scale shows the maximal baryon number density through a slice of the 
simulation box, in $log_{10}(n \left[\mbox{cm}^{-3} \right])$.
\newline
\emph{left}: 
The same slice as shown in Figure \ref{fig:map3D}, 
with dimensions $100\mbox{ Mpc} \times 100\mbox{ Mpc} \times 10\mbox{ Mpc}$. 
\newline
\emph{right}: 
Enlarged region in the left image, marked there by a bright square, 
with dimensions $20\mbox{ Mpc} \times 20\mbox{ Mpc} \times 5\mbox{ Mpc}$. 
}
\label{fig:ShockMaps}
\end{figure}

% IC Spectrum:
% ------------
\begin{figure}
\plotone{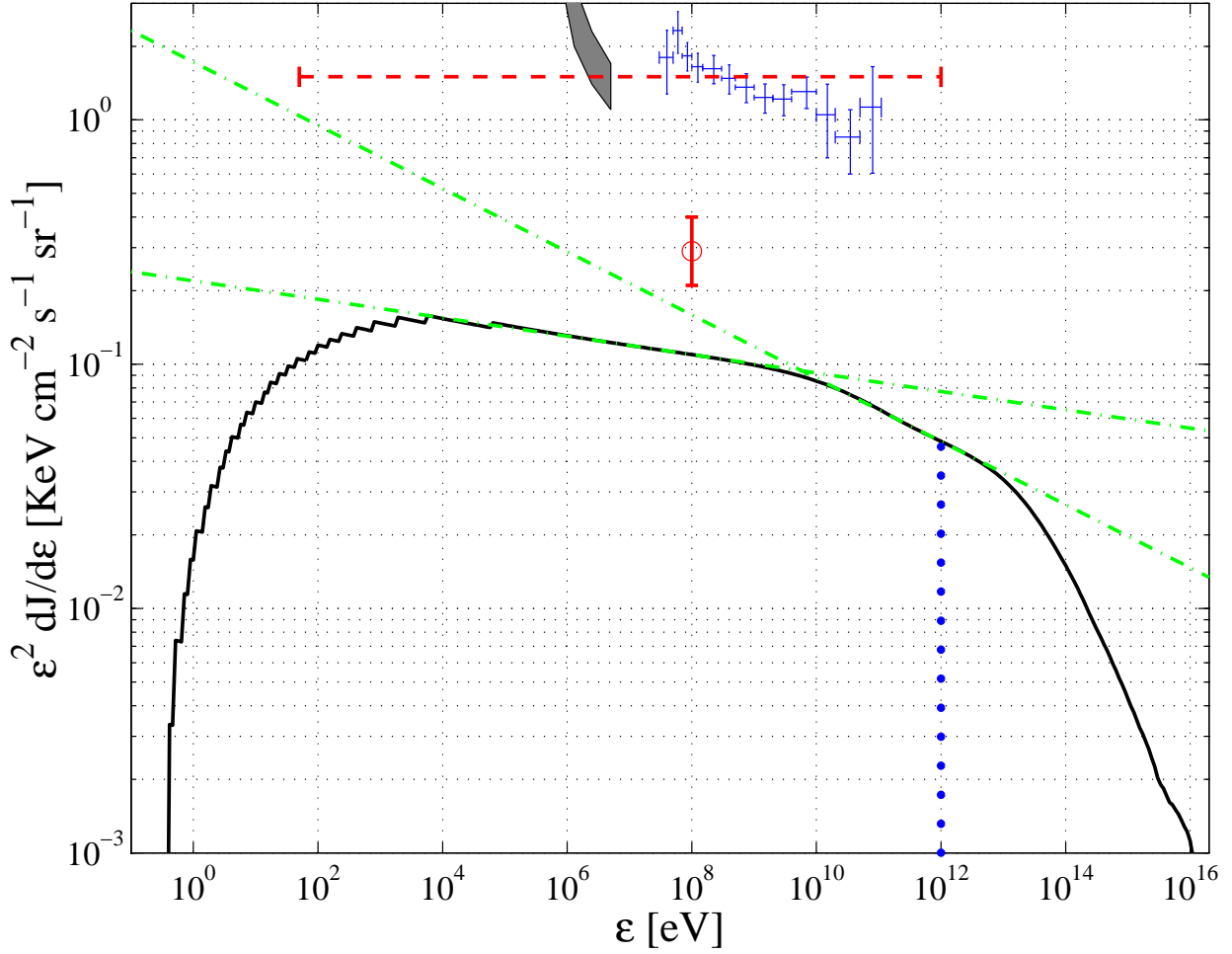}
\caption{ 
Inverse-Compton spectrum (solid line), contributing $\sim10\%$ of the 
EGRET-measured flux (error bars, Sreekumar et al. 1998) 
and the estimates made in \S \ref{sec:EGRB_diffuse_radiation} (dashed line). 
Below $10\mbox{ MeV}$ the observed flux increases significantly: 
the shaded region represents results of the Solar Maximum Mission 
(SMM, Watanabe et al. 1997).
The expected contribution from unresolved point sources, based on empirical 
modeling of the luminosity function of blazars \cite{Mukherjee}, is shown 
(circle with error bar). 
Pair production cascade effectively cuts off the spectrum at photon energy 
$\epsilon \simeq 1\mbox{ TeV}$ (dotted line), 
slightly enhancing the spectrum below this energy. 
The approximated spectra, equation~(\ref{eq:IC_best_fit}) and 
equation~(\ref{eq:IC_best_fit_GeV}), are also shown (dash-dotted lines).
The step-like feature at low energies is a consequence of the temporal 
resolution of selected snapshots 
(and corresponding $\gamma_{\tny{min}}$ values). 
}
\label{fig:IC_results}
\end{figure}

% Full sky map:
% -------------
\begin{figure}
\plotone{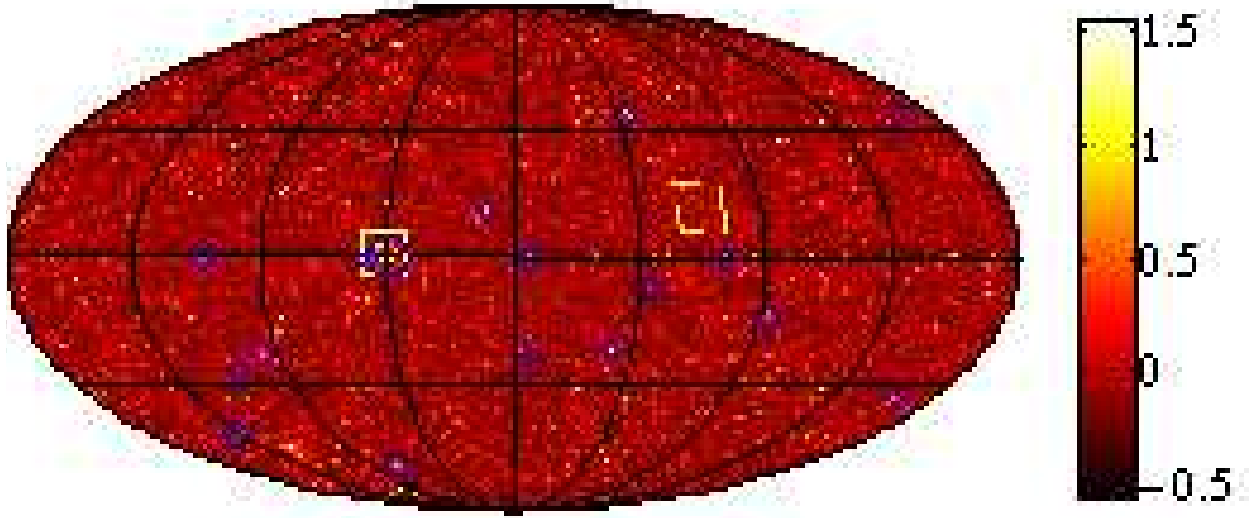}
\caption{ 
Full $\gamma$-ray sky map at photon energies above $100\mbox{ MeV}$. 
Color scale represents $\log_{10} \left( J / \bar{J} \right)$, where 
$\bar{J}=1.1\times 10^{-6} \mbox{ cm}^{-2} \mbox{ s}^{-1}\mbox{ sr}^{-1}$ 
is the average photon number flux at these energies. 
Longitudes and latitudes (dark lines) are plotted $45^{\circ}$ apart. 
Bright contours enclose regions magnified in Figure \ref{fig:regional_maps}. 
Blue circles show simulated sources that will be well-resolved by GLAST 
in photon energies above $1\mbox{ GeV}$ (for $\xi_e=0.05$). 
Angular resolution is $\sim 42^\prime$. 
}
\label{fig:FullSkyMap}
\end{figure}

% Partial sky maps:
% -----------------
\begin{figure}
\epsscale{1}
\plottwo{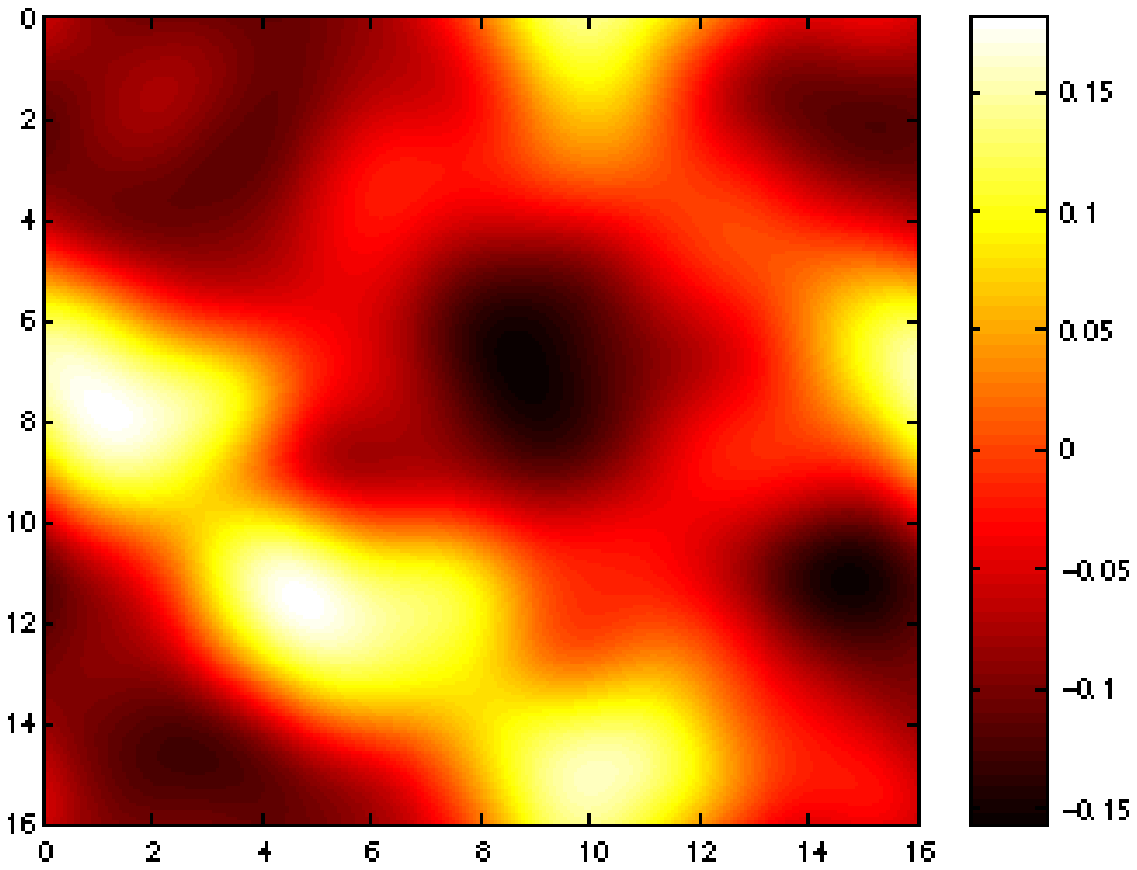}{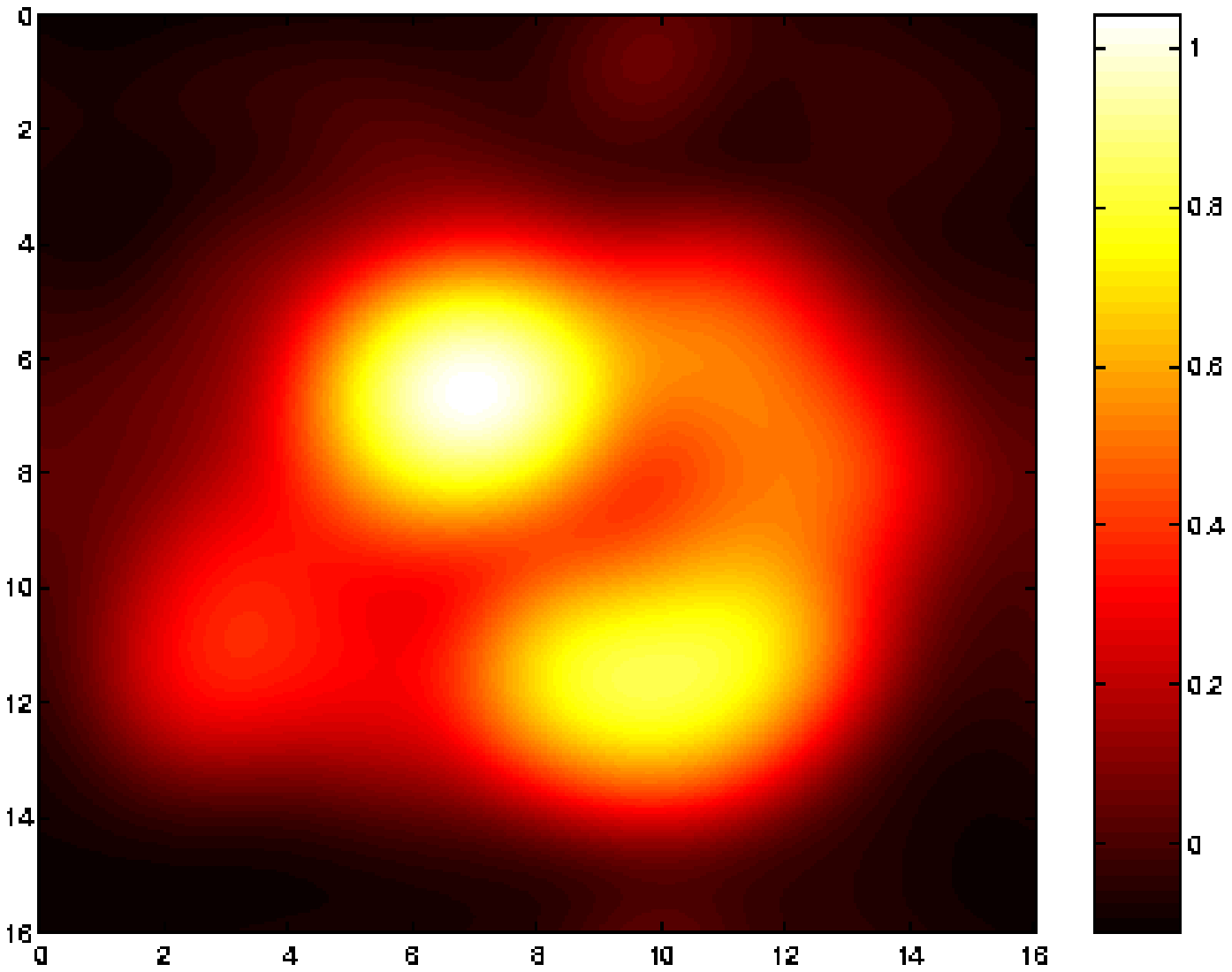}
\epsscale{2.25}
\plottwo{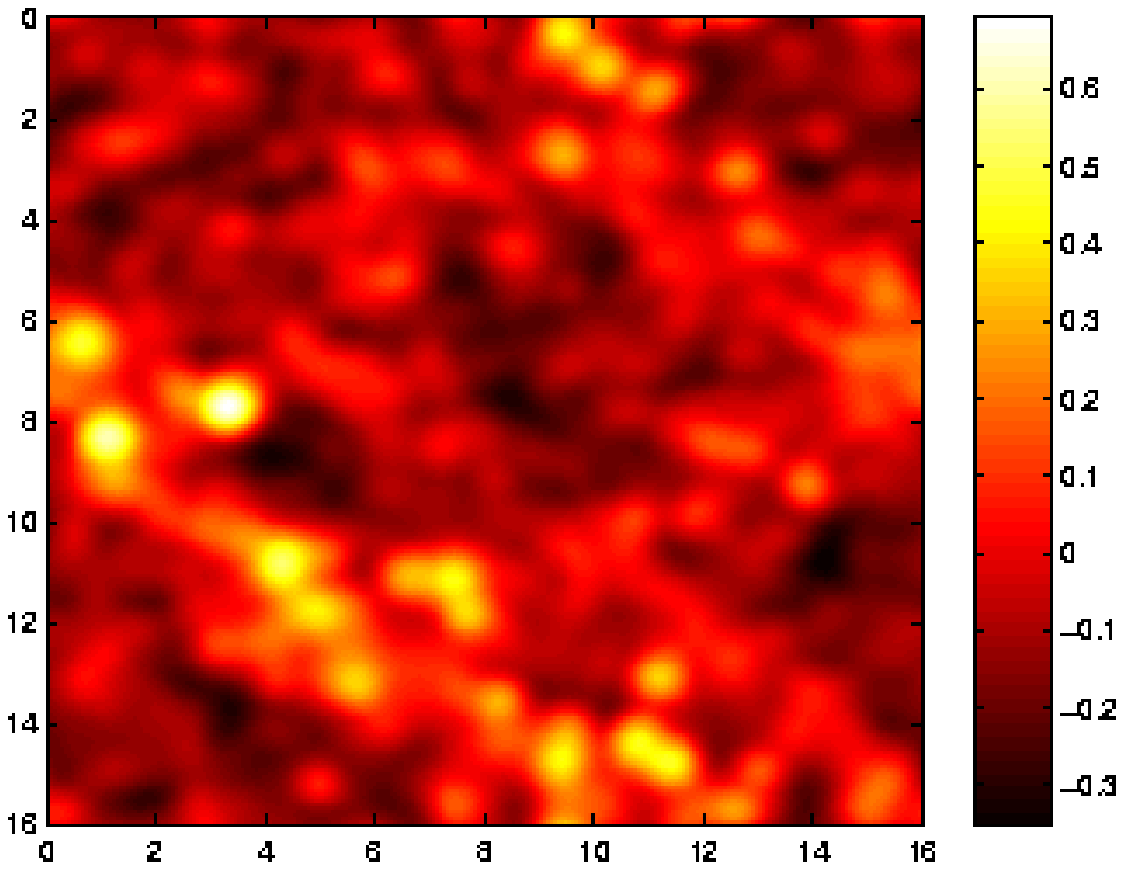}{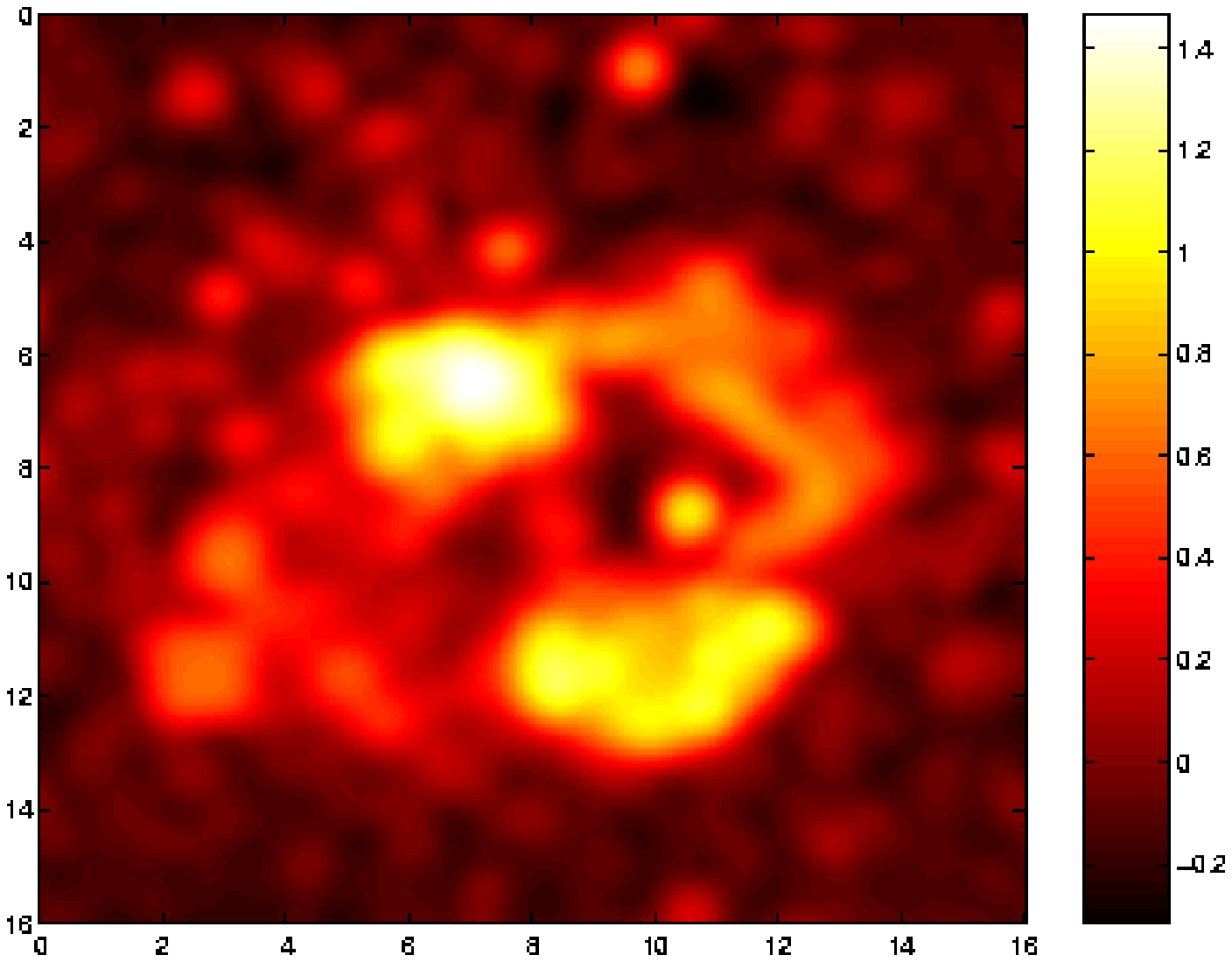}
\caption{ 
Photon number flux above $1\mbox{ GeV}$ from two $16^{\circ}\times 16^{\circ}$ 
regions in the sky, marked in Figure \ref{fig:FullSkyMap} by dashed 
(\emph{left images}) and solid (\emph{right images}) bright contours. 
Images are convolved with Gaussian filter functions, simulating photon 
angular spread of standard deviation 
$1.5^{\circ}$ (for EGRET, \emph{upper images}) and 
$0.42^{\circ}$ (for GLAST, \emph{bottom images}). 
Color scale represents $\log_{10} \left( J/ \bar{J} \right)$, 
where $\bar{J} = 9.9 \times 10^{-8} 
\mbox{ cm}^{-2} \mbox{ s}^{-1} \mbox{ sr}^{-1}$ 
is the average photon number flux above $1\mbox{ GeV}$. 
}
\label{fig:regional_maps}
\end{figure}

% Interesting nearby structure:
% -----------------------------
\begin{figure}
\epsscale{1}
\plottwo{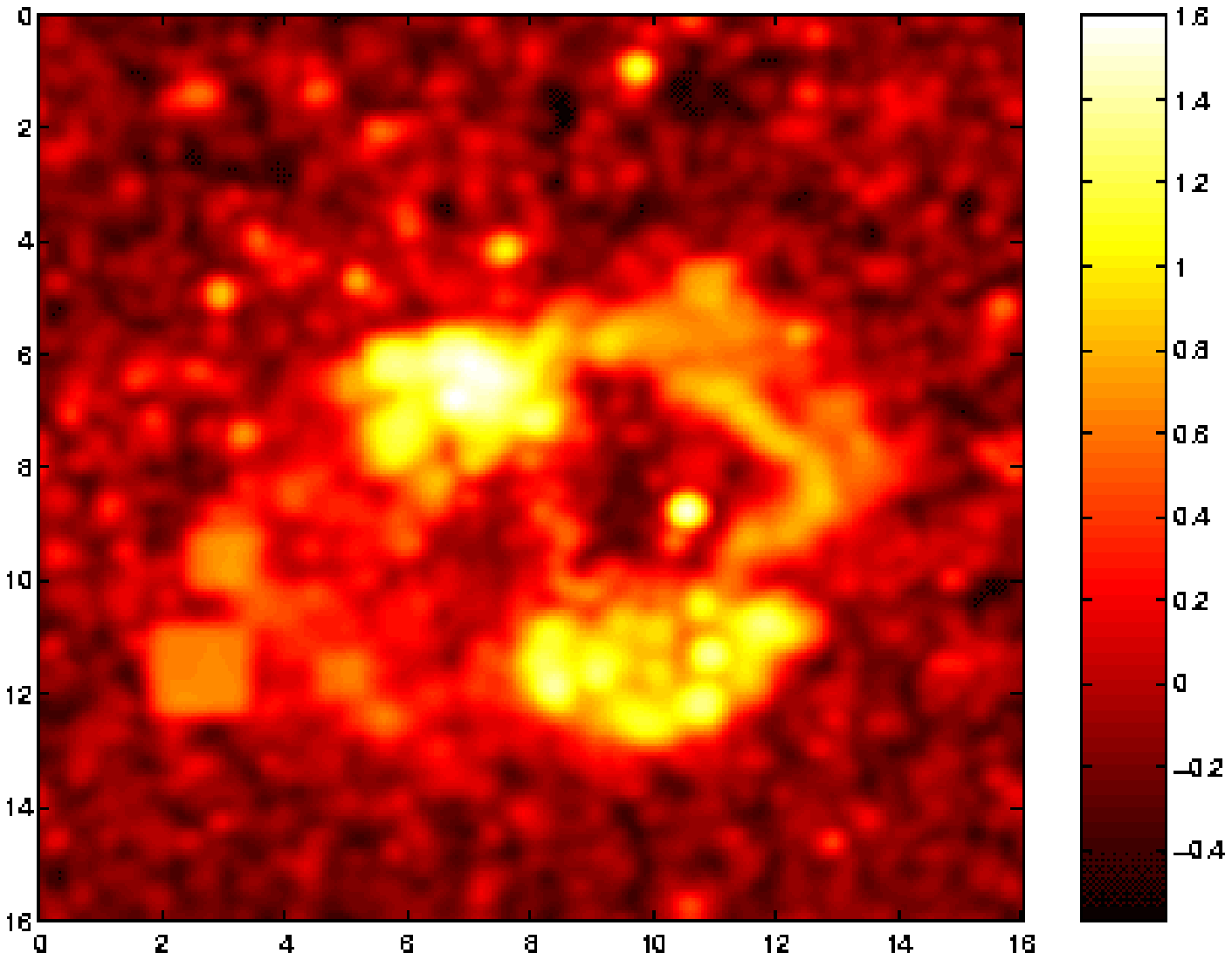}{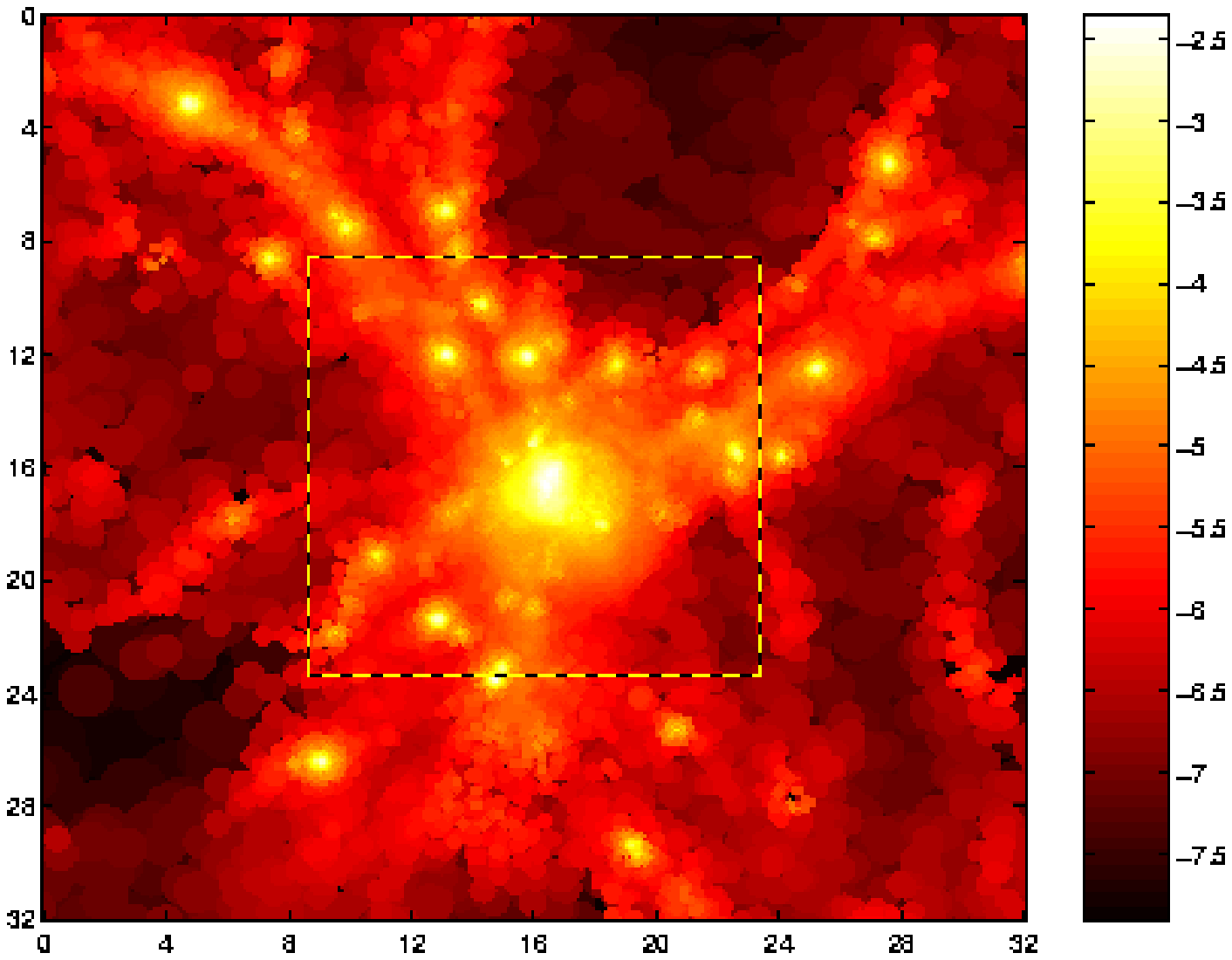}
\epsscale{1.35}
\plotone{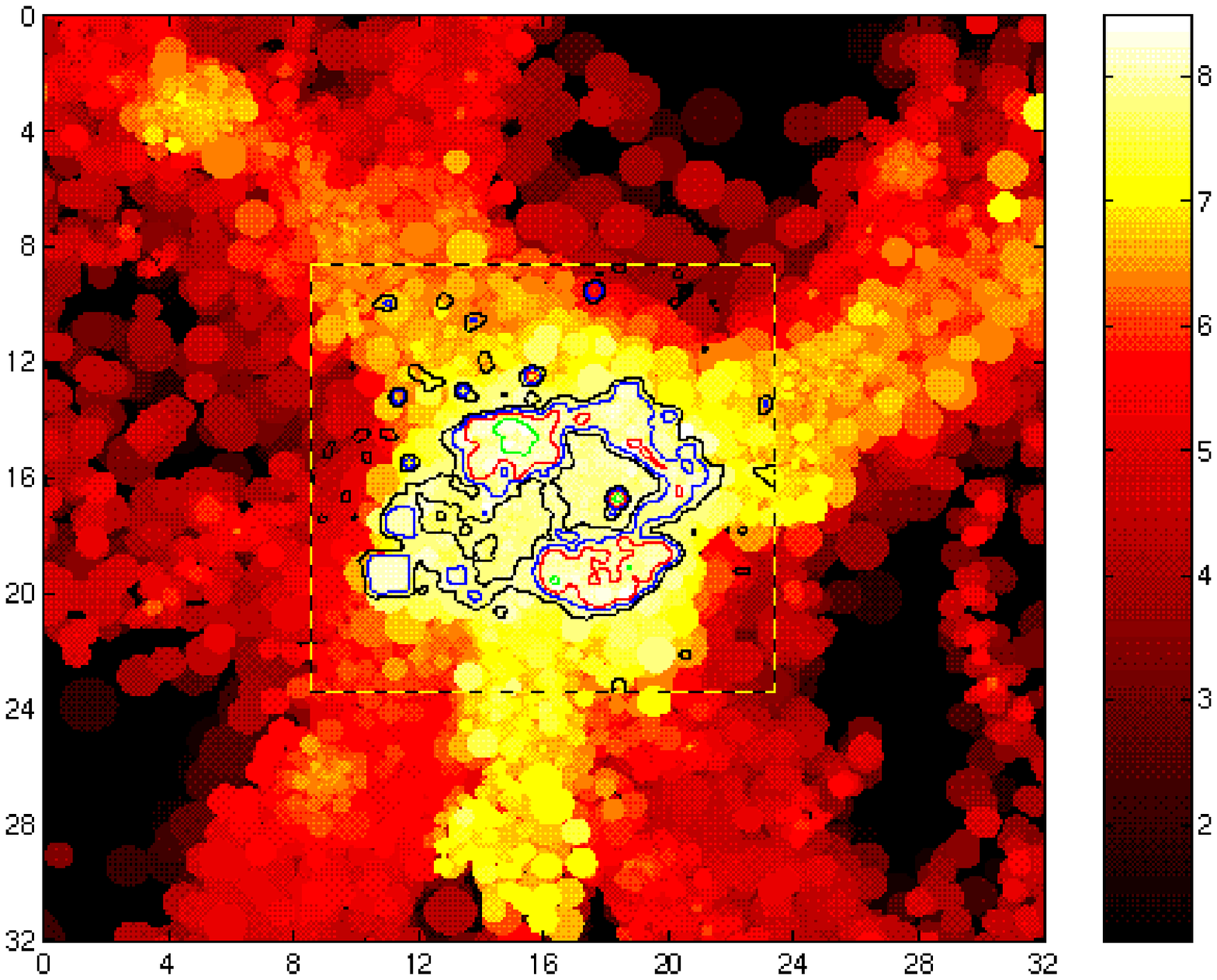}
\caption{ \small
Emission from a nearby rich galaxy cluster.
\newline
\emph{upper, left:}
Photon number flux above $10 \mbox{ GeV}$ from the 
$16^{\circ}\times 16^{\circ}$ region shown in 
Figure \ref{fig:regional_maps} (right images). 
The image is convolved with a Gaussian filter function of standard deviation 
$0.2^{\circ}$, the expected angular resolution of MAGIC in these energies. 
Color scale: $\log_{10} \left( J/ \bar{J} \right)$, where 
$\bar{J} = 8.2 \times 10^{-9} \mbox{ cm}^{-2} \mbox{ s}^{-1} \mbox{ sr}^{-1}$ 
is the average photon number flux above $10\mbox{ GeV}$. 
\newline
\emph{upper, right:}
Maximal baryon number density through a thick slice of the simulation box at 
$z=0$, showing a rich galaxy cluster and connected galaxy filaments. 
The dashed frame marks the central $\sim 15\mbox{ Mpc} \times 15 \mbox{ Mpc}$ 
region, the emission from which is shown in the \emph{upper left} image. 
The slice, 
of dimensions $32\mbox{ Mpc}\times32\mbox{ Mpc}\times 8\mbox{ Mpc}$, 
is oriented perpendicular to the line of sight, with its center located 
$53\mbox{ Mpc}$ ($z\simeq 0.012$) from the observer. 
Color scale: $\log_{10}(n\left[\mbox{cm}^{-3}\right])$.
\newline 
\emph{bottom:}
Maximal temperature through the slice. 
Color scale: $\log_{10}(T\left[\mbox{K}\right])$. 
Contours trace lines of constant number flux above $10\mbox{ GeV}$, 
higher than the average flux in this energy range by a factor ranging 
from 1.7 (black lines) to 22 (green lines). 
}
\label{fig:near_structure}
\end{figure}

% Source number counts:
% ---------------------
\begin{figure}
\epsscale{1.0}
\plottwo{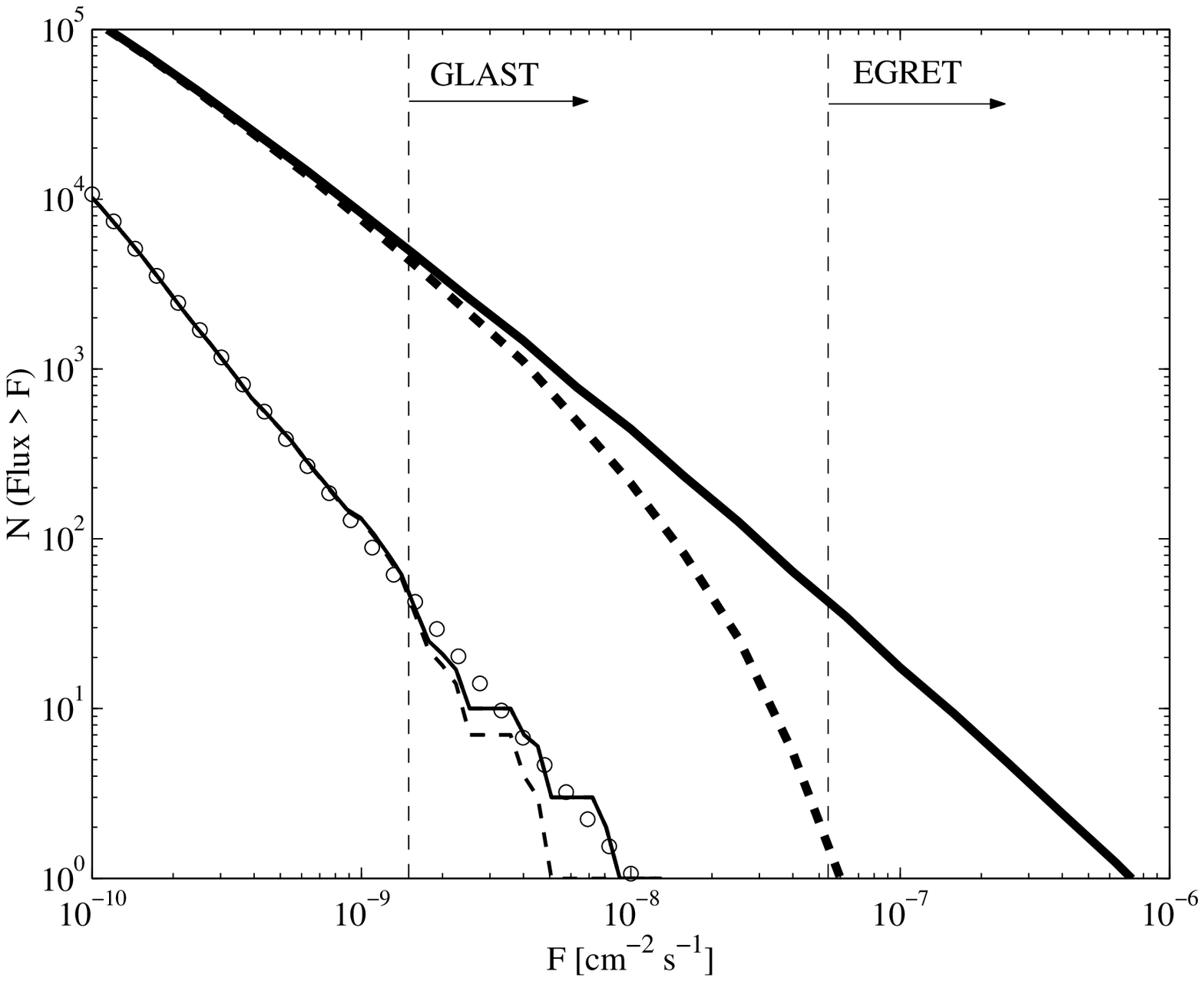}{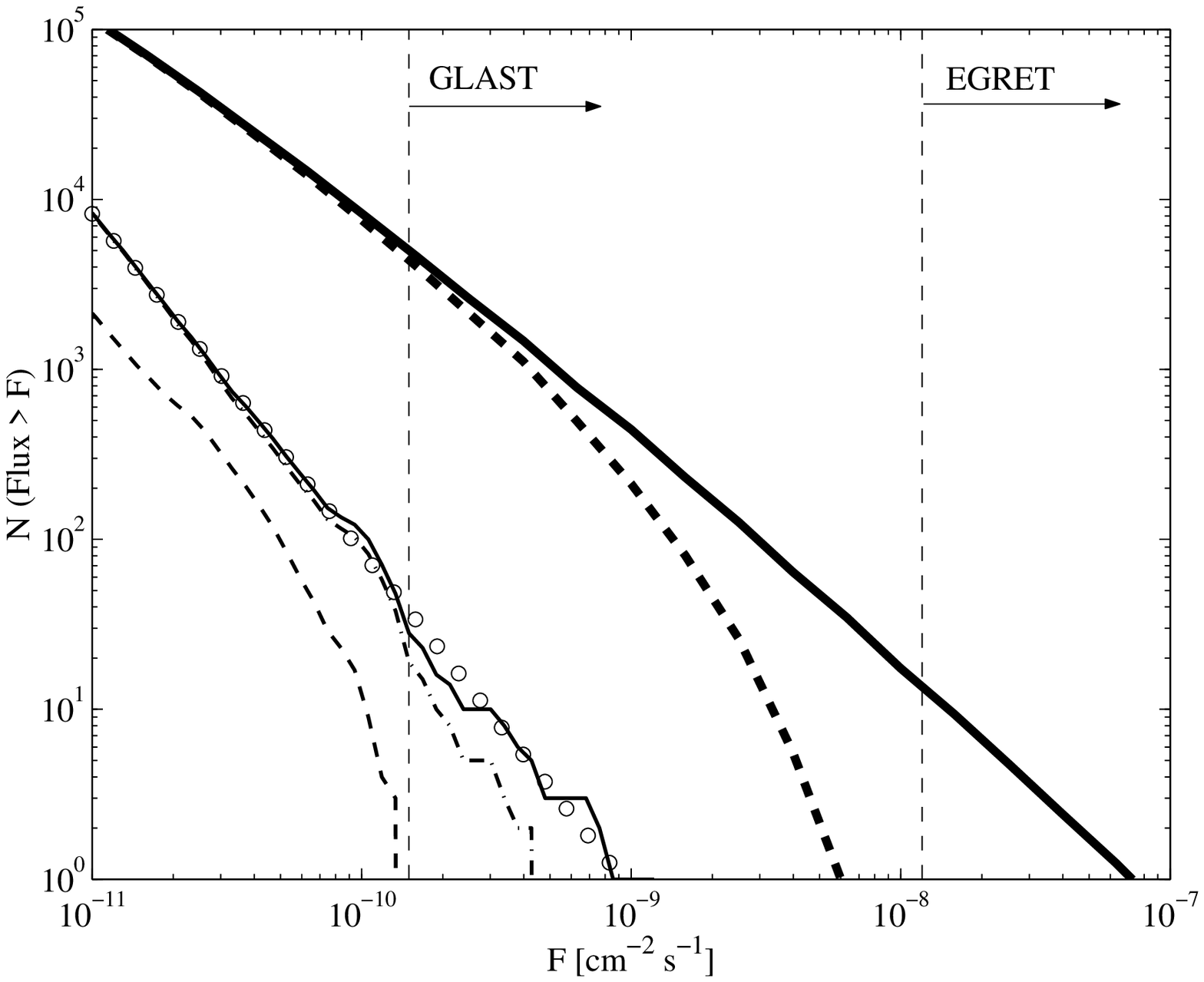}
\caption{ 
Cumulative all-sky number of $\gamma$-ray sources with observed photon 
number flux exceeding $F$, for photon energies above $100\mbox{ MeV}$ 
(\emph{left panel}) and above $1\mbox{ GeV}$ (\emph{right panel}).
Plotted are the number of all sources (thin solid line) and the number of  
sources with angular size smaller than the $1\sigma$ photon-arrival spread 
of EGRET (dash-dotted line) and of GLAST (thin dashed line). 
Estimates based on the Press-Schechter mass function \cite{WaxmanLoeb2000} 
predict a higher number of bright sources (thick solid line) and bright 
sources with angular size $<1^{\circ}$ (thick dashed line). 
In both energy ranges, the number of the brightest sources as a function of 
flux is well-approximated by a power law $N(>F)\sim F^{-2}$ (open circles), 
steeper than the Press-Schechter based estimates $N(>F)\sim F^{-1.38}$. 
Also plotted are the detection flux thresholds of EGRET and GLAST 
(vertical dashed lines). 
}
\label{fig:number_count}
\end{figure}

% Cosmological parameters:
% ------------------------
\begin{figure}
\epsscale{1.0}
\plottwo{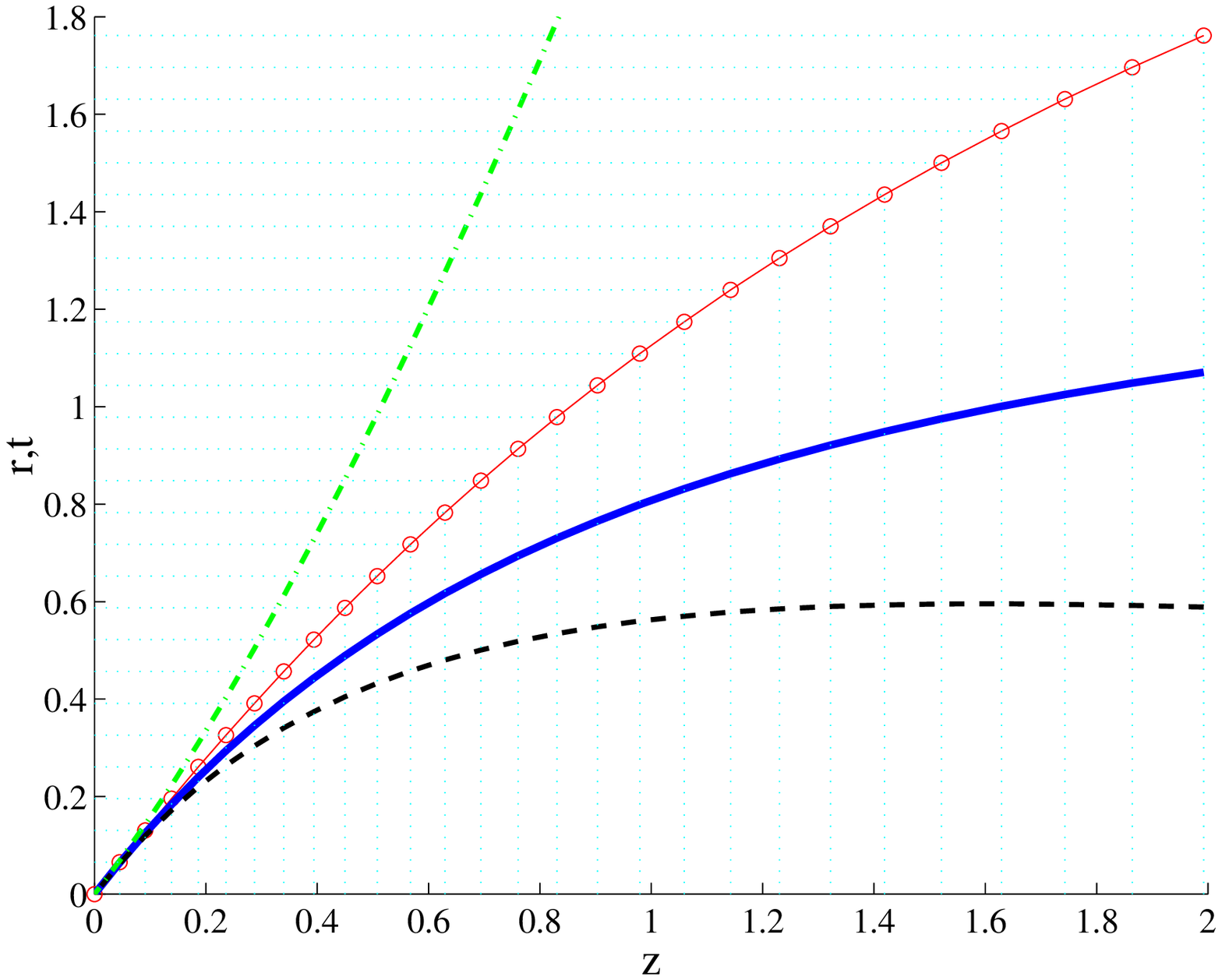}{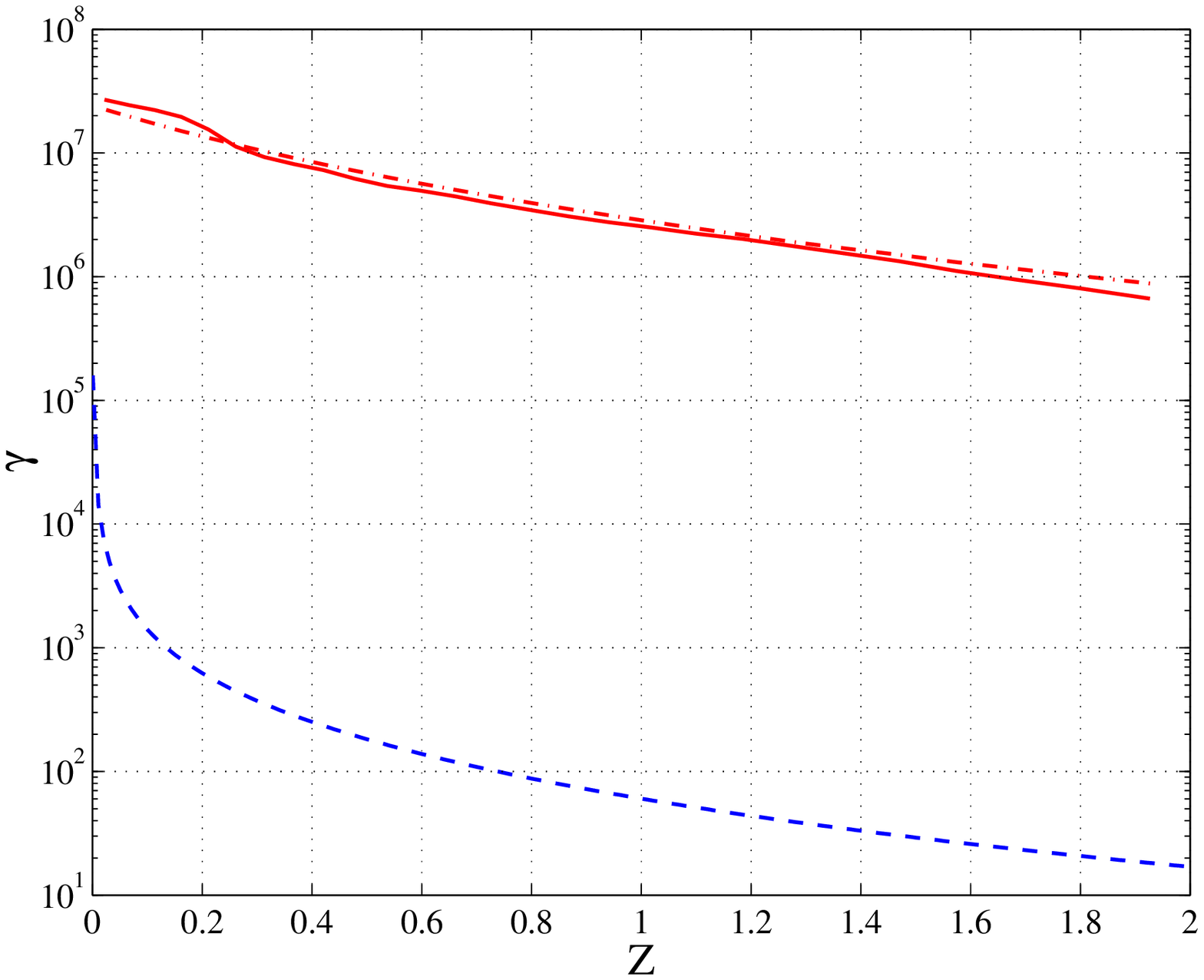}
\caption{ 
Redshift dependence of the parameters, required for EGRB integration. 
\newline
\emph{left}: 
Coordinate distance ($r R_0$, thin solid line) between a source emitting a 
photon at redshift $z$ and an observer detecting the photon today 
(in $10^{10}\mbox{ ly}$ units),
the angular diameter distance ($r R_0 / \left[1+z\right]$, dashed line) 
and the luminosity distance ($r R_0 \left[1+z\right]$, dash-dotted line) 
between them. 
The time that elapsed between emission and detection is given in 
$10^{10}\mbox{ yr}$ units (thick solid line). 
Circles and dotted lines illustrate the choice of simulation snapshots 
examined, separated by the comoving simulation-box light-crossing time.
\newline
\emph{right}:
Minimal Lorentz factor of electrons, emitting at least half their energy 
between redshift $z$ and the present (dashed line), 
and maximal Lorentz factor reached by the shocked gas according to the 
simulation (solid line) and according to equation~(\ref{eq_gamma_max}) 
(dash-dotted line), where a post-shock magnetic field of magnitude 
$B=0.1\mbox{ }\mu\mbox{G}$ was assumed. 
}
\label{fig:Vals}
\end{figure}

% Source identification algorithm:
% --------------------------------
\begin{figure}
\epsscale{1.0}
\plottwo{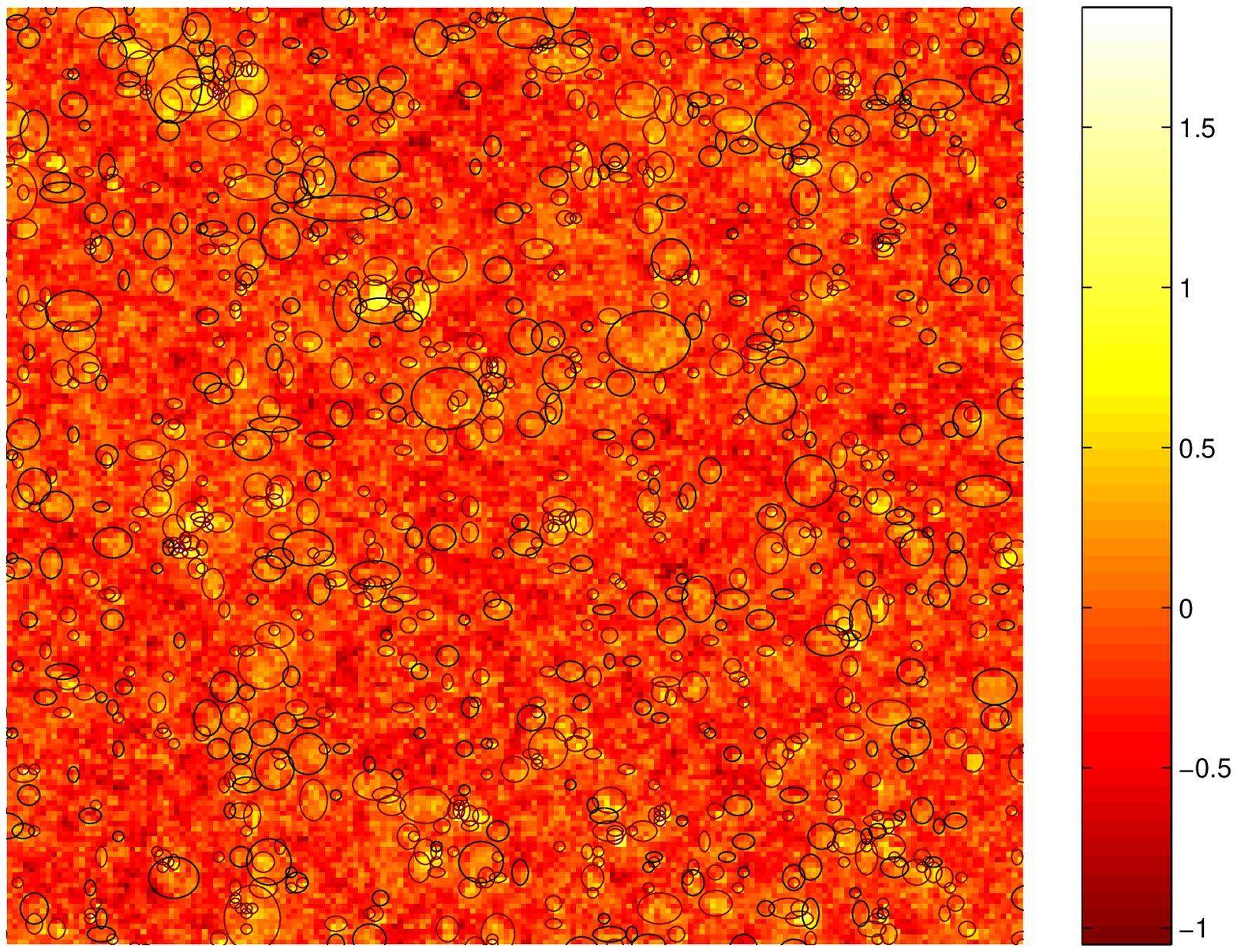}{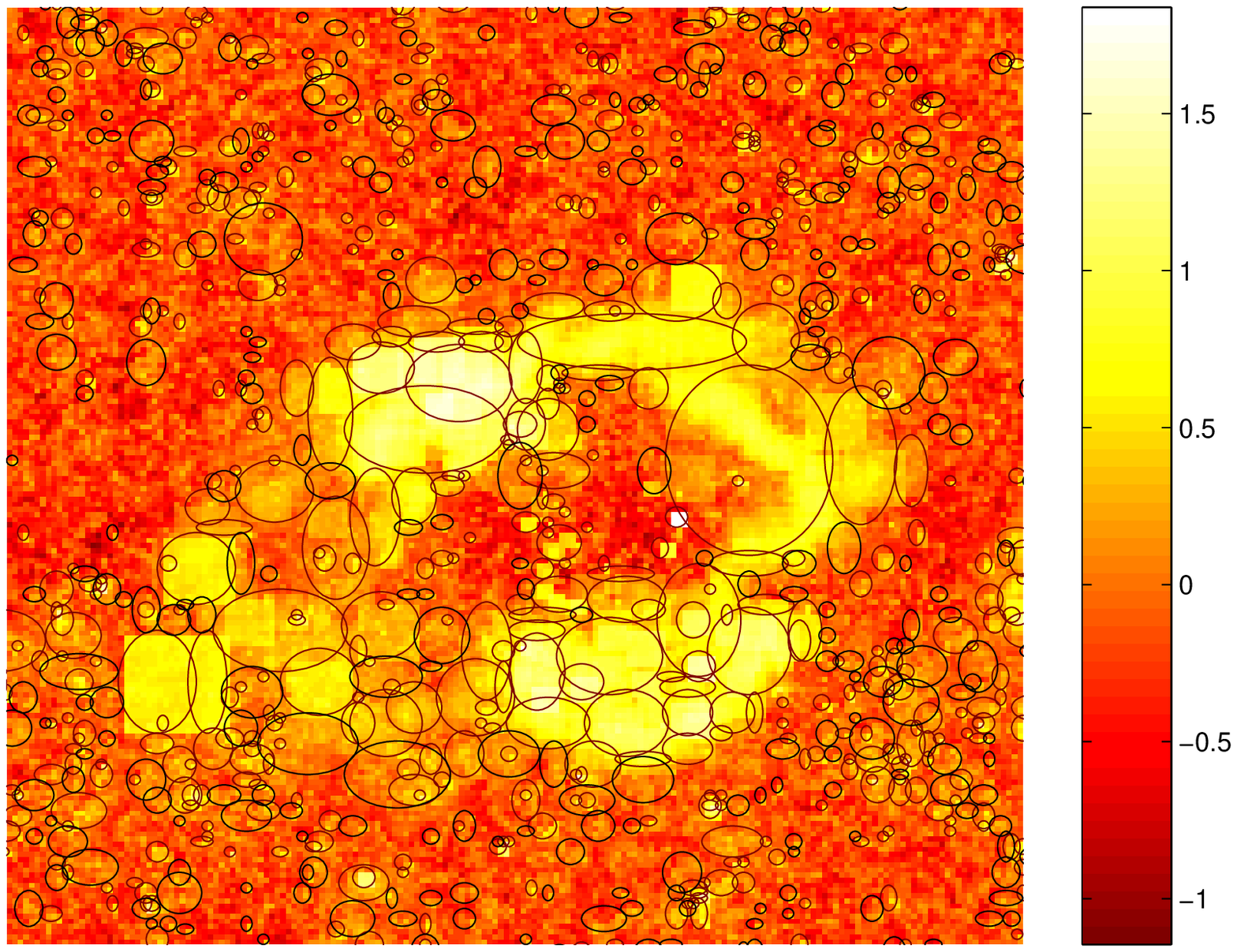}
\caption{ 
Results of the source-identification algorithm, when
applied to two regions in the $\gamma$-ray sky.  
Both images are of angular size $16^{\circ} \times 16^{\circ}$, 
and are similar (but not identical) to those presented in 
Figure \ref{fig:regional_maps}. 
Color scale is identical to the one shown in Figure \ref{fig:regional_maps}.  
}
\label{fig:SIA}
\end{figure}

\clearpage

% ------------------------------------------------------------------------
% Tables:
% ------------------------------------------------------------------------

\begin{deluxetable}{lll}
\tablecaption{\label{tab:CosmoParams}
Parameters of the cosmological model and linear structure formation theory.}
\tablewidth{0pt}
\tablehead{ \em Parameter & \em Meaning & \em Value }
\startdata
$h$           & Hubble parameter & 0.67 \\ 
$k$           & Curvature & 0 \\
\tableline 
$\Omega_M$    & Matter energy density & 0.3 \\
$\Omega_{DM}$ & Dark matter energy density & 0.26 \\ 
$\Omega_B$    & Baryon energy density & 0.04 \\ 
$\Omega_\Lambda$ & Vacuum energy density & 0.7 \\
$\chi$      & Hydrogen mass fraction & 0.76 \\ 
\tableline 
$n$           & Fluctuation spectrum slope & 1 \\
$\sigma_8$    & Spectrum normalization & 0.9 \\ 
$z_{\tny{start}}$ & Starting redshift for the simulation & 50 \\
\enddata
\end{deluxetable}

\begin{deluxetable}{lll}
\tablecaption{\label{tab:AlgoParams} Code parameters of the simulation. }
\tablewidth{0pt}
\tablehead{ \em Parameter & \em Meaning & \em Value }
\startdata
$L_{\tny{com}} R(t_0)$ \tablenotemark{a} & Simulation box comoving length & 
$134h^{-1}\mbox{ Mpc} \simeq 200 \mbox{ Mpc}$ \\ 
\tableline
$N_{\tny{DM}}$ & Number of dark matter particles & 
$224^3 \simeq 11 \times 10^6$ \\
$M_{\tny{DM}}$ & Mass of a dark matter particle & 
$2.3\times 10^{10} \mbox{ M}_{\odot}$ \\ 
\tableline
$N_{\tny{SPH}}$ & Number of gas (SPH) particles & 
$224^3 \simeq 11 \times 10^6$ \\
$M_{\tny{SPH}}$ & Mass of an SPH particle & 
$3.55\times 10^{9} \mbox{ M}_{\odot}$ \\ 
$h_{\tny{min}}$ & Minimal SPH softening & 
$6h^{-1} \mbox{ kpc} \simeq 9 \mbox{ kpc}$ \\ 
$N_s$ & Number of SPH neighbors in kernel & $32$ \\
$M_{\tny{res}}$ & Minimal Mass Resolution & $\sim 10^{11}\mbox{ M}_{\odot}$ \\
\tableline
$S_G R(t_0)$ \tablenotemark{a} & Gravitational softening & 
$25 h^{-1} \mbox{ kpc} \simeq 37 \mbox{ kpc}$  \\ 
\enddata
\tablenotetext{a}
{$R(t_0)$ is the present-day value of the cosmological scale factor.}
\end{deluxetable}

\begin{deluxetable}{llc}
\tablecaption{\label{tab:ProgParams}
Comoving energy density $u_{\tny{com}}$ of various simulated components. }
\tablewidth{0pt}
\tablehead{ \em Parameter & \em Component & 
\em $u_{\tny{com}} \left[ \mbox{eV cm}^{-3} \right]$ }
\startdata
$u_i$ & gas, at $z=2$ & $2.77 \times 10^{-5} $ \\ 
$u_f$ & gas, at $z=0$ & $1.72 \times 10^{-4} $ \\ 
$\Delta u = u_f-u_i$ & gain of gas during $0<z<2$ & $1.44 \times 10^{-4} $ \\ 
\tableline 
$u_{\tny{sh}}$ & detected shocks & $9.62 \times 10^{-5} $ \\
\tableline 
$u_e$ & accelerated electrons ($\xi_e=0.05$) & $4.81 \times 10^{-6} $ \\ 
$u_e (\gamma_{\tny{min}}<\gamma_e<\gamma_{\tny{max}})$ & 
electrons with $\gamma_{\tny{min}} < \gamma_e < \gamma_{\tny{max}}$ & 
        $2.34 \times 10^{-6} $ \\       
\tableline 
$u_\gamma$ & radiation, redshifted & $1.40 \times 10^{-6}$ \\ 
\enddata
\end{deluxetable}

\begin{deluxetable}{ccccc} 
\tablecaption{\label{tab:EGRETandGLAST}
Parameters of EGRET and of the conceptual designs of GLAST and of MAGIC.}
\tablewidth{0pt}
\tablehead{
\em Parameter & \em Photon Energy & 
\em EGRET \tablenotemark{a} & 
\em GLAST \tablenotemark{a} & 
\em MAGIC \tablenotemark{b} \\
 & & (one year all-sky) & (one year all-sky) & ($50$ \emph{hours}) }
\startdata
Point source  & $E>100\mbox{ MeV}$ & 
$5.4 \times 10^{-8} $ & 
$1.5 \times 10^{-9} $ & --- \\
-- sensitivity & $E>1\mbox{ GeV}$ & 
$1.2 \times 10^{-8} $ & 
$1.5 \times 10^{-10} $ & --- \\
$\left[\mbox{cm}^{-2}\mbox{ s}^{-1}\right]$ & 
$E>10\mbox{ GeV}$ & 
--- &  $1.0 \times 10^{-10}$ &  $1.0 \times 10^{-10}$ \\
\tableline 
Photon -- & $E>100\mbox{ MeV}$ & 
$5.6^{\circ}$ & $2.5^{\circ}$ & --- \\
position -- & $E>1\mbox{ GeV}$ & 
$1.5^{\circ}$ & $0.42^{\circ}$ & --- \\
error ($1\mbox{ }\sigma$) & $E>10\mbox{ GeV}$ &
--- & $0.1^{\circ}$ & $0.2^{\circ}$ \\ 
\enddata
\tablenotetext{a}{ See http://www-glast.stanford.edu} 
\tablenotetext{b}{ See Gonz\'{a}lez et al. (1997) and 
http://hegra1.mppmu.mpg.de/MAGICWeb}
\end{deluxetable}


\newpage
\begin{thebibliography}{}

\bibitem[Achterberg et al. 1994]{Achterberg94}
	Achterberg, A., Blandford, R. D., \& Reynolds, S. P. 1994, 
	\aap, 351, 330

\bibitem[Aharonian \& Atoyan 1999]{Aharonian99}
	Aharonian, F. A. \& Atoyan, A. M. 1999, \aap, 281, 220

\bibitem[Baring et al. 1999]{Baring99}
	Baring, M. G., Ellison, D. C., Reynolds, S. P., Grenier, I. A.,
	\& Goret, P. 1999, \apj, 513, 311

\bibitem[Barnes \& Hut 1986]{BarnesHut}
        Barnes, J. \& Hut, P. 1986, Nature, 324, 446

\bibitem[Barwick et al. 1998]{Barwick98}
	Barwick, S. W. et al. 1998, \apj, 498, 779	

\bibitem[Blandford \& Eichler 1987]{Blandford87}
        Blandford, R. \& Eichler, D. 1987, \physrep, 154, 1

\bibitem[Cargill \& Papadopoulos 1988]{Cargill88}
	Cargill, P. J. \& Papadopoulos, K. 1988, \apj, 329, L29

\bibitem[Cen \& Ostriker 1999]{CenOstriker99}
        Cen, R. \& Ostriker, P. 1999, \apj, 514, 1

\bibitem[Croft et al. 2001]{Croft01}
        Croft, R. A. C., Di Matteo, T., Dav\'e, R., Hernquist, L.,
        Katz, N., Fardal, M. A., \& Weinberg, D. H. 2001, \apj, 557, 67

\bibitem[Daly \& Loeb 1990]{Daly90}
	Daly, R. A. \& Loeb, A. 1990, \apj, 364, 451

\bibitem[Dav\'{e} et al. 1999]{Dave99}
        Dav\'{e}, R., Hernquist, L., Katz, N., \& Weinberg, D. 1999, 
        \apj, 511, 521

\bibitem[Dav\'{e} et al. 2001b]{Daveb}
        Dav\'{e}, R., Katz, N., Hernquist, L., \& Weinberg, D. 2001b, in
        Sesto 2001-Tracing Cosmic Evolution with Galaxy Clusters
        [astro-ph/0109394]

\bibitem[Dav\'{e} et al. 2001a]{Davea}
        Dav\'{e}, R., Cen, R., Ostriker, J.P., Bryan, G.L., Hernquist, L.,
        Katz, N., Weinberg, D.H., Norman, M.L., \& O'Shea, B. 2001a, 
        \apj, 552, 473

\bibitem[Drury 1983]{Drury83}
	Drury, L. Oc. 1983, Rep. Prog. Phys. 46, 973

\bibitem[Dyer et al. 2001]{Dyer2001}
        Dyer, K. K., Reynolds, S. P., Borkowski, K. J., Allen, G. E., \& 
	Petre, R. 2001, \apj, 551, 439

\bibitem[Eisenstein \& Hu 1998]{EisensteinHu}
        Eisenstein, D. J. \& Hu, W. 1998, \apj, 496, 605

\bibitem[Ellison et al. 2001]{Ellison01}
        Ellison, D. C., Slane, P., \& Gaensler, B. M. 2001, \apj, 563, 191

\bibitem[En\ss{}lin et al. 2001]{Ensslin}
        En\ss{}lin, T. A., Simon, P., Biermann, P. L., Klein, U., 
	Kohle, S., Kronberg, P. P., \& Mack, K. H. 2001, \apj, 549, L39

\bibitem[Furlanetto \& Loeb 2001]{Furlanetto01}
	Furlanetto, S. R. \& Loeb, A., ApJ, 556, 619

\bibitem[Fusco-Femiano et al. 1999]{Fusco99}
        Fusco-Femiano, R. et al. 1999, \apj, 513, L21

\bibitem[Gingold \& Monaghan 1977]{gm77}
        Gingold, R.A. \& Monaghan, J.J. 1977, MNRAS, 181, 375

\bibitem[Gonz\'{a}lez et al. 1997]{Gonzalez}
	Gonz\'{a}lez, J. C., Mirzoyan, R., Fonseca, V., Lorenz, E., 
	in Proceedings of ``Towards a Major Atmospheric Cherenkov Detector-V" 
	Int. Workshop, eds. O. De Jager

\bibitem[Gould \& Schreder 1967]{Gould67}
	Gould, J. \& Schreder, G. 1967, Phys. Rev. 155, 1404

\bibitem[Gruzinov \&  Waxman 1999]{Gruzinov99}
	Gruzinov, A. \& Waxman, E. 1999, \apj, 511, 852

\bibitem[Helfand \& Becker 1987]{Helfand87} 
	Helfand, D. J. \& Becker, R. H. 1987, \apj, 314, 203

\bibitem[Hernquist 1993]{hern93}
        Hernquist, L. 1993, \apj, 404, 717

\bibitem[Hernquist \& Katz 1989]{HernquistKatz}
        Hernquist, L. \& Katz, N. 1989, \apjs, 70, 419

\bibitem[Hutchings \& Thomas 2000]{ht00}
        Hutchings, R. M. \& Thomas, P. A. 2000, MNRAS, 319, 721

\bibitem[Kang et al. 1996]{Kang96}
	Kang, H., Ryu, D., \& Jones, T. W. 1996, \apj, 456, 422

\bibitem[Kawasaki \& Totani 2001]{KawasakiTotani}
        Kawasaki, W. \& Totani, T. 2001, ApJ, 576, 679 

\bibitem[Kulsrud et al. 1997]{Kulsrud97}
	Kulsrud, R. M., Cen, R., Ostriker, J. P., \& Ryu, D. 1997, 
	\apj, 454, 60

\bibitem[Kronberg 1994]{Kronberg94}
        Kronberg, P. P. 1994, Rep. Prog. Phys., 57, 325

\bibitem[Kronberg et al. 1999]{Kronberg99}
	Kronberg, P. P., Lesch, H., \& Hopp, U. 1999, \apj, 511, 56

\bibitem[Landau \& Lifshitz 1959]{Landau}
        Landau, L. D. \& Lifshitz, E. M. 1959, Fluid Mechanics (Pergamon Press)

\bibitem[Loeb \& Waxman 2000]{LoebWaxman2000}
        Loeb, A. \& Waxman, E. 2000, Nature, 405, 156

\bibitem[Lucy 1977]{lucy77}
        Lucy, L.B. 1977, AJ, 82, 1013

\bibitem[Mastichiadis \& de Jager 1996]{Mastichiadis96}
	Mastichiadis, A. \& de Jager, O. C. 1996, \aap, 311, L5

\bibitem[Martel \& Shapiro 2001]{ms01}
        Martel, H. \& Shapiro, P. 2001, in Proceedings of IAU
        Symposium 208, eds. J. Makino \& P. Hut

\bibitem[Medvedev \& Loeb 1999]{Medve99} 
 	Medvedev, M. V. \& Loeb, A. 1999, \apj, 526, 697

\bibitem[Miniati et al. 2000]{Miniati}
        Miniati, F., Ryu, D., Kang, H., Jones, T. W., Cen, R., 
        \& Ostriker, J. P. 2000, \apj, 542, 608

\bibitem[Miniati et al. 2001]{Miniati2001}
        Miniati, F., Jones, T. W., Kang, H. \& Ryu, D. 2001, \apj, 562, 1

\bibitem[Monaghan 1992]{Monaghan}
        Monaghan, J. J. 1992, \araa, 30, 543

\bibitem[Mukherjee \& Chiang 1999]{Mukherjee}
        Mukherjee, R. \& Chiang, J. 1999, Astroparticle Physics, 11, 213

\bibitem[Muraishi et al. 2000]{Muraishi2000}
        Muraishi, H. et al. 2000, \aap, 354, L57

\bibitem[Nikishov 1962]{Nikishov62}
	Nikishov, A. I. 1962, Sov. Phys. JETP 14, 2 

\bibitem[Norman et al. 1995]{Norman95}
	Norman, C. A., Melrose, D. B., \& Achterberg, A. 1995, \apj, 454, 60

\bibitem[Ostriker \& Steinhardt 1995]{OstrikerSteinhardt}
        Ostriker, J. \& Steinhardt, P. J. 1995, Nature, 377, 600

\bibitem[\"{O}zel \& Thompson 1996]{Ozel}
        \"{O}zel, M. E. \& Thompson, D. J. 1996, \apj, 463, 105

\bibitem[Peacock \& Heavens 1990]{Peacock}
        Peacock, J. A. \& Heavens, A. F. 1990, \mnras, 243, 133

\bibitem[Press \& Schechter 1974]{PressSchechter}
        Press, W. H. \& Schechter, P. 1974, \apj, 187, 425

\bibitem[Protheroe \& Stanev 1993]{Protheroe}
        Protheroe, R. J. \& Stanev, T. 1993, \mnras, 264, 191

\bibitem[Rees 1987]{Rees87}
	Rees, M. J. 1987, QJRAS 28, 197

\bibitem[Refregier \& Teyssier 2000]{rt01}
        Refregier, A. \& Teyssier, R. 2000, Phys. Rev. D, submitted
        [astro-ph/0012086]

\bibitem[Reynolds \& Ellison 1992]{Reynolds92}
        Reynolds, S. P. \& Ellison, D. C. 1992, \apj, 399, L75

\bibitem[Refregier et al. 2000]{refetal00}
        Refregier, A., Komatsu, E., Spergel, D. N., \& Pen, U.-L. 
        2000, Phys. Rev. D, 61, 123001

\bibitem[Rybicki \& Lightman 1979]{Rybicki}
        Rybicki, G. B. \& Lightman, A. P. 1979, 
        Radiative Processes in Astrophysics (John Wiley and Sons)

\bibitem[Ryu et al. 1993]{Ryu}
        Ryu, D., Ostriker, J. P., Kang, H., \& Cen, R. 1993, \apj, 414, 1

\bibitem[Sarazin 1999]{Sarazin99}
	Sarazin, C. 1999, \apj, 520, 529

\bibitem[Scharf \& Mukherjee 2002]{Scharf2002}
	Scharf, C. A. \& Mukherjee, R. 2002, ApJ, accepted [astro-ph/0207411] 

\bibitem[Springel \& Hernquist 2002a]{sh01a}
        Springel, V. \& Hernquist, L. 2002a, MNRAS, 333, 649

\bibitem[Springel \& Hernquist 2002b]{sh01b}
        Springel, V. \& Hernquist, L. 2002b, MNRAS, submitted, 
	[astro-ph/0206393]

\bibitem[Springel et al. 2001a]{GADGET}
        Springel, V., Yoshida, N., \& White, D. M. 2001a, NewA, 6, 79

\bibitem[Springel et al. 2001b]{Springel2001}
        Springel, V., White, M., \& Hernquist, L. 2001b, \apj, 549, 681

\bibitem[Sreekumar et al. 1998]{Sreekumar98}
        Sreekumar, P. et al. 1998, \apj, 494, 523

\bibitem[Stecker \& Salamon]{Stecker2001}
	Stecker, F. W. \& Salamon, M. H. 2001, in "Gamma 2001" ed. S. Ritz, 
	N. Gehrels and C. Schrader (New York: Amer. Inst. Phys.), 432

\bibitem[Steinmetz 1996]{Steinmetz}
        Steinmetz, M. 1996, \mnras, 278, 1005

\bibitem[Tanimori et al. 1998]{Tanimori98}
        Tanimori, T. et al. 1998, \apj, 497, L25

\bibitem[Totani \& Kitayama 2000]{Totani2000}
	Totani, T. \& Kitayama, T. 2000, \apj, 545, 572

\bibitem[Totani \& Inoue 2001]{Totani2001}
        Totani, T. \& Inoue, S. 2001, Astroparticle Physics, in
        press [astro-ph/0104072]

\bibitem[Watanabe et al. 1997]{Watanabe97}
	Watanabe, K. et al. 1997, in AIP Conf. Proc. 410, Fourth Compton 
	Symposium, ed. C. D. Dermer, M. S. Strickman, \& J. D. Kurfess 
	(New York: AIP)

\bibitem[Waxman \& Loeb 2000]{WaxmanLoeb2000}
        Waxman, E. \& Loeb, A. 2000, \apj, 545, L11

\end{thebibliography}
\end{document}